\documentclass[]{article}
\usepackage[a4paper,top=2cm,bottom=2cm,left=1.5cm,right=1.5cm,marginparwidth=1.75cm]{geometry}
\usepackage{authblk}
\usepackage[numbers, sort&compress]{natbib}
\usepackage[english]{babel}
\usepackage{url,hyperref,microtype}
\usepackage{amsmath}
\usepackage{amssymb}
\usepackage{mathtools}
\usepackage{graphicx}
\usepackage{xcolor}
\usepackage{arydshln}
\usepackage{pdfpages}
\usepackage{soul}
\usepackage{units}
\usepackage[shortlabels]{enumitem}
\usepackage{bbm}
\usepackage{subcaption}
\usepackage[percent]{overpic}
\usepackage{algorithm}
\usepackage[noEnd=true,endLComment=]{algpseudocodex}
\usepackage{tikz}
\usetikzlibrary{shapes.geometric, arrows}
\usepackage{multirow}
\usepackage{multicol}
\usepackage{rotating}

\newcommand{\ve}[1]{\text{\boldmath${#1}$}} 
\newcommand{\te}[1]{\text{\boldmath$\mathrm{#1}$}} 
\newcommand{\tefour}[1]{{\mathbbm{#1}}} 

\newcommand{\fevec}[1]{\ve{#1}} 
\newcommand{\femat}[1]{\mathrm{#1}} 

\newcommand{\grad}{{\mathrm{grad}\,}}
\newcommand{\grads}{\grad^{\!\mathrm{S}}}
\newcommand{\Grad}{{\mathrm{Grad}\,}}
\newcommand{\Div}{\mathrm{Div}\,}

\renewcommand{\top}{\mathrm{T}} 
\newcommand{\GradT}{\Grad^{\!\top}}	

\newcommand{\id}{\te{I}}
\newcommand{\tr}{\mathrm{tr}\,}
\renewcommand{\det}{\mathrm{det}\,}
\newcommand{\inv}[1]{{{#1}^{-1}}}
\newcommand{\invt}[1]{{{#1}^{-\top}}}

\newcommand{\Jpow}{J^{\nicefrac{-2}{3}}}

\newcommand{\Ot}{{\Omega_t}} 
\newcommand{\OO}{{\Omega_0}} 
\newcommand{\n}{\ve{n}} 
\newcommand{\N}{\ve{N}} 
\newcommand{\B}{\ve{B}} 
\newcommand{\GtD}{{\Gamma_t^\mathrm{D}}}
\newcommand{\GtN}{{\Gamma_t^\mathrm{N}}}
\newcommand{\GOD}{{\Gamma_0^{\mathrm{D}}}}
\newcommand{\GON}{{\Gamma_0^{\mathrm{N}}}}

\newcommand{\gD}{{\ve{g}_{D}}}
\newcommand{\hN}{{\ve{h}_{N}}}

\newcommand{\DDu}{{\mathrm{D}_{\ve{u}}}}

\newcommand{\eqq}{\coloneqq}
\newcommand{\motimes}{{\,\text{\boldmath{$\otimes$}}\,}}
\newcommand{\modot}{{\,\text{\boldmath{$\odot$}}\,}}

\newcommand{\store}[1]{{\color{black}{\underline{#1}}}}
\newcommand{\sstore}[1]{{\color{black}{\underline{\underline{#1}}}}}

\newcommand{\stable}[1]{{\color{black}{#1}}}

\newcommand{\lnp}{\ln_{+1}}
\newcommand{\Jm}{J_{-1}}

\newcommand{\draftAll}{false}
\newcommand{\draftStability}{\draftAll} 
\newcommand{\draftPerformance}{\draftAll} 
\newcommand{\draftIliac}{\draftAll} 
\newcommand{\draftOri}{\draftAll}  

\newtheorem{thm}{Theorem}[section]

\newtheorem{prob}[thm]{Problem}
\AfterEndEnvironment{prob}{\noindent\ignorespaces}

\usepackage{xparse}
\NewDocumentCommand\Cycle{O{} m m m O{} m}{%
	\draw[#1](#2.{#3+asin(#6/(#4*1.41))}) arc (180+#3-45:180+#3-45-270:#6/2) #5;
}

\usepackage[mode=multiuser,status=draft]{fixme} 
\fxsetup{innerlayout=inline}
\fxsetup{targetlayout=colorcb}
\fxsetup{marginface=\linespread{1}\scriptsize}

\fxusetheme{color}
\FXRegisterAuthor{mk}{envmk}{MK}

\title{Matrix-Free Higher-Order Finite Element Methods for Hyperelasticity}
\author[1]{R. Schussnig$^{\star,}$}
\author[2]{N. Fehn}
\author[3,4]{P. Munch}
\author[1]{M. Kronbichler}
\affil[$^\star$]{richard.schussnig@rub.de}
\affil[1]{Faculty of Mathematics, Ruhr University Bochum, Germany}
\affil[2]{Institute of Mathematics, University of Augsburg, Germany}
\affil[3]{Department of Information Technology, Uppsala University, Sweden}
\affil[4]{Institute of Mathematics, Technical University of Berlin, Germany}

\begin{document}

\maketitle


\begin{abstract}
	This work presents a matrix-free finite element solver for finite-strain elasticity adopting an $hp$-multigrid preconditioner. Compared to classical algorithms relying on a global sparse matrix, matrix-free solution strategies significantly reduce memory traffic by repeated evaluation of the finite element integrals. 
	
	Following this approach in the context of finite-strain elasticity, the precise statement of the final weak form is crucial for performance, and it is not clear a priori whether to choose problem formulations in the material or spatial domain. With a focus on hyperelastic solids in biomechanics, the arithmetic costs to evaluate the material law at each quadrature point might favor an evaluation strategy where some quantities are precomputed in each Newton iteration and reused in the Krylov solver for the linearized problem. Hence, we discuss storage strategies to balance the compute load against memory access in compressible and incompressible neo-Hookean models and an anisotropic tissue model. 
	Additionally, numerical stability becomes increasingly important using lower/mixed-precision ingredients and approximate preconditioners to better utilize modern hardware architectures.
	
	Application of the presented method to a patient-specific geometry of an iliac bifurcation shows 
	significant speed-ups, especially for higher polynomial degrees, when compared to alternative approaches with matrix-based geometric or black-box algebraic multigrid preconditioners.
	\\
	\textbf{Keywords: }finite-strain problem, matrix-free, finite-element method, hyperelasticity, geometric multigrid
\end{abstract}




\section{Introduction}

Implicit numerical solvers for nonlinear PDE problems in structural mechanics typically spend most of the time in the solution of linear systems of equations, arising from the linearization in Newton's method for nonlinear problems.
Classical matrix-based finite element solvers assemble a (sparse) system matrix and then, in a separate step, solve the linear system. During the solution phase, matrix-vector products (or related triangular matrix solves) are usually the most important computational kernels.
On current hardware, this is rooted in high memory traffic from loading matrix entries into compute units performing the arithmetic work. For higher polynomial degrees $p\geq2$, the increased number of entries per row leads to reduced throughput per unknown. Matrix-free algorithms, on the contrary, avoid storing the system matrix explicitly to reduce memory traffic, typically yielding improved performance for $p\geq2$. Whenever the numerical method (or the iterative techniques involved) need to evaluate the nonlinear or linearized operator, the matrix-free approach re-evaluates the spatial integrals of the finite element discretization by numerical quadrature rules, which we call on-the-fly evaluation in this text.
State-of-the-art matrix-free methods might further incorporate sum-factorization~\cite{Orszag1980}, SIMD vectorization (Single Instruction Multiple Data) over multiple elements~\cite{KronbichlerKormann2012, KronbichlerKormann2019}, and parallelization via domain decomposition~\cite{Burstedde2011}. This can possibly lead to higher performance, as memory traffic and operation counts are significantly reduced for higher-order methods. Comparing matrix-based operator evaluation with its matrix-free counterpart adopting sum factorization with polynomial degree $p$ and dimension $d$, operation counts and memory traffic reduce from an estimated $\mathcal{O}(p^{2d})$ to $\mathcal{O}(p^{d+1})$ and $\mathcal{O}(p^{d})$, respectively.


In structural mechanics, recent developments include the work by Davydov~et~al.~\cite{Davydov2020}, who considered a purely displacement-based approach for finite-strain hyperelasticity in a $h$-multigrid (geometric multigrid) setup. Brown~et~al.~\cite{Brown2022} and Mehraban~et~al.~\cite{Mehraban2021} in comparison adopt a $p$-multigrid (polynomial) approach with algebraic multigrid (AMG) coarse grid solvers, where the former leverages GPU acceleration. In~\cite{Fabien2020}, Fabien presents a hybridizable discontinuous Galerkin solver for linear elasticity (in first order form), combining GPU acceleration, $p$-multigrid and an AMG coarse grid solver. Kiran~et~al.~\cite{Kiran2023, Kiran2024a, Kiran2024b} tackle elastoplasticity with a focus on GPU implementations, whereas the resulting linear systems from lower-order discretizations are solved via \texttt{Ginkgo}~\cite{Ginkgo} or \href{http://cusplibrary.github.io/}{\texttt{Cusp}}.

The multigrid strategy used herein to precondition an outer Krylov method builds upon our previous developments~\cite{KronbichlerWall2018, Fehn2020, Fehn2021b, Munch2023}, using $hp$-multigrid with matrix-free operator evaluation in single precision, and an AMG-preconditioned Krylov solver as coarse-level solver. The implementations of the methods presented in this work are carried out in the software project \href{https://github.com/exadg/exadg}{\texttt{ExaDG}}~\cite{ExaDGgithub} (see \cite{Fehn2021b} for a comprehensive overview and \cite{Arndt2020} for the exa-scale project as a whole), which implements numerical solvers for many PDE model problems in computational fluid and structural dynamics and is based on the \texttt{deal.II}~\cite{dealII95} finite-element library and in particular its matrix-free infrastructure~\cite{KronbichlerKormann2012, KronbichlerWall2018}.

Matrix-free finite element solvers for hyperelastic problems bear great potential to speed up simulations in the context of biomechanics. In fact, soft biological tissue often serves as the prime example for anisotropic hyperelastic continua, and hence, advances in solver design can significantly reduce the simulation times for patient-specific geometries of the aorta or other vessels. This in turn immediately impacts medical device design, surgery planning or enables studies on virtual cohorts to derive statistically sound biomarkers, rendering solver design and advancements highly relevant for such applications.



Constitutive modeling plays a central role in these biomedical applications. The model by Holzapfel~et~al.~\cite{Holzapfel2015b} is considered herein. It incorporates non-symmetrically dispersed collagen fiber families reinforcing a nearly incompressible neo-Hookean ground material, which is key to capture the load-bearing behavior of arterial tissue and additionally confronts us with new challenges regarding fast integration due to its constituents. We adopt a purely displacement-based formulation similar to~\cite{Davydov2020, Brown2022}, where we enforce the incompressibility constraint via a penalty term. This standard approach might suffer from locking for high bulk moduli using linear finite elements. Higher-order finite element methods significantly reduce this problem~\cite{Heisserer2008, Suri1995, Suri1996, Radtke2017} and are hence particularly relevant for the target applications in biomechanics, as thin-walled, anisotropic and nearly incompressible structures are especially prone to locking. More involved alternative approaches are available with mixed displacement-pressure formulations~\cite{Stenberg1996, Canga1999, Maniatty2002, Farrell2021}, enhanced strain methods~\cite{Simo1990, Simo1992}, local pressure-projection methods~\cite{Chen1996a, Yu2012}, or non-conforming finite elements such as Crouzeix--Raviart or DG formulations~\cite{Falk1991, TenEyck2006, Kabaria2015, DiPietro2014}.

The contributions of this work are three-fold: First, we extend numerically stable formulations from Shakeri~et~al.~\cite{Shakeri2024} for the compressible and nearly incompressible neo-Hookean model to the tissue model and present forward stability test results. This is motivated by current hardware trends shifting towards mixed/low-precision arithmetic and the mixed-precision multigrid strategy used herein. Reformulating the weak form of standard structural mechanics problems in spatial configuration results in different formulations, which might simplify the terms to be integrated or allow for storing less linearization data at integration points. This leads to the second contribution, which is on analyzing these alternative formulations of linear(-ized) operators in terms of precomputing and storing linearization data at quadrature points for the compressible and nearly incompressible constitutive models and a fiber-reinforced tissue model~\cite{Holzapfel2015b}. Third, these ingredients are embedded into an $hp$-multigrid framework with matrix-free smoothing and level transfer, whose performance is compared to a matrix-based AMG preconditioner in a practically relevant problem. This last aspect thus bridges the gap between theoretical performance improvements and numerical stability considerations to the practically relevant setting, where the applicability of the introduced concepts is demonstrated.

This paper is organized as follows: Sec.~\ref{sec:continuum_mechanics} introduces finite-strain elasticity in the classical Lagrangian setting and an alternative formulation integrating over the spatial configuration. Thereafter, the relevant material models are introduced in Sec.~\ref{sec:constitutive_modeling}. The related weak forms are discussed in Sec.~\ref{sec:stable_numerics} in terms of their numerical stability, while precomputing strategies are detailed in Sec.~\ref{sec:storage_strategies}. The matrix-free $hp$-multigrid preconditioner adopting the proposed ingredients is discussed in Sec.~\ref{sec:matrix-free_solver}. Numerical tests then demonstrate improved stability for the proposed weak forms in Sec.~\ref{sec:results_stability}, evaluate the linear(-ized) operator in terms of its throughput and memory traffic in Sec.~\ref{sec:results_performance_block}, and showcase the framework's applicability to a patient-specific iliac bifurcation with physiological parameters in Sec.~\ref{sec:results_iliac}. A summary and conclusions are then given in Sec.~\ref{sec:summary_conclusion}.

\section{Continuum mechanics and finite element solver}
\label{sec:continuum_mechanics}

This section presents standard relations from continuum mechanics to derive the nonlinear boundary value problem related to finite-strain elasticity and follows classical literature (see, e.g.,~\cite{Holzapfel2000, Bonet2008}). 
We are interested in finding the map from the material to the spatial configuration $\ve{\phi} : \OO \rightarrow \Ot$. It connects points $\ve{X} \in \OO$ to points $\ve{x} \in \Ot$, that is, maps material coordinates in a body's material or undeformed configuration $\OO$ to spatial coordinates in the body's spatial or deformed configuration $\Ot$. Introducing a displacement field $\ve{u}$, we can express the map as
\begin{equation*}
\ve{\phi}(\ve{X}) = \ve{X} + \ve{u}(\ve{X})
,
\quad
\text{with}
\quad
\ve{u}(\ve{X}) = \ve{x}(\ve{X}) - \ve{X}
.
\end{equation*}
The boundaries of both material and spatial configurations with unit outward normals $\N$ and $\n$, respectively, are decomposed into non-overlapping Dirichlet and Neumann parts, $\partial\OO = \GOD \cup \GON$ and $\partial \Ot = \GtD \cup \GtN$. We further introduce the deformation gradient $\te{F}$ and its determinant, referred to as the Jacobian $J$,
\begin{align*}
    \te{F} \eqq \id + \Grad \ve{u} 
    ,
    \quad
    J\eqq \det{\te{F}}
    .
\end{align*}
Within this contribution, we denote with $\Grad(\cdot)$ the gradient with respect to the material coordinates $\ve{X}\in\OO$, while its spatial counterpart, $\grad(\cdot)$, denotes the gradient with respect to spatial coordinates $\ve{x}\in\Ot$. 

Following the classical Lagrangian approach, the static linear momentum balance in material configuration $\OO$ reads
\begin{equation}
    \label{eqn:mom_bal_strong}
 	- \Div \te{P} 
 	= 
 	\B(\ve{X}) \quad \text{in }\OO
 	,
\end{equation}
with $\Div(\cdot)$ being the divergence with respect to $\ve{X}$, a given body force $\B\in [L^2(\OO)]^d$, the first Piola--Kirchhoff stress tensor $\te{P} \eqq \te{F}\,\te{S}$, and the second Piola--Kirchhoff stress tensor $\te{S}$. 
Eqn.~\eqref{eqn:mom_bal_strong} is independent of the material model and the underlying constitutive relation. We further introduce strain measures being the right Cauchy--Green tensor $\te{C}$ and the Green--Lagrange strain tensor $\te{E}$,
\begin{align*}
	\te{C} \eqq \te{F}^\top\te{F}
	,
	\quad 
	\te{E} \eqq \nicefrac{1}{2} \left(\te{C} - \te{I}\right)
	,
\end{align*}
to define $\te{S}(\te{E})$ or $\te{S}(\te{C})$, capturing the material behavior (see Sec.~\ref{sec:constitutive_modeling}). Finally, Eqn.~\eqref{eqn:mom_bal_strong} is equipped with suitable Dirichlet and Neumann boundary conditions, $\ve u\rvert_{\GOD} = \gD \in [H^{\nicefrac{1}{2}}(\GOD)]^d$ and $\te P \ve N \rvert_\GON = \hN \in [H^{-\nicefrac{1}{2}}(\GON)]^d$, to close the system.
%
%
Employing a standard displacement-based finite element formulation, the residual in weak form then reads
\begin{gather}
		\label{eqn:mom_bal_nonlinear_material_residual}
		r_\OO(\ve v, \ve u )
		\eqq
		\left(
		\Grad\ve{v}
		,
		\te{P}
		\right)_\OO
		-
		\left(
		\ve{v}
		,
		\hN 
		\right)_\GON
		-
		\left(
		\ve{v}
		,
		\B
		\right)_\OO
		,
\end{gather}
where $(\cdot, \cdot)_\OO$ denotes the standard inner product of the two arguments integrated over the given domain. To state the weak forms related to the nonlinear finite strain elasticity problem, define the vector-valued Sobolev spaces
\begin{align*}
	H^1_{\gD}(\OO) 
	\eqq
	\left\{ \ve{v} \in [H^1(\OO)]^d : \ve{v}\rvert_{\GOD} = \gD\right\}
	\qquad\text{and}\qquad
	H^1_0(\Omega) 
	\eqq
	\left\{ \ve{v} \in [H^1(\OO)]^d : \ve{v}\rvert_{\GOD} = \ve{0}\right\}
\end{align*}
of square integrable functions with square integrable first derivatives on $\OO$. Similar definitions for respective counterparts defined on $\Ot$ are omitted for brevity. In material configuration, this leads to
\begin{prob}
	\label{prob:nonlinear_material_residual}
	Find $\ve u \in H^1_{\gD}(\OO)$, such that
	\begin{gather}
			\label{eqn:mom_bal_nonlinear_material}
			r_\OO(\ve v , \ve u)
			=
			0
			\quad \forall \ve v \in H^1_0(\OO)
			.
	\end{gather}
\end{prob}
Regarding well-posedness of Problem~\ref{prob:nonlinear_material_residual} under suitable assumptions, we refer the reader to~\cite{Ball1976, Ciarlet2002, Ruas1989}, noting that the existence of minimizers 
of the related energy functional cannot be guaranteed for general hyperelastic materials and load configurations. %
Herein, we employ standard $C^0$-continuous finite-dimensional subspaces 
and Newton's method to solve the nonlinear problem~\eqref{eqn:mom_bal_nonlinear_material}. In Newton's method, each iteration updates the initial guess $\fevec{u}_0$ via $\fevec{u}_{k+1} = \fevec{u}_k + \Delta \fevec{u}$, $k=0,\dots, N_\mathrm{max}$
starting from $\fevec{u}_0$ with $\ve{u}_0\rvert_\GOD = \gD$. The increment $\Delta \fevec{u}$ is obtained by solving
\begin{prob}
	\label{prob:linear_material_Newton_step}
	Find $\Delta \fevec{u} \in H^1_0(\OO)$ given the previous iterate $\fevec{u}_k \in H^1_\gD(\OO)$ by solving
	\begin{equation}
			\label{eqn:newton_material}
			\left.
			\DDu r_\OO(\ve{v}, \Delta \ve{u})
			\right\rvert_{\ve{u}_k}
			=
			-r_\OO(\ve{v}, \ve{u}_k)
			\quad \forall \ve v \in H^1_0(\OO)
			.
	\end{equation}
\end{prob}
The directional derivative corresponding to Eqn.~\eqref{eqn:mom_bal_nonlinear_material_residual} reads
\begin{gather}
	\label{eqn:DDu_residual_material}
	\left.
	\DDu r_\OO(\ve{v}, \Delta \ve{u})
	\right\rvert_{\ve{u}_k}
	=
	\left(
	\Grad \ve{v}
	,
	(\DDu \te{F}) \, {\te{S}}
	+
	{\te{F}} \, \DDu\te{S}
	\right)_\OO
	,
\end{gather}
where all terms are evaluated using $\ve{u}_k$. A basic Newton method is provided in Alg.~\ref{alg:newton_solver}, which for the sake of brevity does not contain a line-search algorithm or load stepping procedure. It highlights the update of stored quadrature point data of central interest within this work, see Sec.~\ref{sec:storage_strategies}.
\begin{algorithm}
	\begin{algorithmic}[1]
		\caption{Generic Newton's method to solve $\fevec{r}(\fevec{u}) = \fevec{0}$ with quadrature point data update.}
		\label{alg:newton_solver}
		\Function{NewtonSolver}{$\fevec{u}_0$, $\epsilon_\mathrm{abs}$, $\epsilon_\mathrm{rel}$, $N_\mathrm{max}$}
		\State $\fevec{u}_0 \gets \gD$ on $\GOD$
		\Comment{initialize iterate and enforce Dirichlet conditions}
		\State $k = 0$
		\Comment{initialize counter}
		%
		%
		\While {$||\fevec{r}(\fevec{u}_k)|| > \epsilon_\mathrm{abs}$
			and
			$||\fevec{r}(\fevec{u}_k)|| > \epsilon_\mathrm{rel} \, ||\fevec{r}(\fevec{u}_0)||$
			and
			$k<N_\mathrm{max}$
		}
		\State update quadrature point data
		\Comment{see Sec.~\ref{sec:storage_strategies}}
		\State solve for Newton update: $\femat{K} \, \Delta \fevec{u} = - \fevec{r}(\fevec{u}_k)$
		\State $\fevec{u}_k \gets \fevec{u}_k + \Delta \fevec{u}$
		\Comment{update iterate}
		\State $k\gets k +1$
		\Comment{update iteration counter}
		\EndWhile
		\State \Return $\fevec{u}_k$, $k-1$
		\Comment{return last iterate and number of iterations}
		\EndFunction
	\end{algorithmic}
\end{algorithm}

The integrals in the weak form of the Newton update step~\eqref{eqn:newton_material} can be equivalently written as integrals over the spatial configuration $\Ot$. Specifically, one may rewrite the domain integral involving the stress, $\left(\Grad \ve{v}, \te{P}\right)_\OO$, keeping the Neumann data and body force unchanged. Introducing the Kirchhoff stress $\te{\tau}\eqq\te{F}\,\te{S}\,\te{F}^\top$ yields
\begin{gather*}
	\left(
	\Grad \ve{v}
	,
	\te{P}
	\right)_\OO
	=
	\left(
	\Grad \ve{v}
	,
	\te{\tau}
	\,
	\invt{\te{F}}
	\right)_\OO	
	=
	\left(
	\left(
		\Grad \ve{v}
	\right)
	\inv{\te{F}}
	,
	\te{\tau}
	\right)_\OO	
	=
	\left(
	\grad \ve{v}
	,
	\te{\tau}
	\right)_\OO
	=
	\left(
	\nicefrac{1}{J}\,\,
	\grad \ve{v}
	,
	\te{\tau}
	\right)_\Ot	
	=
	\left(
	\nicefrac{1}{J}\,\,
	\grads \ve{v}
	,
	\te{\tau}
	\right)_\Ot	
	,
\end{gather*}
using $\te{P}=\te{F}\,\te{S}$, $\grad \ve{v} = \left(\Grad \ve{v}\right) \inv{\te{F}}$, the notation $(\cdot)^\mathrm{S} \eqq \nicefrac{1}{2}\left[(\cdot) + (\cdot)^\top\right]$ for the symmetric part of a tensor, and symmetry of $\te\tau$ in the last step. The residual then reads
\begin{gather}
		\label{eqn:mom_bal_nonlinear_spatial_residual}
		r_\Ot(\ve v, \ve u )
		\eqq
		\left(
		\nicefrac{1}{J}\,\,
		\grads \ve{v}
		,
		\te{\tau}
		\right)_\Ot
		-
		\left(
		\ve{v}\left(\ve{\phi}\left(\ve{X}\right)\right)
		,
		\hN
		\right)_\GON
		-
		\left(
		\ve{v}\left(\ve{\phi}\left(\ve{X}\right)\right)
		,
		\B
		\right)_\OO
		,
\end{gather}
where $\ve{v}\left(\ve{\phi}\left(\ve{X}\right)\right)$ is equivalent to the test functions in the reference domain $\OO$ and independent of $\ve{u}$.
The spatial counterpart of the weak form of the nonlinear finite strain elasticity problem, Problem~\ref{prob:nonlinear_material_residual}, reads	
	\begin{prob}
		Find $\ve u \in H^1_\gD(\Ot)$, such that
		\begin{gather*}
				r_\Ot(\ve v , \ve u)
				=
				0
				\quad \forall \ve v \in H^1_0(\Ot).
		\end{gather*}
	\end{prob}
Integrating over the spatial domain $\Ot$, the increment $\Delta \fevec{u}$ is obtained in each iteration of Newton's method by solving
	\begin{prob}
		\label{prob:linear_spatial_Newton_step}
		Find $\Delta \fevec{u} \in H^1_0(\Ot)$ given the previous iterate $\fevec{u}_k \in H^1_\gD(\Ot)$ by solving
		\begin{equation*}
				\left.
				\DDu r_\Ot(\ve{v}, \Delta \ve{u})
				\right\rvert_{\ve{u}_k}
				=
				-r_\Ot(\ve{v}, \ve{u}_k)
				\quad \forall \ve v \in H^1_0(\Ot)
				.
		\end{equation*}
	\end{prob}
The directional derivatives of the stress term are given by
\begin{gather*}
	\DDu 
	\left(
	\nicefrac{1}{J}\,\,
	\grads \ve{v}
	,
	\te{\tau}
	\right)_\Ot	
	=
	\DDu
	\left(
	\grads \ve{v}
	,
	\te{\tau}
	\right)_\OO
	=
	\left(
	\grads \ve{v}
	,
	\DDu
	\te{\tau}
	\right)_\OO
	+
	\left(
	\left(
	    \Grad \ve{v}
	\right)
	\DDu 
	\inv{\te{F}}
	,
	{\te{\tau}}
	\right)_\OO
	.
\end{gather*}
Introducing the contravariant push-forward of the fourth-order material part of the stiffness tensor denoted as $J\tefour{c}$ \cite{Wriggers2008, Holzapfel2000} and transformed onto $\Ot$, this can be rewritten as
\begin{gather}
	\left(
	{\grad}\ve{v}
	,
	J \tefour{c} : \grads \Delta \ve{u}
	\right)_\OO
	+
	\left(
	{\grad}\ve{v}
	,
	\left(
		{\grad}\Delta \ve{u}
	\right)
	{\te{\tau}}
	\right)_\OO
	=
	\left(
	\nicefrac{1}{J}\,\,
	{\grad}\ve{v}
	,
	J \tefour{c} : \grads \Delta \ve{u}
	+
	\left(
		{\grad}\Delta \ve{u}
	\right)
	{\te{\tau}}
	\right)_\Ot
	.
	\label{eqn:stress_term_in_Ot_with_JC}
\end{gather}
Note that the actual fourth-order tensor $\tefour{c}$ need not be computed, but rather its action on a symmetric second-order tensor to evaluate $J\tefour{c}:(\cdot)^\mathrm{S}$ in Eqn.~\eqref{eqn:stress_term_in_Ot_with_JC}. Again, \eqref{eqn:mom_bal_nonlinear_spatial_residual} and~\eqref{eqn:stress_term_in_Ot_with_JC} are independent of the material model, which enters via $\te{\tau}$ and $\tefour{c}$ to be discussed in Sec.~\ref{sec:constitutive_modeling}.

Following the strategy of spatial integration comes at the cost of updating the vertex positions of the finite element mesh and related data. When updating the quadrature point data in Alg.~\ref{alg:newton_solver}, the spatial grid, which approximates the body in its deformed configuration $\Ot$, is updated simultaneously.

\section{Constitutive modeling}
\label{sec:constitutive_modeling}

The second Piola--Kirchhoff stress tensor in the residuals~\eqref{eqn:mom_bal_nonlinear_material_residual} or \eqref{eqn:mom_bal_nonlinear_spatial_residual} is defined in terms of a strain measure, i.e., $\te{S}=\te{S}(\te{C})$ or $\te{S}=\te{S}(\te{E})$. For hyperelastic continua, the constitutive relation is expressed in terms of the strain-energy density $\Psi$ (per unit reference volume),
\begin{equation*}
		\inv{\te{F}}\te{P}
		=
		\te{S} 
		\eqq 
		\frac{\partial \Psi\left(\te{E}\right)}{\partial \te{E}} 
		= 
		2 \frac{\partial \Psi\left(\te{C}\right)}{\partial \te{C}}
		.
\end{equation*}
A compressible neo-Hookean model (cNH) is given by~\cite{Treloar1976}, 
\begin{align}
	\label{eqn:NH_compressible}
	\Psi_\mathrm{cNH}(\te{C}) 
	= 
	\nicefrac{\mu}{2}\left( 
	I_1 - \tr\id - 2 \ln J
	\right)
	+
	\lambda\ln^2 J
	,
	\quad
	\te{S}_\mathrm{cNH}
	=
	\mu \id - \left( \mu - 2\lambda \ln J\right) \inv{\te{C}}
	,
\end{align}
with the first invariant $I_1 \eqq \tr \te{C}$, and the material parameters being the shear modulus $\mu$ and Lam\'e coefficient $\lambda$. For the nearly incompressible neo-Hookean model (iNH) yielding $J \approx 1$, the deformation gradient is split into isochoric and volumetric parts according to Flory~\cite{Flory1961}, yielding
\begin{align}
	\label{eqn:NH_nearly_incompressible}
	\Psi_\mathrm{iNH}(\te{C})
	=
	\nicefrac{\mu}{2}
	\left(
	\Jpow I_1 - \tr\id
	\right)
	+
	\nicefrac{\kappa_b}{4}
	\left(
	J^2-1-2\ln J
	\right)
	,	
	\quad
	\te{S}_\mathrm{iNH}
	=
	\mu \Jpow \id + \left[ \nicefrac{\kappa_b}{2} (J^2-1) - \nicefrac{\mu}{3} \Jpow I_1 \right] \inv{\te{C}}
	,
\end{align}
with bulk modulus $\kappa_b$ enforcing $J=1$ as $\kappa_b\rightarrow\infty$, acting as a penalty term. 

Now, for the target applications in biomedical engineering and medicine, more involved constitutive relations are required to capture the material behavior. Aortic tissue of prime interest within this work shows an anisotropic stiffening effect under large strains due to collagen fibers reinforcing the ground material. The model by Holzapfel~et~al.~\cite{Holzapfel2015b}, herein simply referred to as the fiber model (fiber), adds exponential terms to the strain-energy density. The main motivation for this choice lies in the fact that we aim to showcase and analyze the potential performance impact in realistic scenarios. Our performance improvements hence directly translate to problems of high practical relevance, which has not been addressed in literature so far.
The model includes two fiber families and allows accounting for the non-symmetric fiber dispersion, which is more significant in the tangential plane compared to the out-of-plane direction~\cite{Schriefl2012a}. The strain-energy density combines the neo-Hookean ground material $\Psi_\mathrm{iNH}$ given in Eqn.~\eqref{eqn:NH_nearly_incompressible} and collagen fiber contributions via $\Psi_\mathrm{c}$ as
\begin{align}
	\label{eqn:fiber_Psi}
	\Psi_\mathrm{fiber}(\te{C},\te{H}_i)
	=
	\Psi_\mathrm{iNH}(\te{C})
	+
	\sum_{i=4,6}
	\Psi_\mathrm{c}(\te{C}, \te{H}_i)
	=
	\Psi_\mathrm{iNH}(\te{C})
	+
	\sum_{i=4,6}
	\begin{cases}
		\frac{k_1}{2 k_2} 
		\left[ \exp\left(k_2 E_i^2\right) - 1 \right] & \text{if $I_i^\star>1$},\\
		0 & \text{else},
	\end{cases}
\end{align}
where $k_1$ is a stiffness-like parameter and $k_2$ a dimensionless shape parameter. 
The strain energy related to collagen fibers $\Psi_\mathrm{c}$ only contributes to the total energy $\Psi$, when the squared fiber stretches defined as 
\begin{align*}
		I_i^\star
		=
		(\ve{M}_1 \otimes \ve{M}_1) : \te{C},
		\quad
		i = 4,6,
\end{align*}
signal tension, i.e., when $I_4^\star > 1$ or $I_6^\star > 1$, while fiber bundles with compressed mean fiber buckle immediately (cf.~\cite{Holzapfel2015}).
The symmetric structure tensor $\te{H}_i$ and strain-like quantity $E_i$ are defined as
\begin{align}
	\label{eqn:fiber_Ei}
	\te{H}_i 
	=
	H_{11} \ve{M}_1\otimes\ve{M}_1 
	+ 
	H_{22} \ve{M}_2\otimes\ve{M}_2
	+ 
	H_{33} \ve{M}_3\otimes\ve{M}_3
	,
	\quad
	E_i 
	= \te{H}_i : (\te{C} - \te{I})
	= \tr (\te{H}_i \te{C}) - 1
	.
\end{align}
For the targeted applications in biomechanics and medicine, the orthonormal basis spanned by $\ve{M}_1$, $\ve{M}_2$ and $\ve{M}_3$ is related to tailored material coordinate systems in the reference configuration~\cite{Schussnig2022_a, Schussnig2021_a, Schussnig2021_b}. Given a suitable material coordinate system, the mean in-plane and out-of-plane angles $\Phi$ and $\Theta = 0$ then yield
\begin{align*}
	\ve{M}_1 = \ve{E}_1 \cos\Phi - \ve{E}_1 \sin\Phi
	,
	\quad
	\ve{M}_2 = \ve{E}_2 \sin\Phi + \ve{E}_2 \cos\Phi
	,
	\quad
	\ve{M}_3 = \ve{E}_3
	,
\end{align*}
describing the classical helical patterns of collagen fibers in vascular tissue. Here, $\ve{E}_1$ aligns with the circumferential direction, $\ve{E}_2$ with the longitudinal direction, and $\ve{E}_3$ with the radial direction. 
This specific representation of $\te{H}_i$ in~\eqref{eqn:fiber_Ei} assumes a bivariate von Mises distribution of the fiber density~\cite{Holzapfel2015b}. Multiplicative decomposition then leads to only three nonzero components of the generalized structure tensor in the reference configuration,
\begin{gather*}
	H_{11} 
	= 
	\frac{1-H_{33}}{2} 
	\left(  
	1 + \frac{\mathcal{I}_1(a)}{\mathcal{I}_0(a)} 
	\right)
	,
	\quad
	H_{22} 
	=
	\frac{1-H_{33}}{2} 
	\left(  
	1 - \frac{\mathcal{I}_1(a)}{\mathcal{I}_0(a)} 
	\right) 
	,
	\quad 
	H_{33} 
	= 
	\frac{1}{4 b} - \frac{\exp(-2b)}{\sqrt{2\pi b} \, \mathrm{erf}(\sqrt{2b})}
	,
\end{gather*}
where $\mathcal{I}_0$ and $\mathcal{I}_1$ denote the Bessel functions of the first kind of orders 0 and 1, respectively, and $\mathrm{erf}(\cdot)$ is the error function, see~\cite{Wollner2020} for details. Thus, the parameters $a$ and $b$ together with the mean in-plane angle $\Phi$ describe the dispersion of the collagen fibers based on the material coordinate system spanned by $\ve{E}_1$, $\ve{E}_2$ and $\ve{E}_3$. 

This leads to the second Piola--Kirchhoff stress tensor $\te{S}_\mathrm{fiber}$ as a sum of the collagen fiber $\te{S}_\mathrm{c}$ and nearly incompressible neo-Hookean $\te{S}_\mathrm{iNH}$~\eqref{eqn:NH_nearly_incompressible} contributions, that is
\begin{align*}
	\te{S}_\mathrm{fiber}
	=
	\te{S}_\mathrm{iNH}(\te{C})
	+
	\te{S}_\mathrm{c}(\te{C},\te{H}_i)
	=
	\te{S}_\mathrm{iNH}(\te{C})
	+
	\sum_{i=4,6}
	2 k_1 \exp\left(k_2 E_i^2\right) E_i \, \te{H}_i
	,
\end{align*}
where physiological tissue parameters used in this work~\cite{Weisbecker2012, RolfPissarczyk2021b} are summarized in Tab.~\ref{tab:fiber_parameters}.
\begin{table}
	\centering
	\caption{Parameters for the nearly incompressible fiber model~\cite{Holzapfel2015b} fit to material tests of aortic medial tissue, taken from~\cite{Weisbecker2012, RolfPissarczyk2021b}. The bulk modulus $\kappa_b$, i.e., the penalty term in purely displacement-based formulations corresponds to a Poisson's ratio of $0.49$.}
	{
		\scriptsize
		\label{tab:fiber_parameters} 
		\begin{tabular}{||c c c c c c c c c c||}
			\hline
			&&&&&&&&&\\[-1.5ex]
			$\mu$ [kPa] & $\kappa_b$ [kPa] & $a$ [-] & $b$ [-] & $k_1$ [kPa] & $k_2$ [-] & $\Phi$ [$^\circ$]& $H_{11}$ [-] & $H_{22}$ [-]& $H_{33}$[-]\\[0.75ex]
			\hline\hline
			&&&&&&&&&\\[-1.25ex]
			62.1 & 3084.3 & 3.62 & 34.3 & 1.4 & 22.1 & 27.47 & 0.9168 & 0.0759 & 0.0073 \\[0.75ex]
			\hline
		\end{tabular}
	}
\end{table}
The second Piola-Kirchhoff stress tensor and the related directional derivative are the only ingredients in the Newton solver specific to the constitutive model and can be found in App.~\ref{sec:appendix}.

\section{Stable and Fast Numerics for Hyperelasticity}
\label{sec:stable_numerics}

The momentum balance residuals~\eqref{eqn:mom_bal_nonlinear_material_residual} and~\eqref{eqn:mom_bal_nonlinear_spatial_residual} and their respective directional derivatives feature terms that may suffer from significant numerical instability. Especially when considering ongoing and expected future hardware-driven developments towards mixed- and low-precision strategies~\cite{Abdelfattah2021}, the classical formulations have to be reviewed. In the following, we extend numerically stable forms of the strain energy density and the stresses from Shakeri~et~al.~\cite{Shakeri2024} towards the fiber model and furthermore present stable forms of the directional derivatives, which have not been reported in literature yet. Furthermore, we present a fast evaluation scheme for $\Jpow$ based on Newton's method.

Starting with the strain measures and their evaluation, the Green--Lagrange strain tensor defined as
\begin{align*}
	\te{E} 
	\eqq 
	\nicefrac{1}{2} \left( \te{C} - \id \right)
	=
	\nicefrac{1}{2} \left( \te{F}^\top \te{F} - \id \right)
	,
\end{align*}
shows cancellation in the small strain regime since components of $\te{F}$, which is close to the unit tensor $\id$, are used in floating-point operations, followed by a subtraction of the unit tensor. To reduce loss of accuracy, we evaluate $\te{E}$ as 
\begin{align}
	\label{eqn:E_stable}
	\te{E}
	 = 
	\nicefrac{1}{2}\left( \Grad\ve{u} + \GradT\ve{u} + \GradT\ve{u} \,\, 
	\Grad\ve{u}\right)
	.
\end{align}
Similar reasoning lies behind computing the Green--Euler strain tensor according to
\begin{align*}
	\te{\tilde{b}} 
	\eqq 
	\nicefrac{1}{2} \left(\te{b} - \id\right)
	=
	\nicefrac{1}{2} \left(\te{F}\,\te{F}^\top - \id\right)
	=
	\nicefrac{1}{2} \left(\Grad \ve{u} + \GradT \ve{u} + \Grad \ve{u} \,\, \GradT \ve{u}\right)
	,
\end{align*}
where the left Cauchy--Green tensor $\te{b} \eqq \te{F}\,\te{F}^\top$, suffering from similar cancellation for $\te{F} \approx \id$, is avoided. Reformulations of the strain measures also affect the fiber invariants, which are evaluated as
\begin{gather*}
		E_i 
		\eqq 
		\te{H}_i : (\te{C} - \id)	
		= 
		\stable{ 
			2 \, \te{H}_i : \te{E} 
		}
		,
		\quad
		I_i^\star 
		\eqq 
		\left(\ve{M}_1 \otimes \ve{M}_1\right) : \te{C} 
		=
		\stable{
			2 \left(\ve{M}_1 \otimes \ve{M}_1\right) : \te{E} + \tr{\left(\ve{M}_1 \otimes \ve{M}_1\right)}
		}
		.
\end{gather*}

Further recurring terms in finite-strain (hyper-)elasticity are, e.g., the inverse right Cauchy--Green tensor $\inv{\te{C}}$. Cancellation can be reduced by inverting $\te{F}$, where $\mathrm{cond}(\te{F}) = \sqrt{\mathrm{cond}\left(\te{C}\right)}$, and computing
$\inv{\te{C}} 
=
\inv{\te{F}}
\,
\invt{\te{F}}$.
rather than inverting $\te{C}$.
Note that $\inv{\te{F}}$ is often needed anyways, such that this approach also requires fewer arithmetic operations.

The stress tensor of nearly incompressible continua and its derivative also contain $(J-1)$ or similar terms. Since we aim to enforce $J = 1$ through a penalty term scaled by the bulk modulus $\kappa_b$ as in Eqn.~\eqref{eqn:NH_nearly_incompressible}, we inevitably have $J\approx 1$ as the bulk modulus increases. As pointed out by Shakeri~et~al.~\cite{Shakeri2024}, numerical stability can be improved significantly by introducing $\Jm \eqq J-1 = \det(\te{F}) - 1$ and avoiding forming $\det(\te{F})$ first and subtracting 1 from the result, but instead using the components of $\Grad\ve{u}$ directly.
This can further be exploited to rewrite $(J^2-1)$ in a more stable manner as $\Jm (\Jm + 2)$.

Compressible and (nearly) incompressible hyperelastic continua also contain more complex functions taking the Jacobian as an argument. Within the current work, $\ln(J)$ and $\Jpow$ are of interest both in terms of numerical stability {and} computational efficiency. Herein, SIMD vectorization is enabled straight-forwardly via polynomial or rational approximations to process data on all lanes of a SIMD vector. In addition to that, restricting input argument ranges allows for further optimizations. For $\ln(J)$, we define~\cite{Shakeri2024, Beebe2017}
\begin{align*}
	\ln(J) 
	= 
	\ln(\Jm + 1)
	=
	\lnp(\Jm)
	,
	\quad
	\lnp(x) 
	\eqq 
	2 \sum_{n=0}^\infty 
	\frac{1}{2n+1}
	\left(
		\frac{x}{2+x}
	\right)^{2n+1}
	.
\end{align*}
Note that the sum only contains odd powers of $x\in(-1,\infty)$, such that only terms with the same sign are added, guaranteeing numerical stability when summing a fixed number of terms small to large.

For (nearly) incompressible continua, the term $\Jpow$ plays a central role, as its evaluation can be costly. Possible options are: i) looping over SIMD vectors and resorting to standard techniques for scalar types, ii) exploiting the floating point representation and approximation via summation (see \cite{Schraudolph1999}), or iii) a Newton solver given a good initial guess $J\approx1 \Leftrightarrow \Jpow \approx 1$. Depending on the storage strategy used to evaluate the integrals, we consider iii) for on-the-fly evaluation, but i) or ii) in case $\Jpow$ is stored anyways, see Sec.~\ref{sec:storage_strategies}.  

A Newton solver for $x = \Jpow
\Leftrightarrow f(x) 
\eqq 
x^{-3} - J^2 
= 
0$ uses $f^\prime(x) = -3x^{-4}$, leading to
\begin{align}
	\label{eqn:Jpow_Newton}
	x_{k+1} = x_{k}
	-
	\frac{f(x_{k})}{f^\prime(x_{k})}
	= 
	\nicefrac{1}{3}
	\left(
		4 x_{k} - J^2 x_{k}^4
	\right)
\end{align}
for $k=0,\dots,N$ starting from $x_0 = 1$ or some previously computed $\Jpow$. Since for the undeformed initial state of the elastic structure we have 
$\ve{u}=\ve{0}\Rightarrow\te{F}=\id \Rightarrow J=1$
and due to incompressibility, $J \approx 1$ holds throughout the entire motion, we employ a fixed number of 3 Newton iterations according to Alg.~\ref{alg:Jpow_Newton}. This is justified by local quadratic convergence of Newton's method and an excellent initial guess exploiting $\Jpow\approx1$.

\begin{algorithm}
	\begin{algorithmic}[1]
		\caption{Approximation of $\Jpow$ using $N$ Newton steps~\eqref{eqn:Jpow_Newton}}
		\label{alg:Jpow_Newton}
		\Function{FastApproxJpow}{$\Jm$, $N$}
		\State $\alpha = \nicefrac{1}{3}\left[\left(\Jm + 2\right) \Jm + 1\right]$
		\Comment{store $\nicefrac{1}{3}\,J^2$}
		\State $\beta = \nicefrac{4}{3} - \alpha$
		\Comment{assignment is first iteration w. $x_{0} = 1$}
		\For {$n=1,\dots,N-1$}
		    \State $\gamma = \beta^2\,\beta^2$
		    \Comment{$\gamma = x_k^4$ by repeated squaring}
		    \State $\beta \gets \nicefrac{4}{3} \, \beta - \alpha \gamma$
		\EndFor
		\State \Return $\beta$
		\EndFunction
	\end{algorithmic}
\end{algorithm}

Similar to $\ln(J)$ and $\Jpow$, the last ingredient required with regards to the material models considered within this work is a SIMD-compatible $\exp(x)$ appearing in the fiber model~\eqref{eqn:fiber_Psi}. Here, we adopt the approach by Proell~et~al.~\cite{Proell2024b}, which is based on~\cite{Schraudolph1999, Malossi2015, Perini2018}, and exploits the floating point representation for fast evaluation.

\section{Storage Strategies for Hyperelasticity}
\label{sec:storage_strategies}

Contrary to matrix-based algorithms, the integrals involved in the residual and its linearization are evaluated repeatedly. For achieving optimal performance, the memory access and arithmetic workload of these operations need to be compared to the capabilities of the underlying hardware. Depending on the material models' complexity, precomputing and storing data on the integration point level might be beneficial if the in-core resources such as arithmetic units or available registers are highly busy. More precisely, we aim to identify certain data that we compute once per nonlinear Newton iteration, store it at the integration points, and load it during operator evaluation, see Alg.~\ref{alg:newton_solver}, whereas other quantities are computed repeatedly in each matrix-vector product.
Based on the observations from Davydov~et~al.~\cite{Davydov2020}, we introduce three stages in this regard: i) on-the-fly integral evaluation of all terms, ii) precompute and store scalar quantities where useful, and iii) precompute and store scalars and second-order tensors where useful. Additionally, the symmetry of stresses and strains and certain intermediate quantities is exploited both in terms of memory consumption and when performing operations such as addition, multiplication, double contraction, push-forward and others.

In the following, the final weak forms as derived in App.~\ref{sec:appendix} are presented in stable form besides the classical form, while quantities to be stored are highlighted as $\store{(\cdot)}$ if they are scalar, and as $\sstore{(\cdot)}$ if they are second-order (possibly symmetric) tensors. Here, the stable form of the fiber model, the linearizations using integration over the spatial configuration, and the storage strategies in particular are novel contributions. Common to all constitutive models is the Newton update step in Alg.~\ref{alg:newton_solver}. In material configuration, we seek $\Delta \ve{u} \in H^1_0(\OO)$, such that there holds
\begin{gather*}
	\left(
	\Grad \ve{v}
	,
	\left( \Grad \Delta \ve{u}\right) \sstore{\te{S}}
	+
	\sstore{\te{F}} \,
	\DDu \te{S}
	\right)_\OO
	=
	\left(\ve{v},\hN\right)_\OO
	-
	\left(
	\Grad\ve{v},
	\sstore{\te{F}}
	\, \sstore{\te{S}}
	\right)_\OO
	+
	\left(\ve{v},\B\right)_\OO
	\quad
	\forall \ve{v} \in H^1_0(\OO)
	,
\end{gather*}
with the directional derivative $\DDu \te{S}$ defined shortly.
As indicated, the tensors $\te{S}$ (symmetric) and $\te{F}$ can be precomputed and stored. Note that the additional tensors and local operations only involve a modest number of simple operations, such as multiplications and additions, to be performed in local variables, typically mapped to registers for execution on hardware. These are cheaper than memory accesses on all temporary hardware architectures. The alternative approach integrating over the spatial domain requires solving for $\Delta \ve{u} \in H^1_0(\Ot)$, such that there holds
\begin{gather*}
	\left(
	\store{\nicefrac{1}{J}}
	\,\,
	{\grad}\ve{v}
	,
	J \tefour{c} : \grads \Delta \ve{u}
	+
	{\grad}\Delta \ve{u}
	\,\,
	\sstore{\te{\tau}}
	\right)_\Ot
	=
	\left(
	\ve{v}(\inv{\ve{\phi}})
	,
	\hN
	\right)_\GON
	-
	\left(
	\store{\nicefrac{1}{J}}
	\,\,
	\grads \ve{v}
	,
	\sstore{\te{\tau}}
	\right)_\Ot
	+
	\left(
	\ve{v}(\inv{\ve{\phi}})
	,
	\B
	\right)_\OO
	\quad
	\forall \ve{v}\in H^1_0(\Ot)
	,
\end{gather*}
where we store the scalar quantity $\nicefrac{1}{J}$ as well as the symmetric tensor $\te{\tau}$. The stress tensors and their derivatives depend on the material model given in the following.
\paragraph{Compressible neo-Hookean model in material configuration}
{\small
\begin{align}
	{\te{S}_\mathrm{cNH}}
	&
	=
	\mu \id - \left( \mu - 2\lambda \store{\ln J} \right) \inv{\te{C}}
	\stackrel{d=3}
	{=}
	\stable{
		\inv{\te{C}} \left( 
		2 \mu \te{E} + 2 \lambda \, \id \, \store{\lnp \Jm}
	 	\right)
	}
	\label{eqn:newton_storage_material_compressible_NH_S}
	\\
	\DDu\te{S}_\mathrm{cNH} 
	&
	= 
	\left(
	\mu-2\lambda \store{\ln {J}}
	\right)
	2 
	\left(
	\sstore{\inv{\te{F}}} 
	\left(\Grad\Delta\ve{u}\right) \sstore{\inv{\te{C}}} 
	\right)^\mathrm{S}
	+
	2 \lambda \,
	\tr\left(\sstore{\inv{{\te{F}}}} \Grad \Delta \ve{u}\right)
	\sstore{\inv{{\te{C}}}}
	\nonumber
	\\
		&
	= 
	\left(
	\mu-2\lambda \store{\lnp{\Jm}}
	\right)
	2 
	\left(
	\sstore{\inv{\te{F}}} 
	\left(\Grad\Delta\ve{u}\right) \sstore{\inv{\te{C}}} 
	\right)^\mathrm{S}
	+
	2 \lambda \,
	\tr\left(\sstore{\inv{{\te{F}}}} \Grad \Delta \ve{u}\right)
	\sstore{\inv{{\te{C}}}}
	\nonumber
\end{align}
}%
The notation $\stackrel{d=3}{=}$ indicates that we assume $d=3$ for the sake of presentation, since a term of the form $\id \left(1-\nicefrac{d}{3}\right)$ cancels here and at similar places in this manuscript. 
\paragraph{Spatial integration of the compressible neo-Hookean model}
{\small
\begin{align}
	{\te{\tau}}_\mathrm{cNH}
	&
	=
	\mu \, \te{b} - \left( \mu - 2\lambda \store{\ln J} \right) \id
	\stackrel{d=3}
	{=}
	\stable{
		2\mu 
		\,
		\te{\tilde{b}}
		+ 2 \lambda \, \id \, \store{\lnp \Jm }
	}
	\label{eqn:tau_cNH}
	\\
	J\tefour{c}_\mathrm{cNH} : (\cdot)^\mathrm{S}
	&
	=
	2
	\left(
	\mu - 2 \lambda \store{\ln {J}}
	\right)
	(\cdot)^\mathrm{S}
	+
	2 \lambda \,
	\tr(\cdot)
	\,
	\id
	=
	2
	\left(
	\mu - 2 \lambda \store{\lnp \Jm}
	\right)
	(\cdot)^\mathrm{S}
	+
	2 \lambda \,
	\tr(\cdot)
	\,
	\id
	\nonumber
\end{align}
}%
\paragraph{Nearly incompressible neo-Hookean model in material configuration} 
{\small
\begin{align*}
	{\te{S}}_\mathrm{iNH}
	&
	=
	\mu \store{\Jpow} \id 
	+ 
	\store{c_1}
	\,
	\inv{\te{C}}
	\stable{
		\stackrel{d=3}
		{=}
		\,\,
		\inv{\te{C}}
		\left[
		\nicefrac{\kappa_b}{2} \, \store{\Jm} (\store{\Jm} + 2) \, \id
		+
		2 \mu \store{\Jpow} \left( \te{E} - \nicefrac{1}{3} \, \id \, \tr \te{E} \right)
		\right]
	}
	\\
	\DDu\te{S}_\mathrm{iNH}
	&
	= 
	-
	\nicefrac{2\mu}{3} \store{\Jpow} \left(\nicefrac{1}{{J}} \, \DDu J\right)
	\, 
	\id
	+
	2 
	\store{c_1}
	\left[
	\sstore{\inv{\te{F}}} 
	\left(\Grad\Delta\ve{u}\right) \sstore{\inv{\te{C}}} 
	\right]^\mathrm{S}
	+
	\left[
	\store{c_2}
	\left(
	\nicefrac{1}{{J}}
	\,
	\DDu J
	\right)
	-
	\nicefrac{2\mu}{3} \store{\Jpow}
	\tr 
	\left( 
	\sstore{\te{F}}^\top \Grad\Delta\ve{u} 
	\right)	
	\right]
	\sstore{\inv{{\te{C}}}}
\end{align*}
}%
The additional scalars are
{\small
\begin{align}
	c_1 
	\eqq\,\,
	&
	\nicefrac{\kappa_b}{2} (J^2-1) - \nicefrac{\mu}{3} \store{\Jpow} I_1
	=
	\stable{
		\nicefrac{\kappa_b}{2} \, \store{\Jm} (\store{\Jm} + 2) 
		- \nicefrac{\mu}{3} \, \store{\Jpow} (d + 2 \, \tr \te{E})
	}
	, 
	\label{eqn:c1_def}
	\\
	c_2 
	\eqq\,\,
	&		
	\nicefrac{2\mu}{9} \store{\Jpow}  {I}_1
	+
	\kappa_b {J}^2
	=
	\stable{
		\nicefrac{2 \mu}{9} \store{\Jpow}  (d + 2 \, \tr \te{E})
		+
		\kappa_b \, {J}^2
	}
	.
	\label{eqn:c2_def}
\end{align}
}%
\paragraph{Spatial integration of the nearly incompressible neo-Hookean model}
{\small
\begin{align*}
	{\te{\tau}}_\mathrm{iNH}
	&
	=
	\mu \store{\Jpow} \te{b}
	+ 
	\store{c_1} 
	\id
	\stackrel{d=3}
	{=}
	\stable{
		\nicefrac{\kappa_b}{2} \, \store{\Jm} (\store{\Jm} + 2) \, \id
		+
		2 \mu \store{\Jpow} \left( \te{\tilde{b}} - \nicefrac{1}{3} \, \id \, \tr \te{\tilde{b}} \right)
	}
	,
	\\
	J\tefour{c}_\mathrm{iNH} : (\cdot)^\mathrm{S}
	&
	=
	-
	\nicefrac{4\mu}{3} \store{\Jpow}  
	\tr(\cdot) \, \sstore{\te{C}}
	-
	2\store{c_1}
	(\cdot)^\mathrm{S}
	+
	\store{c_2}
	\tr(\cdot) \, \id
	,
\end{align*}
}%
where we exploit $\tr{\te{C}} = \tr{\te{b}} = 2 \tr{\te{\tilde{b}}} + d$ and use $c_1$~\eqref{eqn:c1_def} and $c_2$~\eqref{eqn:c2_def}.
\paragraph{Nearly incompressible fiber model in material configuration}
\begin{align*}
	{\te{S}}_\mathrm{fiber}
	=
	\te{S}_\mathrm{iNH}
	+
	{
		\sum_{i=4,6}
		\store{c_3}
		\,
		\store{E_i}
		\,
		\sstore{\te{H}_i}
	}
	,
	\quad
	\DDu\te{S}_\mathrm{fiber}
	= 
	\DDu\te{S}_\mathrm{iNH}
	+
	{
		\sum_{i=4,6}
		\store{c_3} 
		\left(2 k_2 \store{E_i}^2 + 1 \right) 
		\left[
		\sstore{\te{H}_i} : 
		\left( 2 \sstore{\te{F}}^\top \Grad \Delta \ve{u}\right)^\mathrm{S}
		\right]
		\sstore{\te{H}_i} 
	}
\end{align*}
with 
an additional scalar
\begin{gather}
	c_3 \eqq 
	\begin{cases}
		2 k_1 \exp\left(k_2 \store{E_i}^2\right) & \text{if $\store{I_i^\star}>1$},\\
		0 & \text{else}.
	\end{cases}
	\label{eqn:c3_def}
\end{gather}

\paragraph{Spatial integration of the nearly incompressible fiber model} 
\begin{align*}
	{\te{\tau}}_\mathrm{fiber}
	&
	=
	{\te{\tau}}_\mathrm{iNH}
	+
	{
		\sum_{i=4,6}
		\store{c_3} \, \store{E_i} \te{F} \sstore{\te{H_i}} \te{F}^\top
	}
	,
	\\
	J\tefour{c}_\mathrm{fiber} : (\cdot)^\mathrm{S}
	&
	=
	J\tefour{c}_\mathrm{iNH} : (\cdot)^\mathrm{S}
	+
		\sum_{i=4,6}
		2 \store{c_3} \left(2 k_2 \store{E_i}^2 + 1 \right)
		\left[
		\left(
		\sstore{
			\te{F} \te{H}_i \, \te{F}^\top
		}
		\right)
		:
		(\cdot)
		\right] \sstore{\te{F} \, \te{H}_i \, \te{F}^\top}
	,
\end{align*}
with $c_3$ defined in Eqn.~\eqref{eqn:c3_def}.

An overview of all the variants and the respective precomputed and stored variables is given in Tab.~\ref{tab:overview_precompute}, which also lists the required memory for storing in double precision
(1 double-precision variable = 8~bytes, 8~B). 
For the fiber model, storing scalar quantities requires 248~B (material configuration) or 328~B (spatial configuration), while storing tensorial quantities requires 488~B (material configuration) or 520~B (spatial configuration) per integration point. Depending on the exact form of the integrals and arithmetic operations to evaluate them, the increased memory traffic storing second-order tensors might pay off. Furthermore, this example also shows that it is not clear per se, if integrating over the material or spatial configuration is favorable in terms of memory traffic, since this depends also on whether or not tensors or scalars are precomputed and stored. For the linearized operator, not all tensors need to be loaded when integrating over the spatial domain.
\begin{table}[h!]
	\centering
	\caption{Precomputed quantities per integration point for the variants storing scalars or scalars \emph{and} tensors with related memory using double precision and $d=3$. The Jacobian matrices of the finite element maps, $\te{J}_0$ and $\te{J}_t$, are stored for each domain. $\te{S}$, $\te{\tau}$, $\te{C}$, $\inv{\te{C}}$, $\te{H}_i$ and $\te{F}\te{H}_i\te{F}^\top$, $i=4,6$ are symmetric. Numbers in brackets indicate contribution to asymptotic memory traffic in linearized operator application where differing from storage requirement.
	}
	{
		\scriptsize
		\label{tab:overview_precompute} 
		\begin{tabular}{||c || l | l | c | c | l | l | c | c||}
			\hline
			& \multicolumn{4}{|c|}{integration over reference domain $\OO$} & \multicolumn{4}{|c||}{integration over spatial domain $\Ot$}\\
			\hline
			& \multicolumn{2}{|c|}{quantities} & \multicolumn{2}{|c|}{memory in B} & \multicolumn{2}{|c|}{quantities} & \multicolumn{2}{|c||}{memory in B} \\
			\hline
			&\multicolumn{1}{|c}{scalar}&\multicolumn{1}{|c}{tensor}&\multicolumn{1}{|c}{scalar}&\multicolumn{1}{|c}{tensor}&\multicolumn{1}{|c}{scalar}&\multicolumn{1}{|c}{tensor}&\multicolumn{1}{|c}{scalar}&\multicolumn{1}{|c||}{tensor}\\
			\hline\hline
			      &&&&&&&& \\[-1.25ex]
			cNH   & $\te{J}_0$, $\ve{u}_k$, $\ln J$ 
			      & $\te{F}, \te{S}$, $\inv{\te{F}}$, $\inv{\te{C}}$ 
				  & 104  
				  & 320 
				  & $\te{J}_0$, $\te{J}_t$, $\ve{u}_k$, $\nicefrac{1}{J}$, $\ln J$ 
				  & $\te{\tau}$ 
				  & 184 
				  & 208~(136) 
				  \\[1.5ex]
			iNH   & $\te{J}_0$, $\ve{u}_k$, $\Jm$, $\Jpow$,
			      & $\te{F}, \te{S}$, $\inv{\te{F}}$, $\inv{\te{C}}$ 
			      & 128~(120)
			      & 344~(336)
			      & $\te{J}_0$, $\te{J}_t$, $\ve{u}_k$, $\nicefrac{1}{J}$, $\Jm$, 
			      & $\te{\tau}$, $\te{C}$
			      & 208 
			      & 280~(200) 
   				  \\
   				  & $c_1$, $c_2$
   				  &
   				  &
   				  &
   				  & $\Jpow$, $c_1$, $c_2$
   				  &
   				  &
   				  &
			      \\[1.5ex]
			fiber & iNH, $c_3$, $I_i^\star$, $E_i$, $\te{H}_i$          
				  & iNH 
				  & 272~(248)
				  & 488~(464) 
				  & iNH, $c_3$, $I_i^\star$, $E_i$, $\te{H}_i$
				  & iNH, $\te{F}\te{H}_i\te{F}^\top$
				  & 352
				  & 520~(328) 
				  \\[0.75ex]
			\hline
		\end{tabular}
	}
\end{table}

\section{Matrix-free Preconditioning}
\label{sec:matrix-free_solver}

The linear systems 
corresponding to the Newton update step are solved with the preconditioned flexible generalized minimal residual method~(FGMRES)~\cite{Saad2003}. We use a flexible formulation since we use a Krylov solver at the coarse level in the multigrid preconditioner, which renders the operation non-stationary. The Krylov solver only requires the action of the operator on a vector, not the explicit entries of the matrix. 
The matrix-free evaluation of the action of the matrix on a vector is realized via numerical quadrature, exploiting the tensor product structure of the shape functions and quadrature rule via sum factorization techniques (see~\cite{KronbichlerKormann2012, KronbichlerKormann2019, Vos2010}) and employing SIMD vectorization over batches of elements~\cite{KronbichlerKormann2012}.

\begin{figure}
    \centering
	\begin{tikzpicture}
		
		\node[circle, draw, fill=lightgray] (D) at (-3,3) {};
		\node[circle, draw, fill=lightgray] (B) at (-1.5,1.5) {};
		\node[circle, draw, fill=lightgray] (A) at (0,0) {};
		\node[circle, draw, fill=lightgray] (C) at (1.5,1.5) {};
		\node[circle, draw, fill=lightgray] (E) at (3,3) {};
		
		\draw[-latex, thick] (D) -- (B); 
		\draw[-latex, thick] (B) -- (A); 
		\draw[-latex, thick] (A) -- (C); 
		\draw[-latex, thick] (C) -- (E); 

		\fill (A) node[fill=lightgray, below,draw]{Coarse solver};
	
		\draw (D) -- (B) node [midway, above, sloped] (leftFineToMid) {\scriptsize restrict};
		\draw (B) -- (A) node [midway, above, sloped] (leftMidToCoarse) {\scriptsize restrict};
		\draw (A) -- (C) node [midway, above, sloped] (rightCoarseToMid) {\scriptsize prolongate};
		\draw (C) -- (E) node [midway, above, sloped] (rightMidTOFine) {\scriptsize prolongate};
		
		\Cycle[-latex]{D}{225}{5mm}[{node[anchor=east,pos=0.5]{\scriptsize pre-smooth}}]{5mm}
		\Cycle[-latex]{B}{225}{5mm}[{node[anchor=east,pos=0.5]{\scriptsize pre-smooth}}]{5mm}		
		
		\Cycle[-latex]{C}{-45}{5mm}[{node[anchor=west,pos=0.5]{\scriptsize post-smooth}}]{5mm}
		\Cycle[-latex]{E}{-45}{5mm}[{node[anchor=west,pos=0.5]{\scriptsize post-smooth}}]{5mm}		
		
		\begin{scope}[xshift=7.5cm, yshift=2.5cm]
			\def\tilt{1.00}
			\def\yfac{0.75}

                        \foreach \x in {0, 0.5}
                          \foreach \y in {0, 0.5}
                            \draw[fill=gray,opacity=0.5] (\x+\tilt*\yfac*\y,\yfac*\y) --
                            (\x+0.5+\tilt*\yfac*\y,\yfac*\y) --
                            (\x+0.5+\tilt*\yfac*\y+0.5*\tilt*\yfac,\yfac*\y+0.5*\yfac) --
                            (\x+\tilt*\yfac*\y+0.5*\tilt*\yfac,\yfac*\y+0.5*\yfac) --
                            (\x+\tilt*\yfac*\y,\yfac*\y);

                        \foreach \x in {0, 0.25, 0.5, 0.75, 1}
                          \foreach \y in {0, 0.25, 0.5, 0.75, 1}
                            \node[circle, draw, fill=white, inner sep=1.5pt, scale=0.75] at (\x+\tilt*\yfac*\y,\yfac*\y) {};

		\end{scope}
		
		\begin{scope}[xshift=6.0cm, yshift=1cm]
			\def\tilt{1.00}
			\def\yfac{0.75}
			
                        \foreach \x in {0, 0.5}
                          \foreach \y in {0, 0.5}
                            \draw[fill=gray,opacity=0.5] (\x+\tilt*\yfac*\y,\yfac*\y) --
                            (\x+0.5+\tilt*\yfac*\y,\yfac*\y) --
                            (\x+0.5+\tilt*\yfac*\y+0.5*\tilt*\yfac,\yfac*\y+0.5*\yfac) --
                            (\x+\tilt*\yfac*\y+0.5*\tilt*\yfac,\yfac*\y+0.5*\yfac) --
                            (\x+\tilt*\yfac*\y,\yfac*\y);

                        \foreach \x in {0, 0.5, 1}
                          \foreach \y in {0, 0.5, 1}
                            \node[circle, draw, fill=white, inner sep=1.5pt, scale=0.75] at (\x+\tilt*\yfac*\y,\yfac*\y) {};
	
		\end{scope}
		
		\begin{scope}[xshift=4.5cm, yshift=-0.5cm]
			\def\tilt{1.00}
			\def\yfac{0.75}
			
			\draw[fill=gray,opacity=0.5] (0, 0) -- (1, 0) -- (1 + \tilt*\yfac, \yfac) -- (\tilt*\yfac, \yfac) -- (0, 0);
			
                        \foreach \x in {0, 1}
                          \foreach \y in {0, 1}
                            \node[circle, draw, fill=white, inner sep=1.5pt, scale=0.75] at (\x+\tilt*\yfac*\y,\yfac*\y) {};
		\end{scope}
				
		\def\tilt{1.00}
		\def\yfac{0.75}
		\node (bottom) at (4.5, -0.5+0.5*\yfac) {};
		\node (middle) at (6.0, 1.0+0.5*\yfac) {};
		\node (top)    at (7.5, 2.5+0.5*\yfac) {};

		\node (bottomright) at (1.75+4.5, -0.5+0.5*\yfac) {};
		\node (middleright) at (1.75+6.0, 1.0+0.5*\yfac) {};
		\node (topright)    at (1.75+7.5, 2.5+0.5*\yfac) {};
		
		\draw[bend right=30, -latex, thick] (top) to node [below left] {} (middle);
		\draw[bend right=30, -latex, thick] (middle) to node [below left] {} (bottom);

		\draw[bend right=30, -latex, thick] (bottomright) to node [right] {{~$h$-multigrid}} (middleright);
		\draw[bend right=30, -latex, thick] (middleright) to node [right] {{~$p$-multigrid}} (topright);
		
	\end{tikzpicture}
    \caption{Visualization of a multigrid V-cycle (left) and an exemplary $hp$-multigrid hierarchy (right), with degrees of freedom indicated by circles.}
    \label{fig:multigrid_hierarchy}
\end{figure}
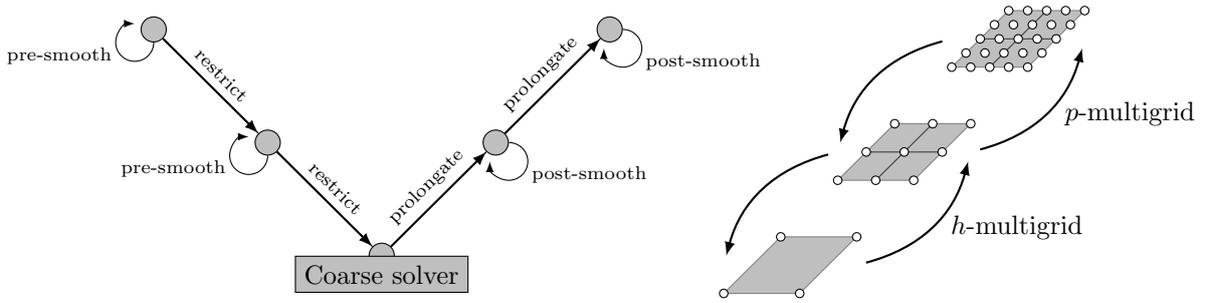

For constructing a preconditioner that is compatible with the matrix-free evaluation, an $hp$-multigrid preconditioning strategy with matrix-free smoothers is adopted from Fehn~et~al.~\cite{Fehn2020}, where first the polynomial degree is lowered recursively from $p$ to $\lfloor \nicefrac{p}{2} \rfloor$,
that is halved and rounded down to the nearest integer, going from level $l$ to $l-1$, after which $h$-coarsening is performed, see Fig.~\ref{fig:multigrid_hierarchy}. 
This strategy is denoted as $ph$-multigrid in~\cite{Fehn2020} and \texttt{ExaDG}.
The outer Krylov solver operates in double precision, while the preconditioner operates in single precision, see~\cite{KronbichlerKormann2012,KronbichlerLjungkvist2019}. The individual $h$-levels are created by uniformly refining an initial coarse grid, where the finest grid is equipped with a potentially higher-order accurate mapping. This higher-order mapping is then interpolated to the coarser levels keeping the order of the mapping constant, i.e., equal to the finest grids' polynomial degree. The mappings are assumed invertible on all levels, which is checked for the grid hierarchies used in the numerical examples.


In a standard multigrid V-cycle (see, e.g.,~\cite{Brandt1977, Hackbusch1985, Trottenberg2001}), smoothers reduce the high-frequency part of the error associated to each level, and restriction and prolongation operators transfer residuals and corrections between the levels, respectively. The smoother chosen is a Chebyshev-accelerated Jacobi scheme~\cite{KronbichlerWall2018,Fehn2020,Adams2003}, which uses the inverse of the matrix diagonal on each level (precomputed before solving) and the level operator application for computing residuals in the Jacobi-type iteration. Note that also the additional fine-scale errors resulting from round-off due to the mixed-precision strategy are well-captured by the multigrid smoothers~\cite{KronbichlerLjungkvist2019}. 
For the coarse solver, we use a conjugate gradient method (CG) preconditioned via AMG from the \texttt{Trilinos ML} package~\cite{Heroux2012, Gee2006}, which runs in double precision only. Employing a constant preconditioner would enable using CG as outer solver. Herein, however, we present a more general approach, noting that it might not be the optimal choice.


Updating the quadrature point data of the multigrid hierarchy means updating data of all operators on all levels using the multigrid transfer operator. Invertibility of the displacement map on all levels is checked, is not violated in the numerical examples discussed in the following, but is not enforced with the present approach.

\section{Numerical Results}

We first conduct forward stability tests comparing double and single precision evaluations of the stress tensor and linearization, whereafter the single-node performance is showcased in a simplified setting. Lastly, we solve a finite-strain elasticity problem using a patient-specific geometry of an iliac bifurcation adopting the fiber model.

\subsection{Forward stability test}
\label{sec:results_stability}

Forward numerical stability of the first and second Piola--Kirchhoff stress tensors and the directional derivatives is investigated by evaluating them with a pseudo-randomly sampled second-order tensor $\te{G}$, whose components fulfill
\begin{align*}
    -\epsilon \leq \te{G}_{ij} \leq -\nicefrac{\epsilon}{10}
    \,\,
    \lor
    \,\,
    \nicefrac{\epsilon}{10} \leq \te{G}_{ij} \leq \epsilon
    \qquad
    i,j = 1,\dots,d
    ,
\end{align*}
with a pseudo-random sign and an additional scale $\epsilon$. This scale is used to emulate strain rates ranging from $\mathcal{O}(10^{-8})$ to $\mathcal{O}(10^2)$ in 200 steps, while
the interval $[-\nicefrac{\epsilon}{10},\nicefrac{\epsilon}{10}]$ is not considered to ensure samples of the desired order of magnitude only.
To illustrate, assume $\te{G}_{11} = 1.0$, while the remaining entries of $\te{G}$ are zero;
this means that $\partial \ve{u}_1 / \partial \ve{x}_1 = 1.0$, that is,
a stretch of 100\% in $\ve{x}_1$-direction, being already well beyond reasonable design limits in most engineering applications.
However, note that the randomly generated tensor $\te{G}$ does in general not fulfill $\det{\left(\id+\te{G}\right)} = 1$.
For each gradient scale, we generate $10^3$ independent samples, set $\Grad\ve{u} = \te{G}$ and evaluate the stress tensors
and the directional derivatives using double and single precision arithmetic.
The relative error $\epsilon_\mathrm{rel}$ between the double precision and single precision representations
is then computed as the maximum over all samples and over all corresponding tensor entries.

From this experiment we can infer forward stability (up to the observed limit) also for double precision arithmetic. It has to be mentioned, however, that this is not intended to be a rigorous analysis of numerical stability, but rather serves to showcase the improvements in the small strain limits adopting the stable formulations. Specifically, no analysis has been performed to quantify the effect of the uniform scaling of the tensor on the obtained results.

\begin{figure}
	\centering
	\begin{subfigure}{.5\textwidth}
		\centering
		\includegraphics[width=1.0\linewidth, draft=\draftStability]{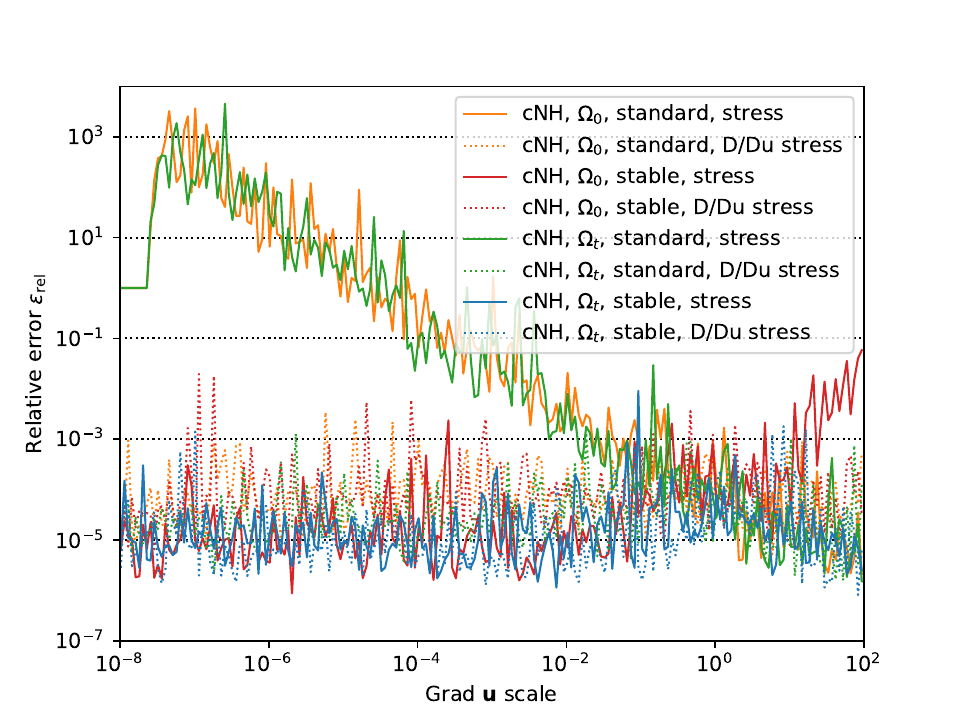}
		\caption{compr. neo-Hookean model (cNH)}
	\end{subfigure}%
	\begin{subfigure}{.5\textwidth}
		\centering
		\includegraphics[width=1.0\linewidth, draft=\draftStability]{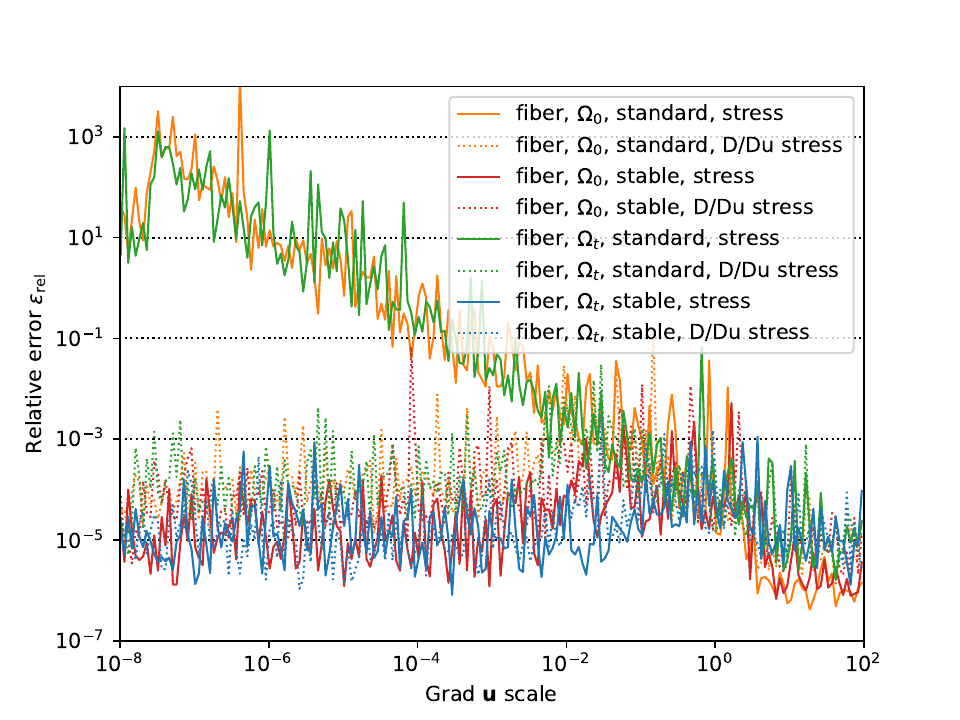}
		\caption{nearly incompr. fiber model (fiber)}
	\end{subfigure}
	\caption{Relative errors in the stress and directional derivative (D/Du stress) for sampled $\Grad\ve{u}$ comparing standard and stable formulations in the material ($\OO$) and spatial ($\Ot$) configurations. The nearly incompressible neo-Hookean and fiber models yield similar results, as the iNH model is one part of the fiber model.}
	\label{fig:stability_results}
\end{figure}

The investigations here extend the work by Shakeri~et~al.~\cite{Shakeri2024}, compared to which we also present results for the directional derivatives, which enter the linear system directly.
Fig.~\ref{fig:stability_results} illustrates the relative error in stresses (stress) and directional derivatives (D/Du stress) adopting the material or spatial integration strategies ($\OO$ or $\Ot$) using stable and standard formulations. The stable formulations yield small relative errors in the small strain limit, while the standard formulations show significant numerical instability. The individual material models also show different behavior in the medium strain range of $10^{-3}$ to $10^0$. Interestingly, the second Piola--Kirchhoff stress tensor of the compressible neo-Hookean model in stable formulation shows numerical instability for (excessively) large strains. This might be related to the stable form in Eqn.~\eqref{eqn:newton_storage_material_compressible_NH_S} using $\inv{\te{C}}\te{E}$ instead of $\inv{\te{C}}$ only in the standard formulation. While the standard formulation is stable for large strains, it is not numerically stable in the small strain limit, such that the stable form is considered regardless. The St.Venant--Kirchhoff model, that is, $\te{S}_\mathrm{VK} \eqq \lambda \tr(\te{E}) \id + 2 \mu \te{E}$ with $\te{E}$ according to Eqn.~\eqref{eqn:E_stable}, for comparison yields relative errors $\epsilon_\mathrm{rel} \in [10^{-5}, 10^{-3}]$ for all strains considered here.

\begin{figure}
	\centering
	\begin{subfigure}{.5\textwidth}
		\centering		\includegraphics[width=1.0\linewidth, draft=\draftStability]{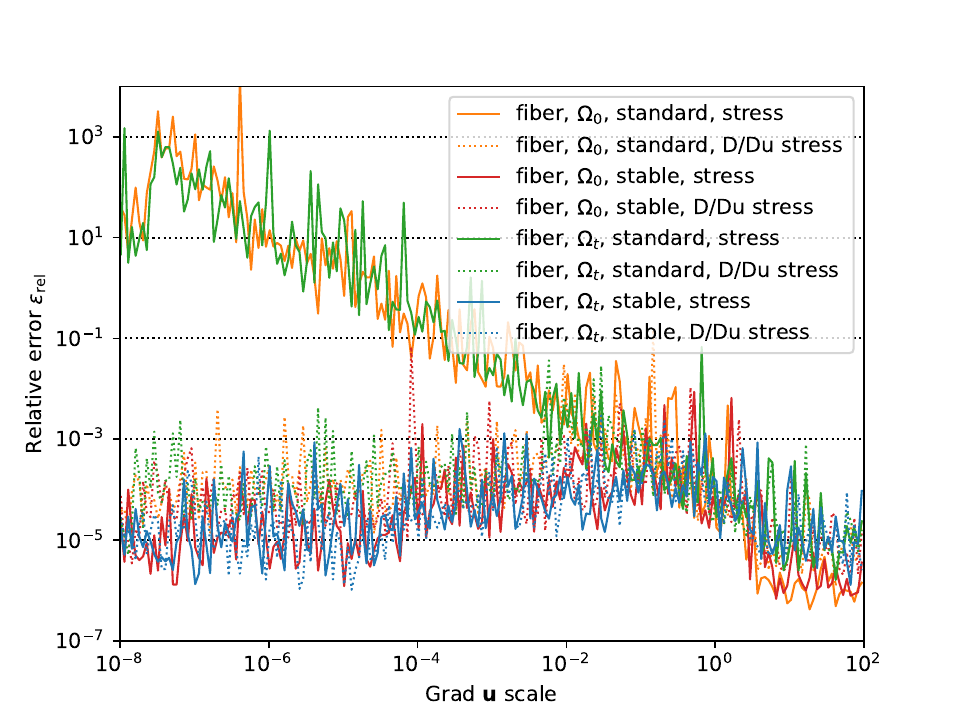}
		\caption{fiber model with fast $\exp()$ evaluation}
	\end{subfigure}%
	\begin{subfigure}{.5\textwidth}
		\centering
		\includegraphics[width=1.0\linewidth, draft=\draftStability]{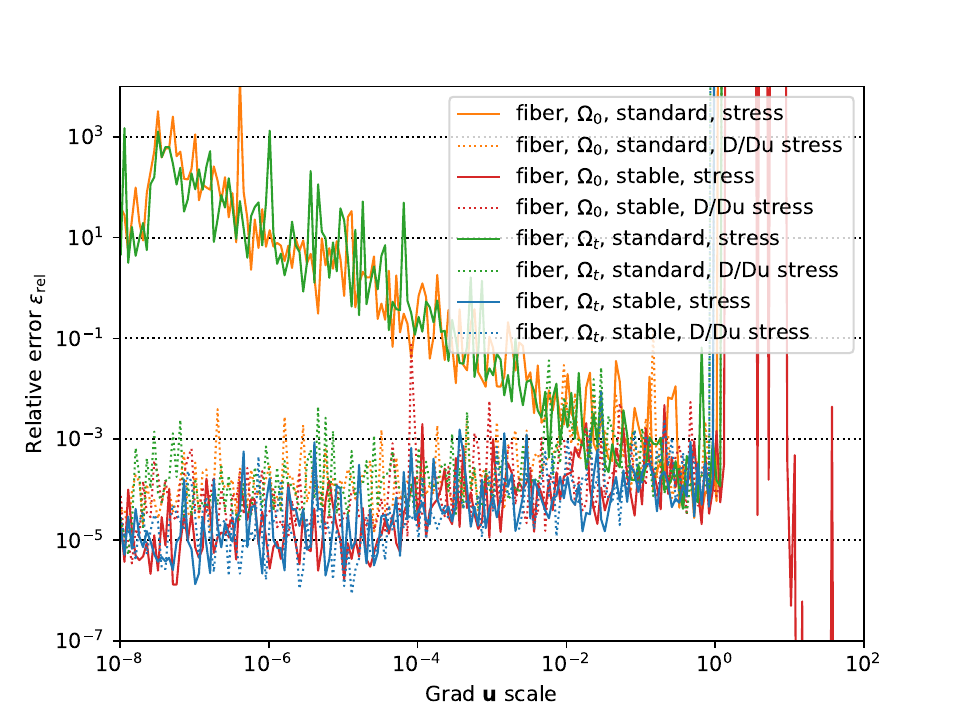}
		\caption{fiber model w. fast $\exp()$ and $\Jpow$ evaluation}
	\end{subfigure}
	\caption{Relative errors in the stress and directional derivative (D/Du stress) for sampled $\Grad\ve{u}$ comparing standard and stable formulations in the material ($\OO$) and spatial ($\Ot$) configurations with fast evaluation strategies.}
	\label{fig:stability_results_iHGO}
\end{figure}

As a next experiment, Fig.~\ref{fig:stability_results_iHGO} presents the accuracy of the fast evaluation of $\Jpow$ and $\exp(x)$ as discussed in Sec.~\ref{sec:stable_numerics}, which are relevant for the nearly incompressible neo-Hookean and fiber models. These results show that the effect of the fast evaluation of $\exp(\cdot)$ exploiting the floating point representation only mildly affects the numerical stability. For large strain scales, we observe large relative errors or even values out of the admissible range for the approximation of $\Jpow$. This is due to the fact that the sampling strategy does not enforce $J\approx1$, and hence the initial guess $\Jpow=1$ used in the Newton solver (see Alg.~\ref{alg:Jpow_Newton}) is inadequate. Note however, that this only occurs for large strain scales, and did not lead to any problems in the results presented within this work. Hence, we employ the Newton solver for $\Jpow$ by default. Fast evaluation of $\Jpow$ has similar effects as shown in Fig.~\ref{fig:stability_results_iHGO}(a) when using the iNH model, such that we omit these results.

Summing up, we do not observe excessive round-off errors in the relevant strain range (0 to 100\%) employing the stable formulations. The fast evaluation strategies as described in Sec.~\ref{sec:stable_numerics} only mildly affect numerical stability in that reasonable range.

\subsection{Node-level performance analysis}
\label{sec:results_performance_block}

The single-node performance of operator evaluation is tested on an Intel Xeon Platinum 8360Y ``Ice Lake'' with 36 physical cores with 2.0~GHz base frequency per socket, and two sockets per node with 256 GB of DDR4-3200 memory and 100 GBit/s full-duplex Infiniband interconnect. The sum of L2 and L3 caches amounts to 200~MB, which corresponds to $25\times10^6$ floating point double-precision numbers. The full width of the AVX-512 instruction set (8 double-precision or 16 single-precision floating point numbers) is used in a vectorization-across-cells strategy~\cite{KronbichlerKormann2019}. The GNU compiler version 12.1.0 with flags ``\texttt{-O3 -march=icelake-server}'' and OpenMPI version 4.1.3 are used.

The grids are constructed by uniformly refining an initial coarse cube of $N_0\times N_0\times N_0$ elements of polynomial degree $p=1,\dots,8$ until 2.5--5$\times10^6$ DoFs are reached. With this DoF count, caches are saturated according to our tests (omitted here for brevity).
In the tests, we analyze the performance for different polynomial degrees $p$ as a parameter. However, the objective of this work is to identify regimes of high throughput, not exploiting high asymptotic convergence rates of higher-order methods for sufficiently smooth solutions. We thus assume similar accuracy per unknown in all examples within this work, noting that general geometries of practical interest and resulting complexities related to mesh construction limit the practically feasible maximum polynomial degree.
Tab.~\ref{tab:performance_element_number} lists the resulting grid parameters, showing discretizations with similar numbers of DoFs,
where the discretizations with degrees $p=5$ and $p=7$ differ by 13\% and 58\%, respectively. To account for non-trivial geometries in a practical setting,
the generated cube is deformed, resulting in a non-constant Jacobian of the isoparametric mapping.
\begin{table}
	\centering
	\caption{Number of continuous finite elements $N_\mathrm{el}$ and DoFs per polynomial degree $p$ on a $l$ times refined initial coarse grid of $N_0\times N_0\times N_0$ elements.}
	{
		\scriptsize
		\label{tab:performance_element_number} 
		\begin{tabular}{||l | c c c c c c c c||}
			\hline
			&&&&&&&&\\[-1.5ex]
			$p$&1&2&3&4&5&6&7&8\\[0.75ex]
			\hline\hline
			&&&&&&&&\\[-1.25ex]
   			$N_0$               &3      &3      &1     &3     &5    &1    &1    &3    \\[0.75ex]
            $l$                 &5      &4      &5     &3     &2    &4    &4    &2    \\[0.75ex]
            $N_\mathrm{el}$     &884736 &110592 &32768 &13824 &8000 &4096 &4096 &1728 \\[0.75ex]
   			MDoFs               &2.74   &2.74   &2.74  &2.74  &3.09 &2.74 &4.33 &2.74 \\[0.75ex]
			\hline
		\end{tabular}
	}
\end{table}

\begin{figure}
	\centering
	\begin{subfigure}[b]{.32\textwidth}
		\centering		
        \begin{overpic}[width=1.0\textwidth,draft=\draftPerformance]{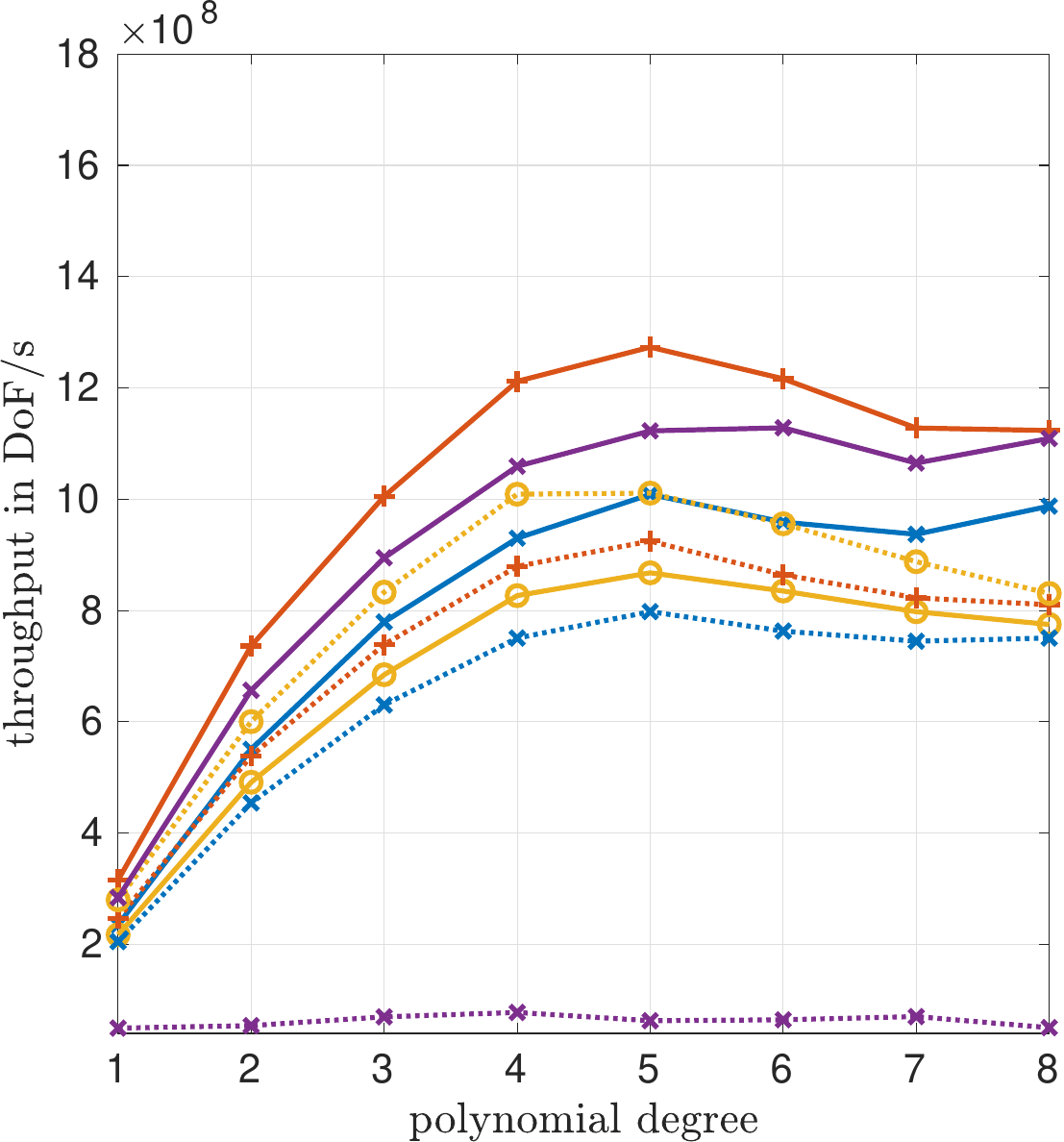}
        \put(17.5,22){\tiny 5--8$\times10^7$~DoF/s}
		\put(35,20){\vector(1,-1){7.5}}
		\end{overpic}
		\caption{fiber}
	\end{subfigure}%
	\hfil
	\begin{subfigure}[b]{.32\textwidth}
		\centering
		\begin{overpic}[width=1.0\textwidth,draft=\draftPerformance]{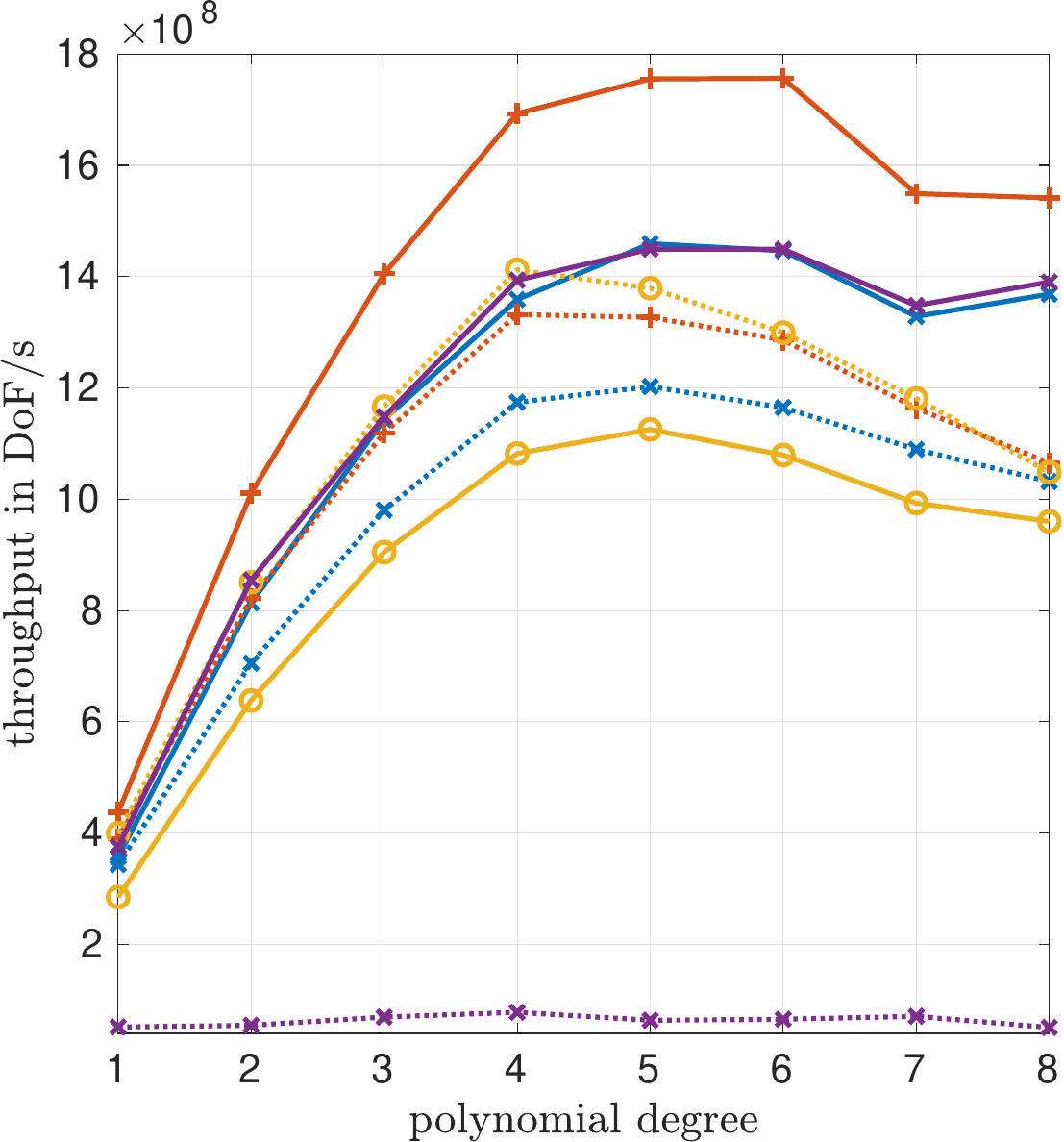}
			\put(17.5,22){\tiny 5--8$\times10^7$~DoF/s}
			\put(35,20){\vector(1,-1){7.5}}
		\end{overpic}
		\caption{incompr. neo-Hookean}
	\end{subfigure}
	\hfil
	\begin{subfigure}[b]{.32\textwidth}
		\centering
		\begin{overpic}[width=1.0\textwidth,draft=\draftPerformance]{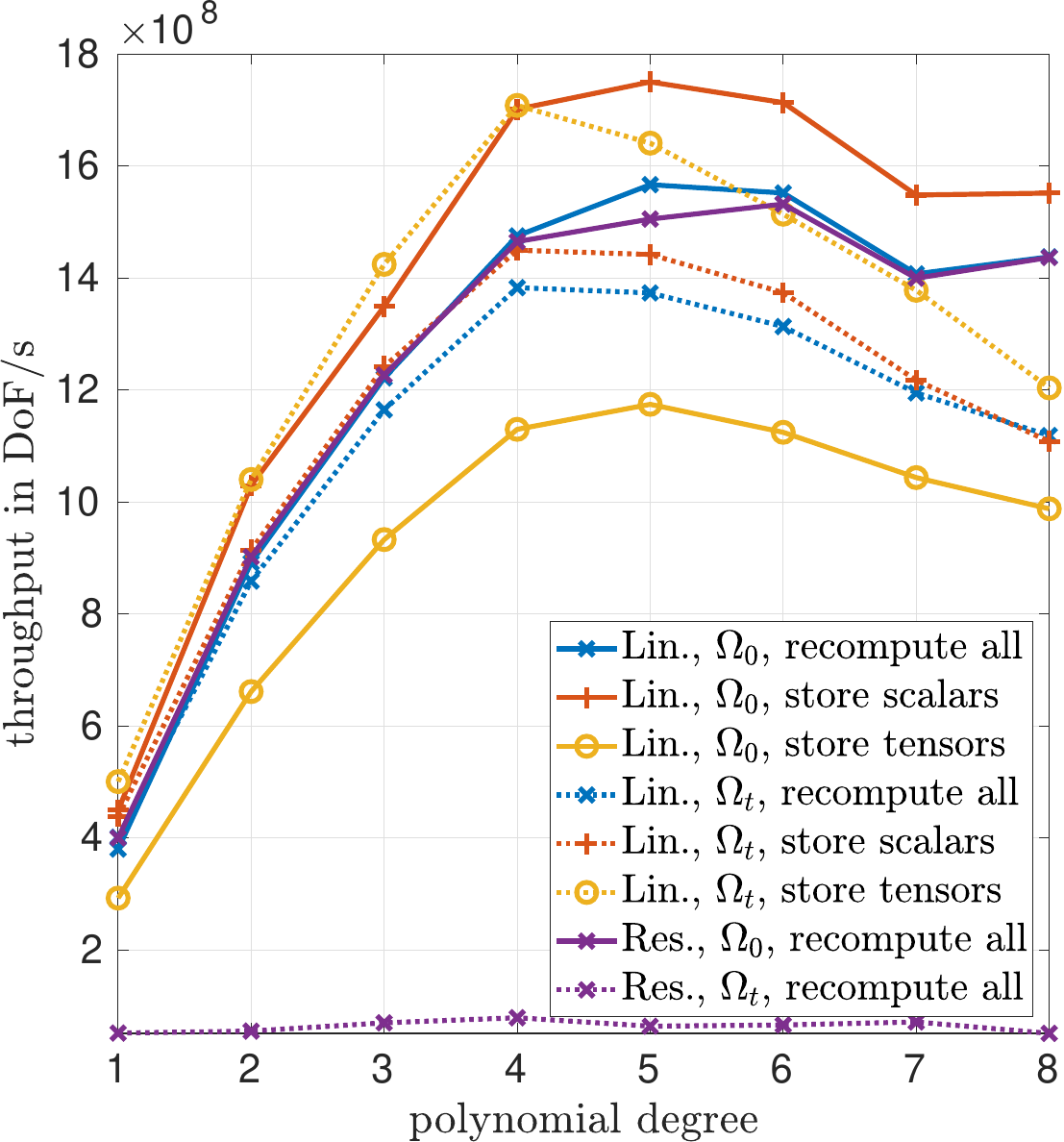}
			\put(17.5,22){\tiny 5--8$\times10^7$~DoF/s}
			\put(35,20){\vector(1,-1){7.5}}
		\end{overpic}
		\caption{compr. neo-Hookean}
	\end{subfigure}
	\\
	\begin{subfigure}[b]{.32\textwidth}
		\centering
		\includegraphics[width=1.0\linewidth, draft=\draftPerformance]{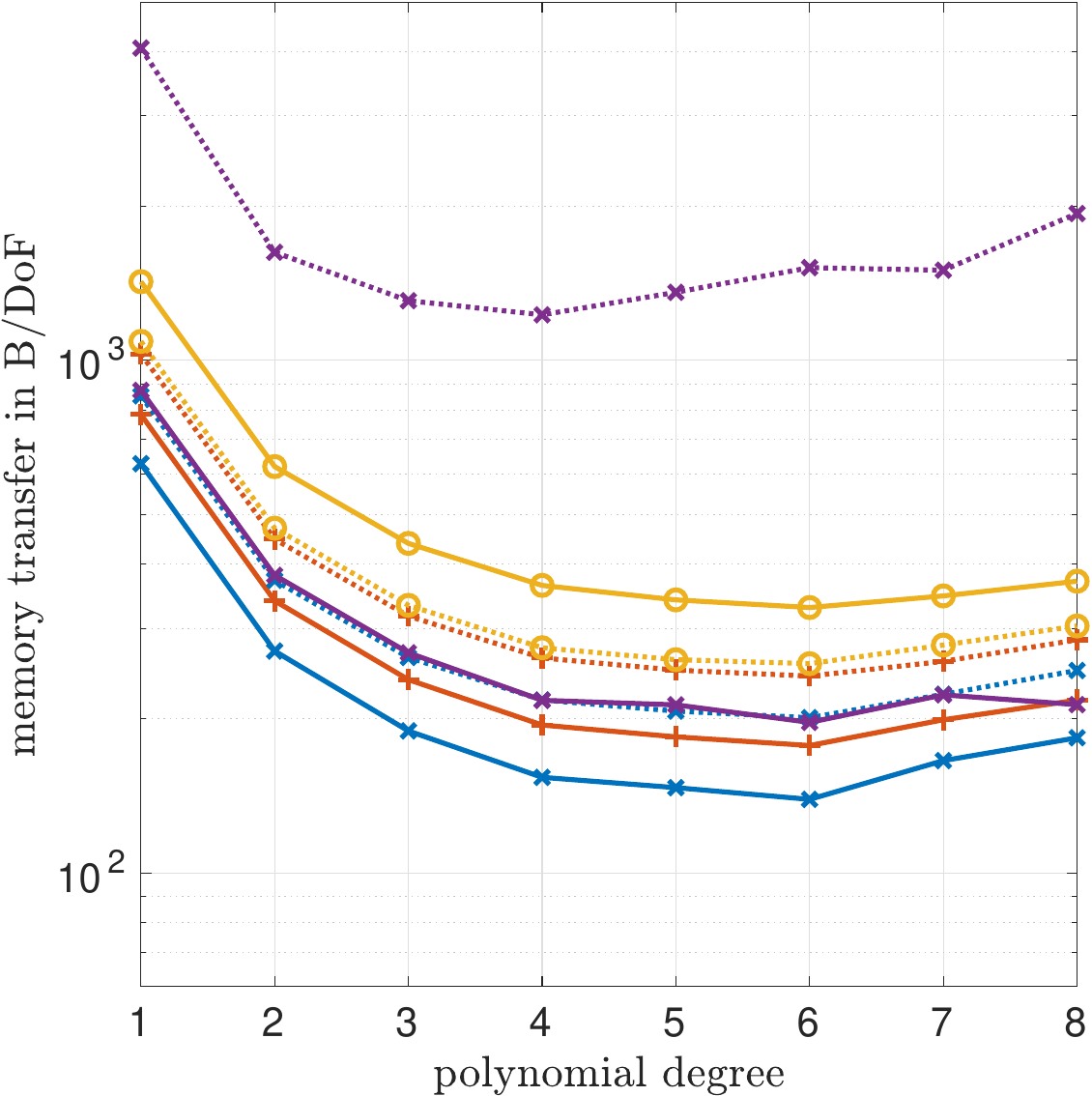}
		\caption{fiber}
	\end{subfigure}%
	\hfil
	\begin{subfigure}[b]{.32\textwidth}
		\centering
		\includegraphics[width=1.0\linewidth, draft=\draftPerformance]{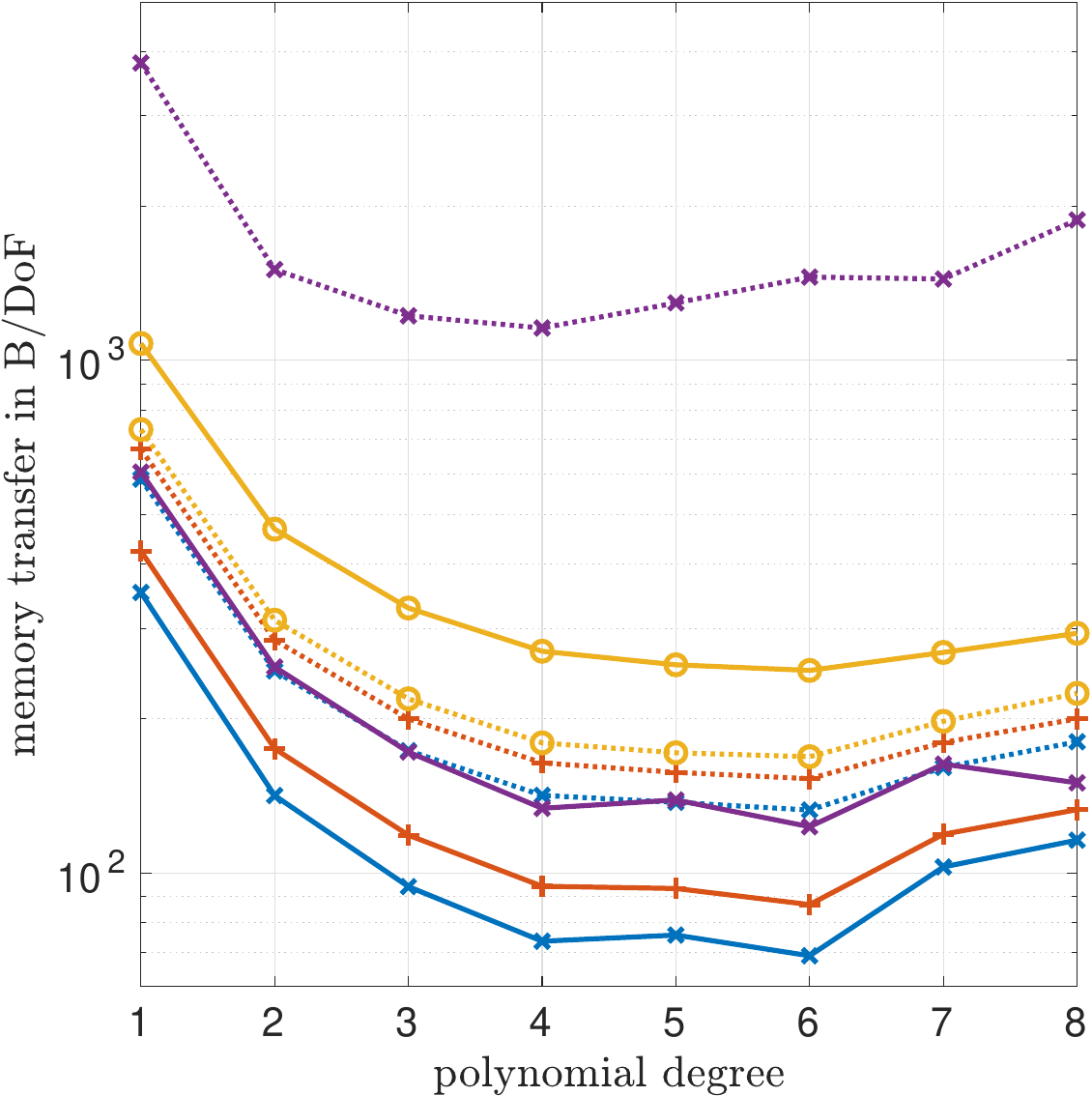}
		\caption{incompr. neo-Hookean}
	\end{subfigure}
	\hfil
	\begin{subfigure}[b]{.32\textwidth}
		\centering
		\includegraphics[width=1.0\linewidth, draft=\draftPerformance]{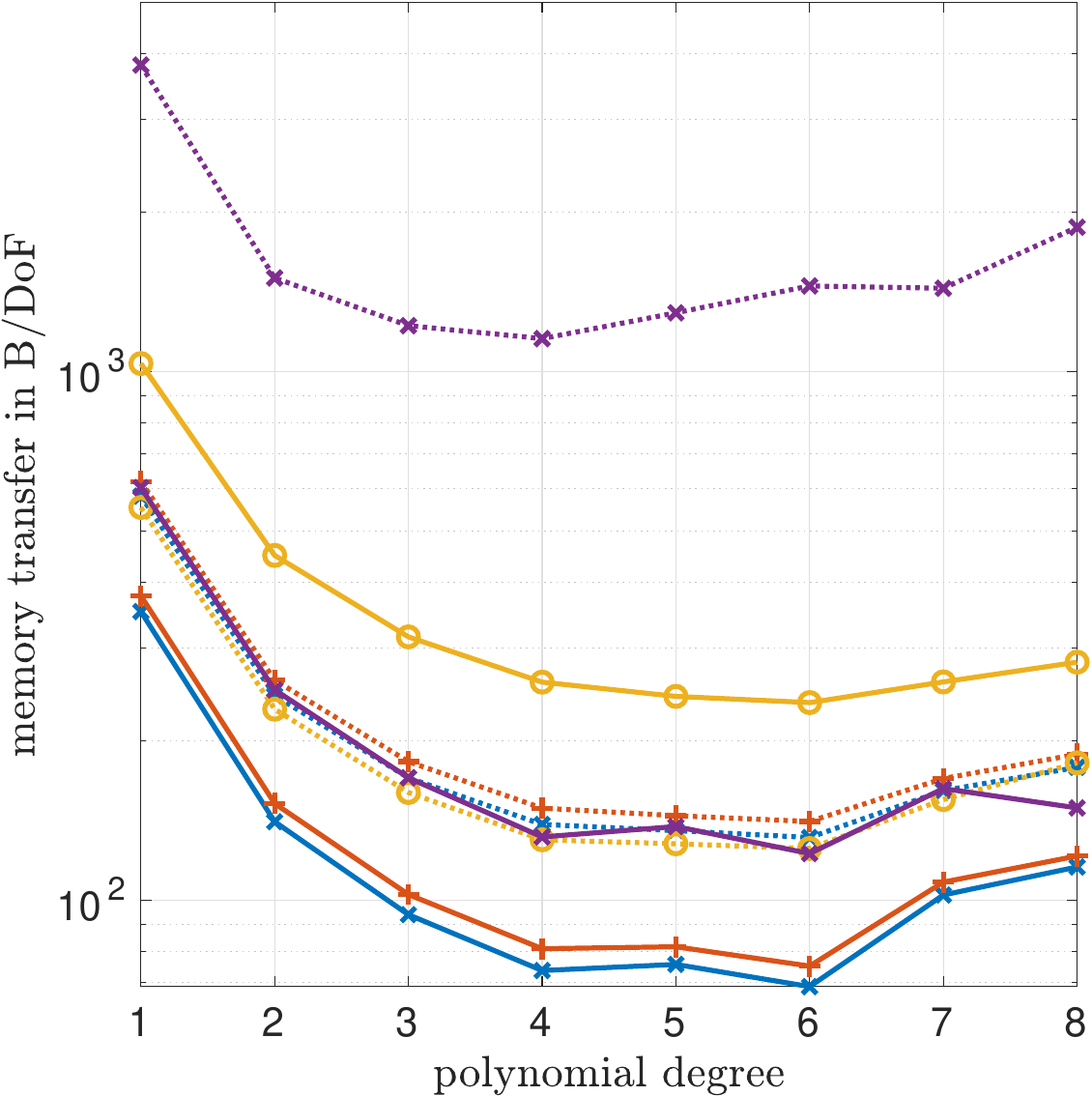}
		\caption{compr. neo-Hookean}
	\end{subfigure}
	\caption{Throughput (top row) and memory transfer (bottom row) using various material models considering integration over the reference configuration ($\OO$) or the spatial configuration ($\Ot$) and precomputing i) no data, ii) scalar values, or iii) tensorial quantities. Throughput for residual evaluation integrating over $\Ot$ yields values of 5--8$\times10^7$~DoF/s.}
	\label{fig:performance_measurements_fiber_only}
\end{figure}

The tests consist of i) applying the linearized operator corresponding to the directional derivative, that is, the left-hand side of the Newton update step~\eqref{eqn:newton_material} on a vector as repeatedly executed within a Krylov solver, and ii) evaluating the nonlinear residual, i.e., the right-hand side of said equations and storing it in a vector. The full node is used with 72 threads and an MPI-only parallelization, measuring the throughput in DoF/s and total memory transfer in B/DoF via \texttt{LIKWID}~\cite{Treibig2010}. The results are derived by measuring multiples of 100 repetitions that are stopped once at least 1 second of execution time has been elapsed. The quadrature point and mapping data are updated only once in the beginning for the linearized operator application, but for every evaluation of the nonlinear residual. This is motivated by the fact that in a Newton scheme, the linearized operator is repeatedly evaluated in a single Newton iteration, but the residual needs to be evaluated with the current iterate. Tests are repeated 10 times, where the best 3 runs are averaged to reduce effects stemming from other jobs on the compute system. Fig.~\ref{fig:performance_measurements_fiber_only} depicts the obtained results for the application of the linearized operator/single evaluation of the residual for various polynomial degrees and switching between the approaches adopting integration over the spatial or material configuration and precomputing strategies. The following observations are made:
\begin{enumerate}[i)]
	\item The residual evaluation involves less terms in the related integrals than the linearized operator application, but requires updating all integration point data. Therefore, integration over the reference or spatial configuration yield significantly different results, where the former competes with the fastest precomputing strategy for the linearized operator application, and the latter requires an update of all geometry-related data structures in the implementation of the deal.II library, including the detection of possible data compression and reuse of metric terms of the underlying finite element grid~\cite{KronbichlerKormann2019, dealII95}, and can thus only achieve a throughput of 5--8$\times10^7$~DoF/s.
    \item Increasing the polynomial degree for linearized operator application reduces the memory transfer per DoF as the ratio~$p^3/(p+1)^3$ between unique DoFs and quadrature points gets more favorable, such that the throughput is significantly~(up to $5$ times) increased up to a polynomial degree $p=4, 5, 6$. For higher polynomial degrees, the memory transfer per DoF increases slightly due to cache effects in the element-wise integration~\cite{KronbichlerKormann2019}.
    \item Regarding the precomputing and integration strategies in linearized operator application,
    we observe that when integrating over $\OO$, storing scalars is faster than recomputing all
    data, which is faster than storing tensors. Integrating over $\Ot$, this trend changes: storing
    tensors is the fastest option, followed by storing scalars, followed by recomputing all data.
    This sequence only mildly depends on the material model, but is of course highly dependent on
    the ratio of the number of operator evaluations to number of data updates. The fastest variant
    for integration over $\OO$ achieves in general better performance than the fastest variant
    for integration over $\Ot$.
    \item The amount of linearization data stored per integration point does not directly translate to memory traffic as already indicated in Tab.~\ref{tab:overview_precompute}. Storing tensors hence might circumvent loading the non-symmetric Jacobians ($\te{J}_0$ and $\te{J}_t$) of the finite element maps and the DoF vector of the linearization point needed otherwise -- see, e.g., Eqn.~\eqref{eqn:tau_cNH} for $\te{\tau}_\mathrm{cNH}$ which is symmetric, but depends on $\Grad{\ve{u}}$ (hence requiring $\te{J}_0$ and $\ve{u}_k$, see Tab.~\ref{tab:overview_precompute}).
    \item Comparing the constitutive models, we see that the fiber model yields the lowest throughput and has the highest memory traffic. However, the peak throughput achieved for each of the constitutive models is $\approx12.5\times10^8~\text{DoF/s}$ (fiber), $\approx17.5\times10^8~\text{DoF/s}$ (iNH and cNH), which is due to the higher complexity of the fiber model.
    \item Using $p=4$, the variant storing tensors is close to saturating the memory
          bandwidth of the machine (approx. 260--280~GB/s), 
          whereas the variant storing scalars reaches a lower memory transfer 
          (approx. 220~GB/s), but is limited by the additional computations 
          and unstructured data access. Recomputing all linearization 
          data yields 140--160~GB/s.
    %
    \item Jacobian-free Newton--Krylov solvers might be an attractive alternative to the present approach. Based on the current results, integration has to be carried out over the reference configuration, or the mapping and/or other integration point data has to be updated selectively.
\end{enumerate}
Given these observations, it is thus not clear a priori, which of the presented strategies reliably delivers the highest throughput. This decision has to be made based on measurements and hence depends on the problem at hand and the target hardware. Additionally, the fraction of updates performed per linearized operator application or residual evaluation impacts the obtained results as well. The choice of executing 100 repetitions here is therefore to be taken into account when interpreting the results. Nonetheless, the data transfer measurements shown in Fig.~\ref{fig:performance_measurements_fiber_only} and theoretical considerations of Sec.~\ref{sec:storage_strategies} provide guidelines for other hardware with possibly different arithmetic performance and memory bandwidth. For the practical influence of these low-level measurements, we apply the proposed schemes to a real-life example in the following section. 

For a broader perspective, we further compare these results to linear elasticity, where $\te{P} \eqq \lambda \, \id \, \Div \ve{u} + 2 \, \mu \, \Grad^\mathrm{S}\ve{u}$, and an alternative matrix-based implementation based on identical integration routines, which assembles and stores the system matrix. Taking the fastest variant of the fiber model, i.e., integrating over the reference configuration and storing scalars, as baseline, Tab.~\ref{tab:performance_matrix_based_matrix_free_linear_easticity} lists the relative throughput and memory transfer. The linear elastic model yields 2.6--3.1 times higher throughput having 0.17--0.36 times the memory transfer, while the matrix based implementation delivers merely 0.71 ($p=1$) to 0.01 ($p=6$) times the throughput. The memory transfer using the matrix-based variant of the fiber model is higher by a factor of 1.40 for $p=1$, which further increases up to 112.0 for $p=6$, taking the matrix-free implementation as baseline.
\begin{table} 
	\centering
	\caption{
		Relative throughput and memory transfer of linearized operator application compared to a baseline method (value 1.0). The baseline method refers to the fiber model, integration over $\OO$ and a matrix-free evaluation that stores scalars. This baseline method is modified in two ways: i) by changing the material model to linear elasticity, and ii) by choosing a matrix-based implementation for the fiber model.}
	{
		\scriptsize
		\label{tab:performance_matrix_based_matrix_free_linear_easticity} 
		\begin{tabular}{||l | r || c c c c c c 
				||}
			\hline
			\multicolumn{2}{||r||}{                     }& & & & & & 
			\\[-1.5ex]
			\multicolumn{2}{||r||}{polynomial degree $p$}&1&2&3&4&5&6
			\\[0.75ex]
			\hline\hline
			&&&&&&&
			\\[-1.25ex]
			\multirow{2}{*}{i) linear elasticity, matrix-free}
			&rel. throughput
			&2.66&2.73&2.69&2.71&2.63&2.93
			\\[0.75ex]
			&rel. memory transfer
			&0.36&0.23&0.25&0.22&0.24&0.17
			\\[0.75ex]
			\hline
			&&&&&&&
			\\[-1.5ex]
			\multirow{2}{*}{ii) fiber model, matrix-based} 
			&rel. throughput
			&0.71&0.15&0.06&0.03&0.02&0.01
			\\[0.75ex]
			&rel. memory transfer 
			&1.39&7.32&20.71&43.47&72.84&112.29
			\\[0.75ex]
			\hline
		\end{tabular}
	}
\end{table}

\subsection{Application to biomechanics: iliac bifurcation}
\label{sec:results_iliac}

\begin{center}
    \begin{figure}
    	\centering
        \begin{subfigure}{0.15\textwidth}
            \centering
            \begin{overpic}[width=1.0\textwidth,draft=\draftIliac]{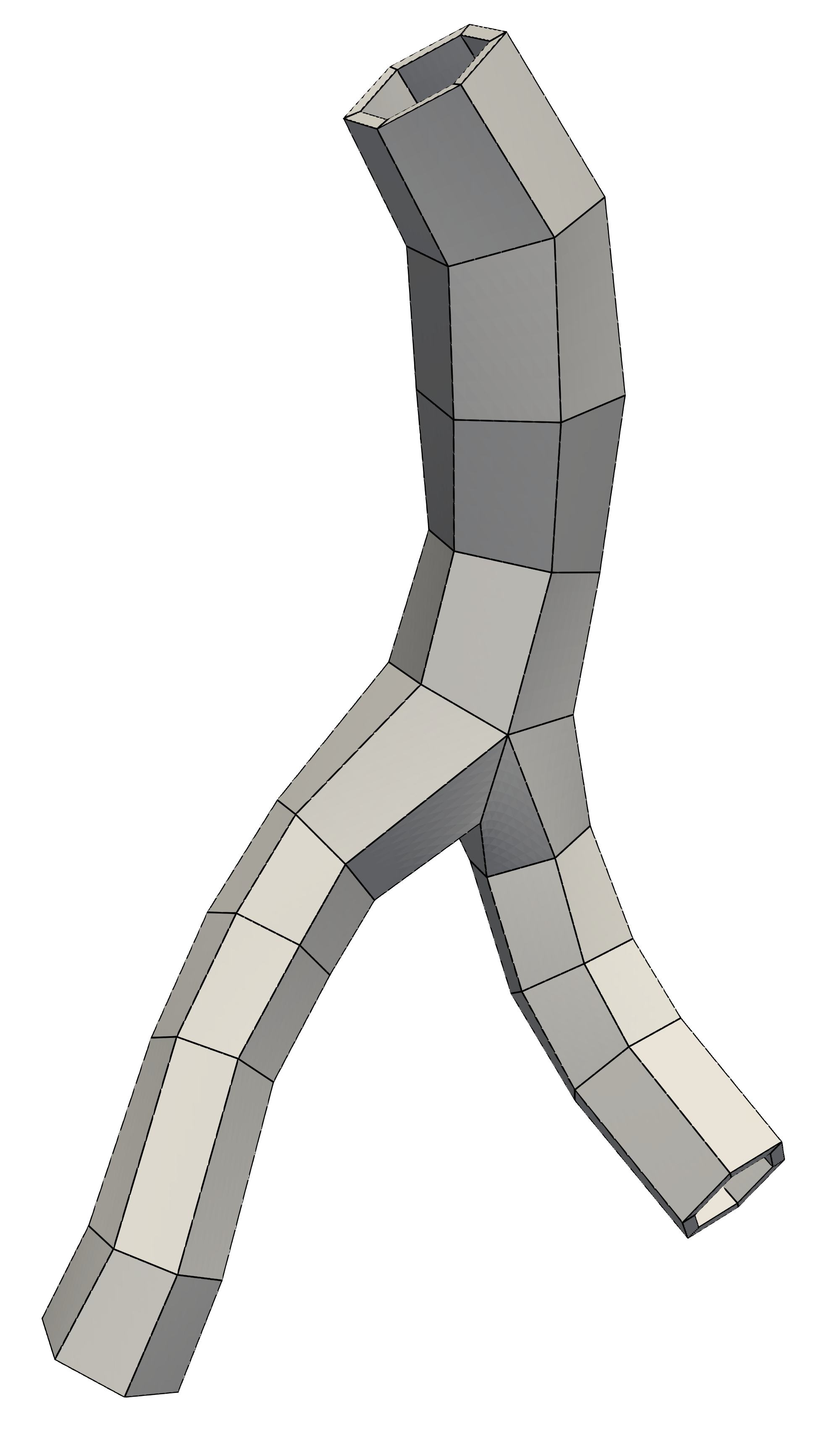}
                \put(-3,57){\frame{\includegraphics[width=0.5\linewidth,draft=\draftIliac]{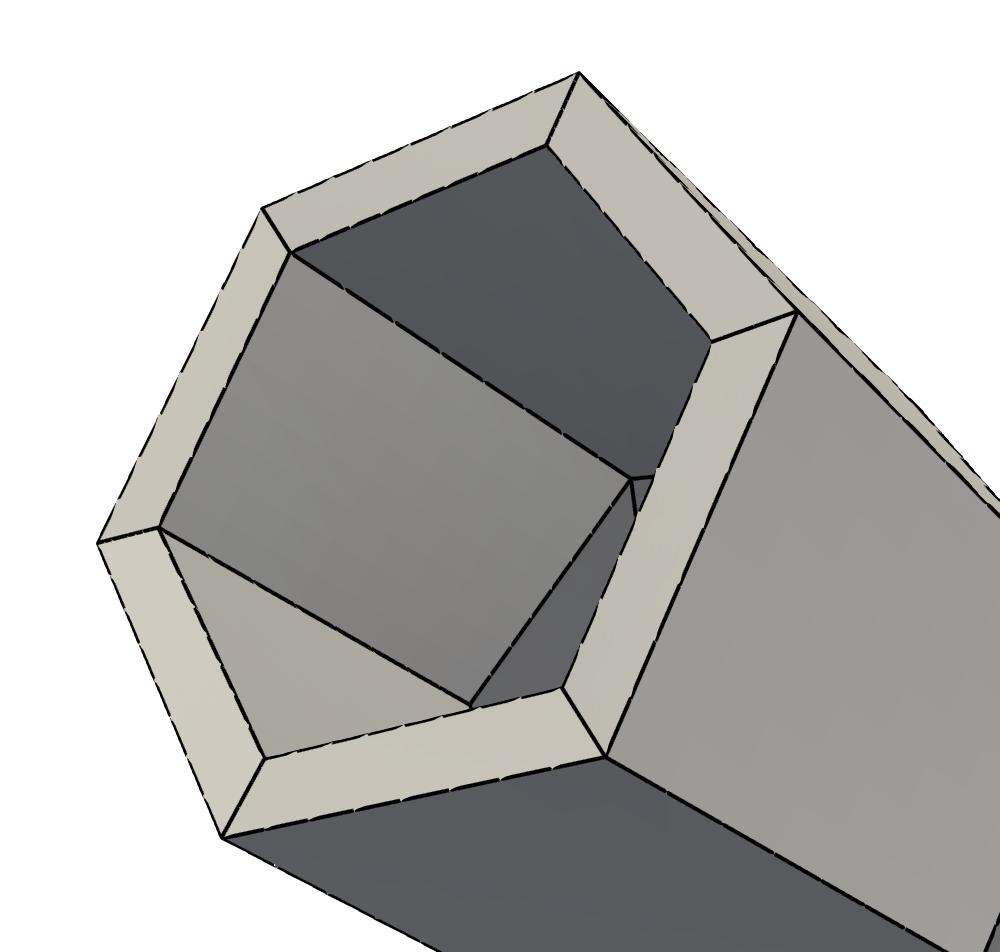}}}
            \end{overpic}
            \caption{$l=0$, $p=1$}
        \end{subfigure}
    	\hfil
        \begin{subfigure}{0.15\textwidth}
            \centering
            \begin{overpic}[width=1.0\textwidth,draft=\draftIliac]{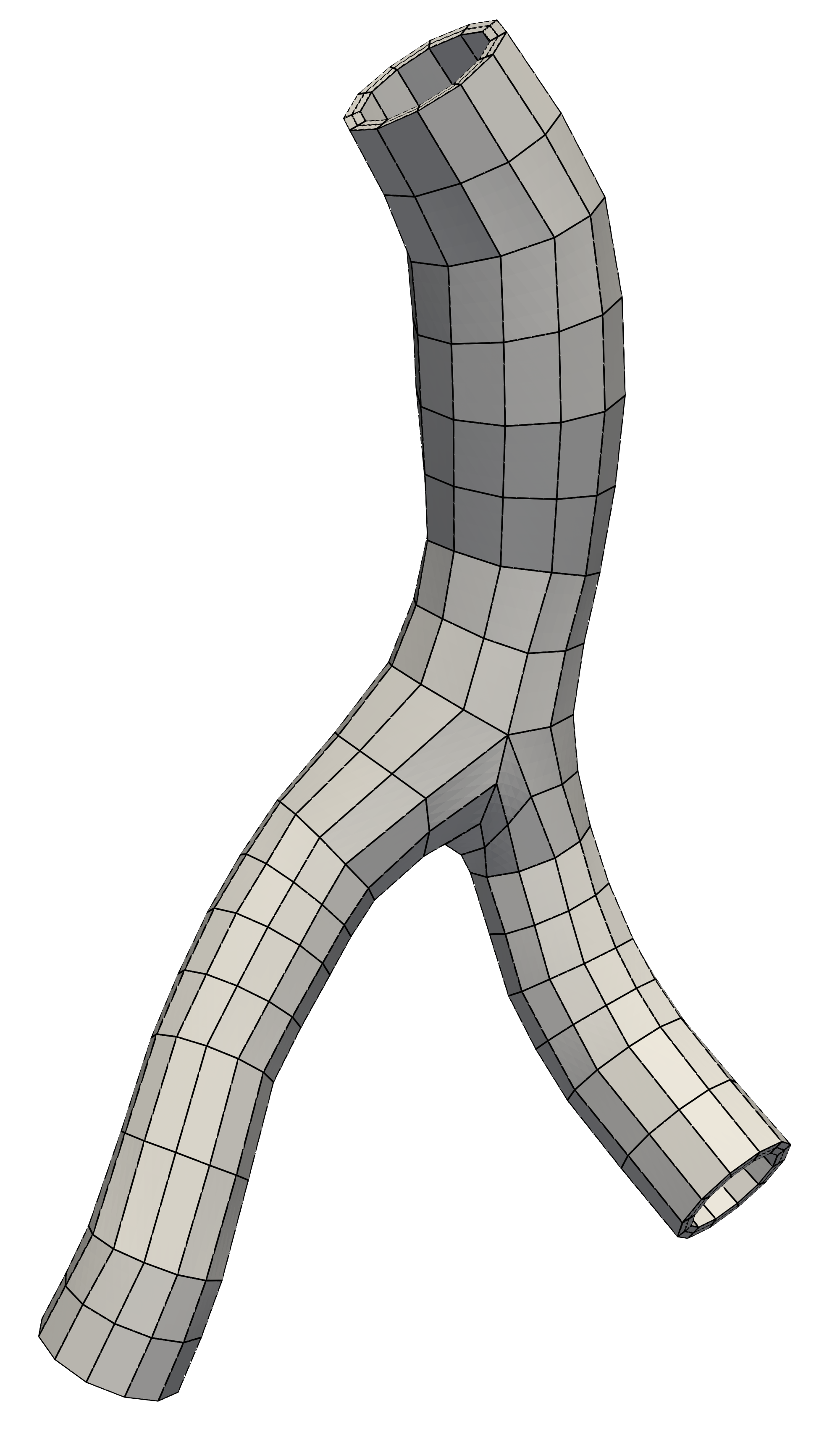}
                \put(-3,57){\frame{\includegraphics[width=0.5\linewidth,draft=\draftIliac]{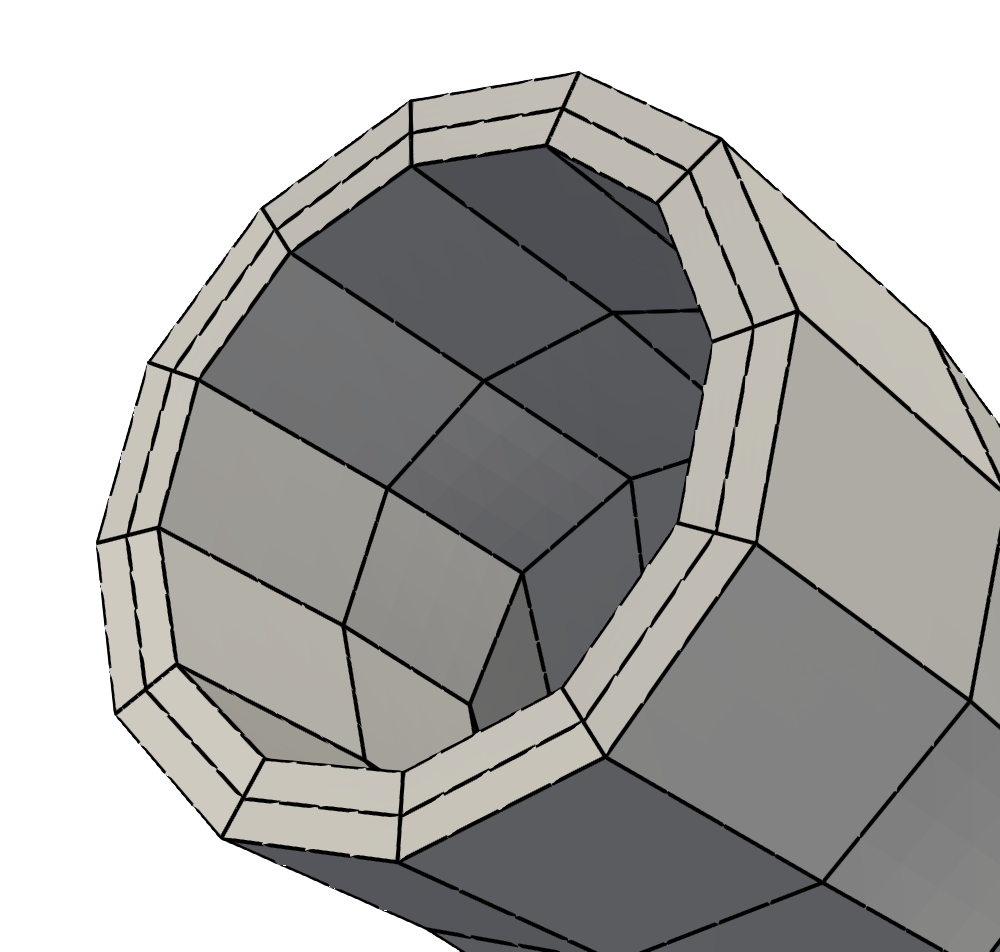}}}
            \end{overpic}
            \caption{$l=1$, $p=1$}
        \end{subfigure}
        \hfil
        \begin{subfigure}{0.15\textwidth}
            \centering
            \begin{overpic}[width=1.0\textwidth,draft=\draftIliac]{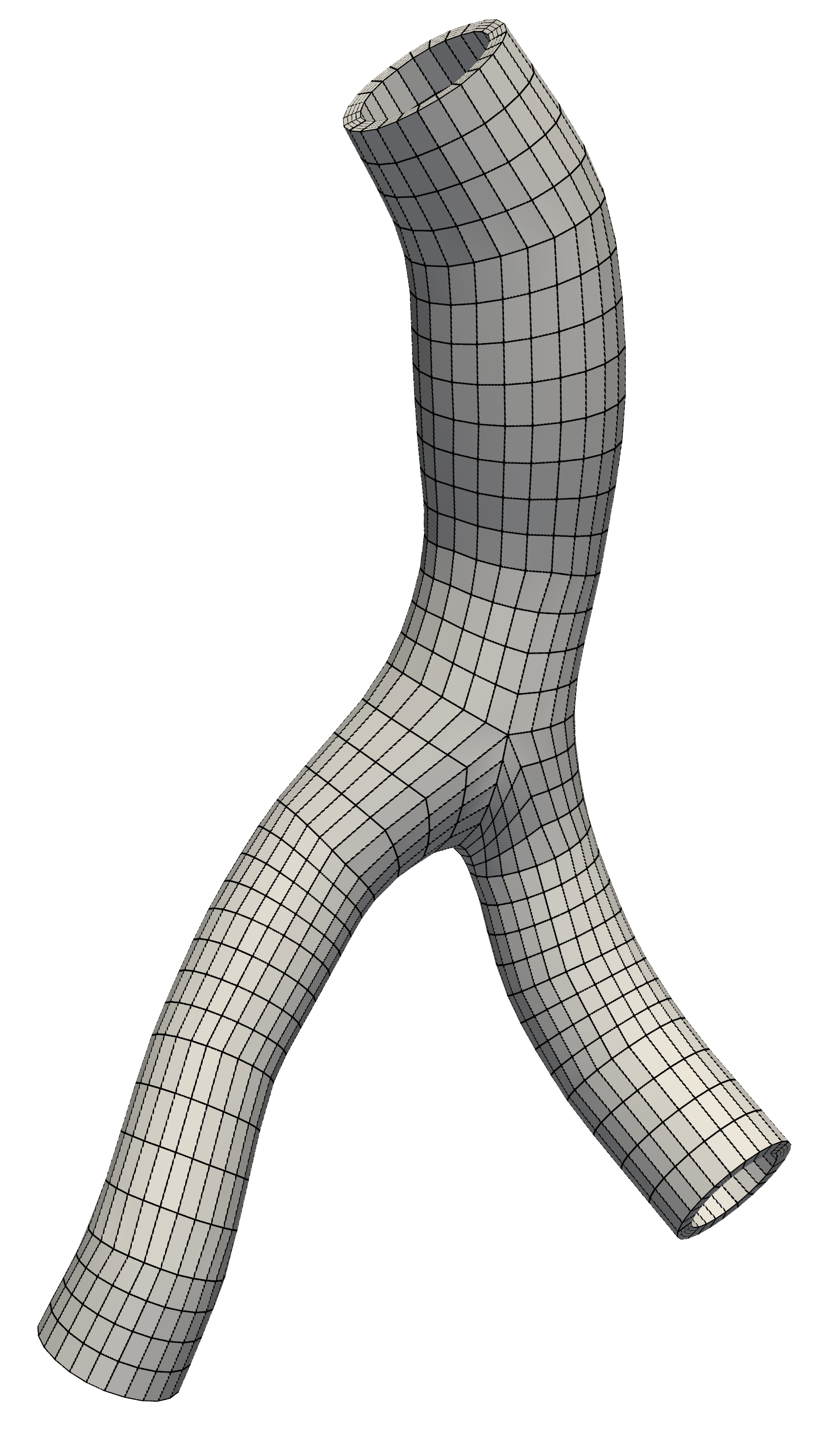}
                \put(-3,57){\frame{\includegraphics[width=0.5\linewidth,draft=\draftIliac]{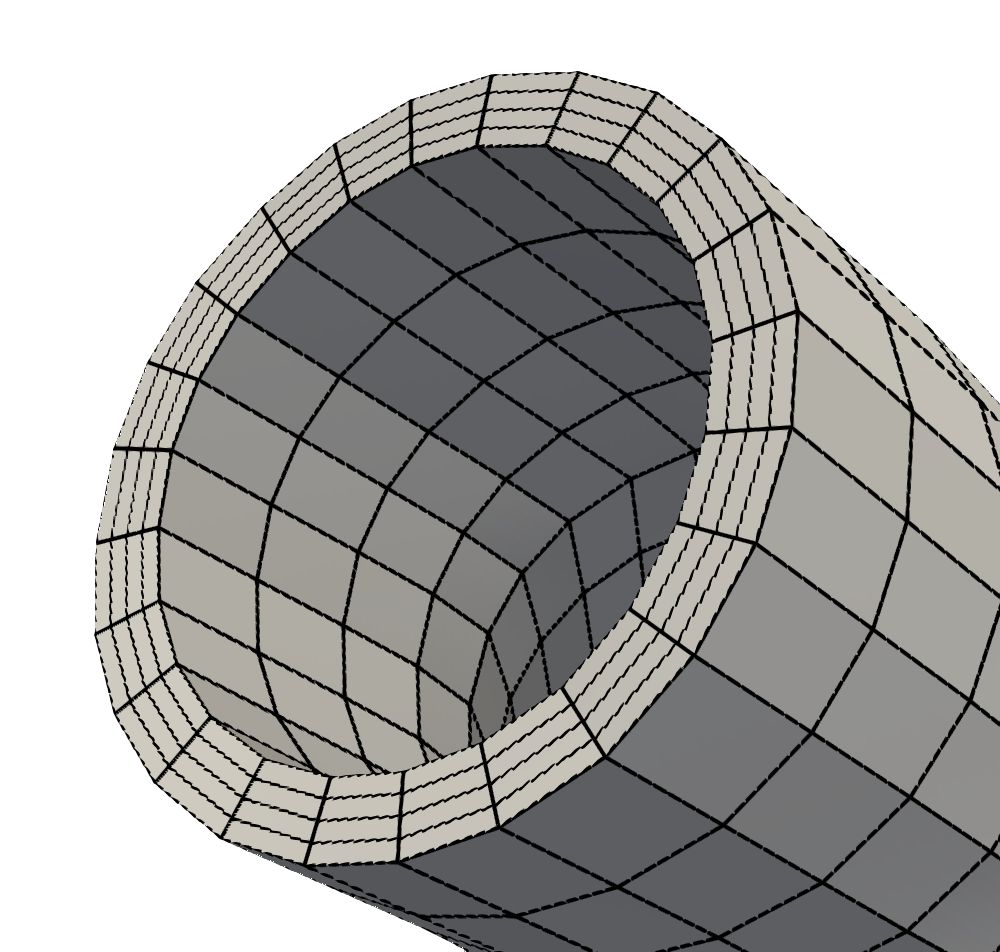}}}
            \end{overpic}
            \caption{$l=2$, $p=1$}
        \end{subfigure}
    	\hfil
        \begin{subfigure}{0.15\textwidth}
            \centering
            \begin{overpic}[width=1.0\textwidth,draft=\draftIliac]{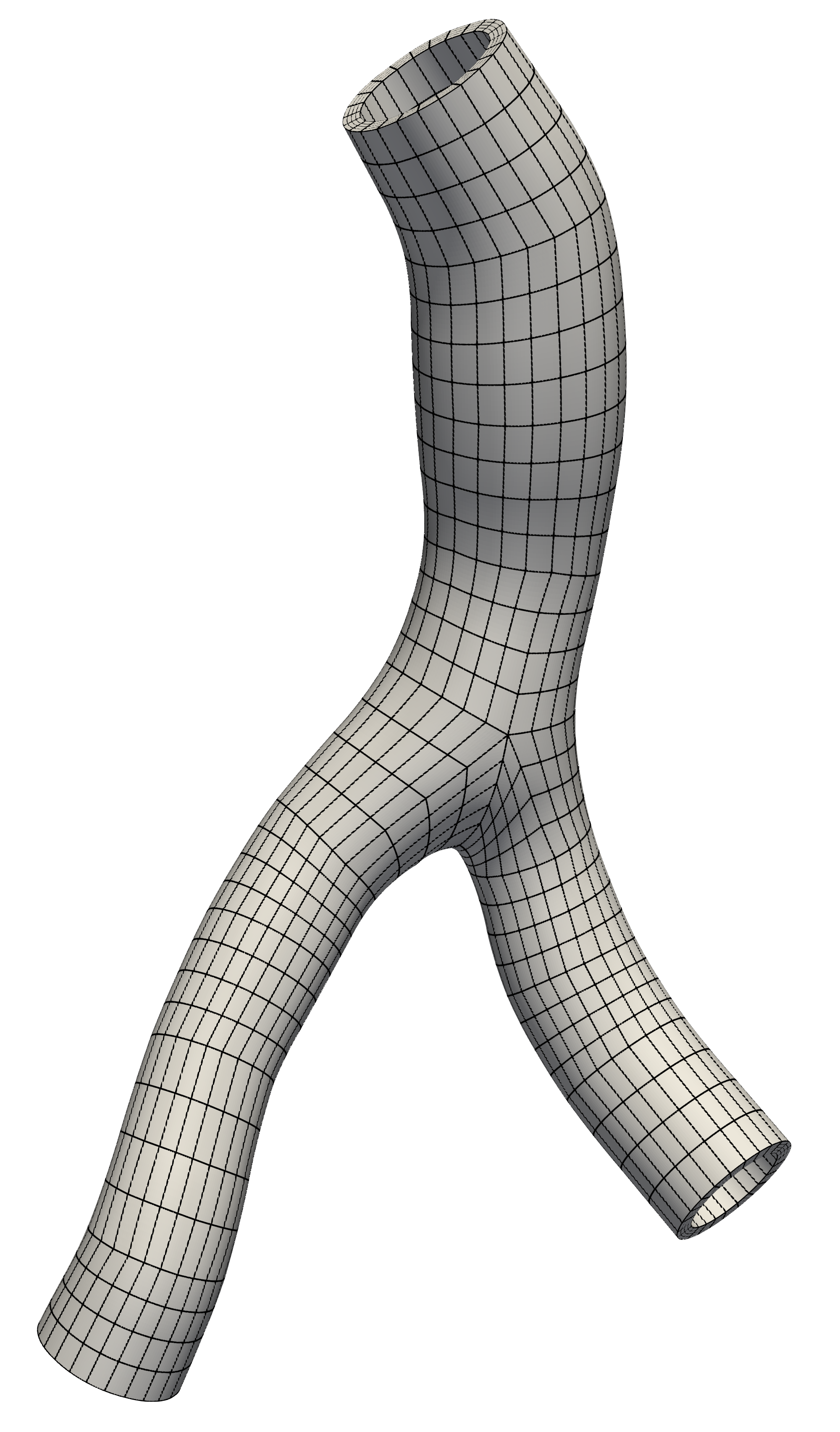}
                \put(-3,57){\frame{\includegraphics[width=0.5\linewidth,draft=\draftIliac]{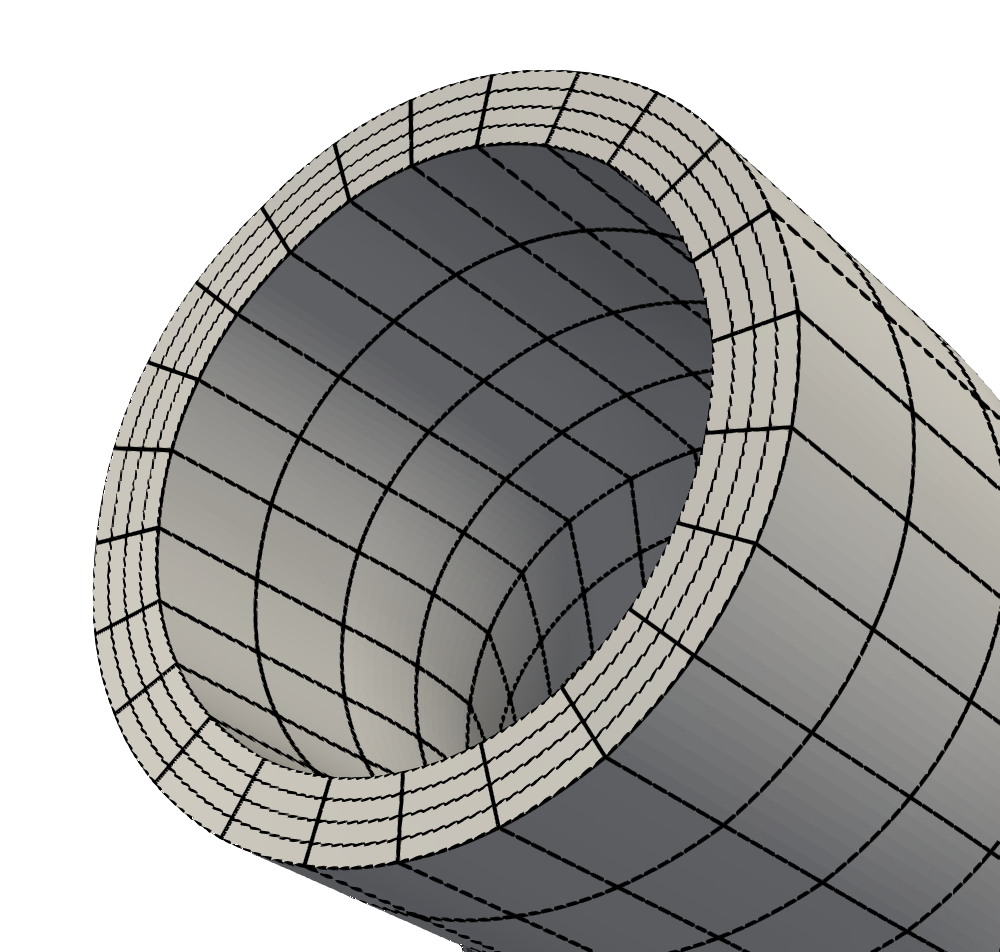}}}
            \end{overpic}
            \caption{$l=2$, $p=2$}
        \end{subfigure}
        \caption{Discretizations of the iliac bifurcation using $l=0,1,2$ refinement levels and mapping degree $p=1,2$.}
	\label{fig:iliac_bifurcation_grids}
    \end{figure}
\end{center}

The numerical results in this section focus on the application to a patient-specific geometry of an iliac bifurcation, which can be considered a prototypical configuration of practical interest. The spatial approximation via a pure-hex finite element grid follows ideas from Bo\v{s}njak~et~al.~\cite{Bosnjak2024}, resulting in higher-order discretizations as shown in Fig.~\ref{fig:iliac_bifurcation_grids} for refinement level $l = 0,1,2$ and mapping degree $p=1,2$.
To demonstrate the advantages of the present approach, we aim for a problem size of $10^6$~DoFs on the finest level, emulating engineering-size structural mechanics problems. The polynomial degree is varied from $p=1$ to $p=5$, executing uniform refinement to generate the nested multigrid hierarchy and reaching the target number of DoFs based on a coarse grid with 78 cells yielding 516 DoFs for $p=1$. Given the fact that the coarse mesh is non-trivial and already contains a certain number of elements to resolve the topology, the resulting discretizations yield slightly differing DoF numbers, see Tab.~\ref{tab:iliac_element_number}. 

Contrary to the previous example, which lies in the regime with saturated caches, 
this practical example of limited size showcases the effects of (partial) caching. 
Such a setup is relevant for practical application, when the fastest time to solution 
given sufficient compute resources is of interest. For the discretizations as listed 
in Tab.~\ref{tab:iliac_element_number} and memory traffic related to integration point data, 
see Tab.~\ref{tab:overview_precompute}, we estimate that roughly 50--70\% of the overall data 
can be cached in some cases. Polynomial degrees $p=2,3,4$ are favorable in this regard, 
but this also depends on the spatial or material integration approach chosen and if linearization data 
is stored. However, even when precomputing scalar or tensorial quantities, 
a significant fraction might be cached as well.
\begin{table}
	\centering
	\caption{Number of continuous finite elements $N_\mathrm{el}$ and DoFs per polynomial degree $p$ on an $l$-times uniformly refined initial coarse grid of the iliac bifurcation with $78$ elements. The discretizations left to the vertical double line are intermediate $h$-refined levels with $p=1$, whereas entries on the right list fine-level discretizations.}
	{
		\scriptsize
		\label{tab:iliac_element_number} 
		\begin{tabular}{||l | c c c c c || c c c c c||}
			\hline
			&&&&&&&&&&\\[-1.5ex]
			$p$&1&1&1&2&1&1&2&3&4&5\\[0.75ex]
			\hline\hline
			&&&&&&&&&&\\[-1.25ex]
            $l$              &0   &1      &2     &2 &3               &4                &3                &2                   &2                &2    \\[0.75ex]
            $N_\mathrm{el}$  &78  &624    &4992 &4992  &39936            &319488           &39936            &4992                &4992             &4992 \\[0.75ex]
   			DoFs         &516 &2961   &19245 &136701 &136701 &1.03$\times10^6$ &1.03$\times10^6$ &442221 &1.03$\times10^6$ & 1.98$\times10^6$  \\[0.75ex]
			\hline
		\end{tabular}
	}
\end{table}

\begin{figure}
   	\centering
    \begin{subfigure}[b]{0.25\textwidth}
        \centering
        \begin{overpic}[width=1.0\textwidth,draft=\draftOri]{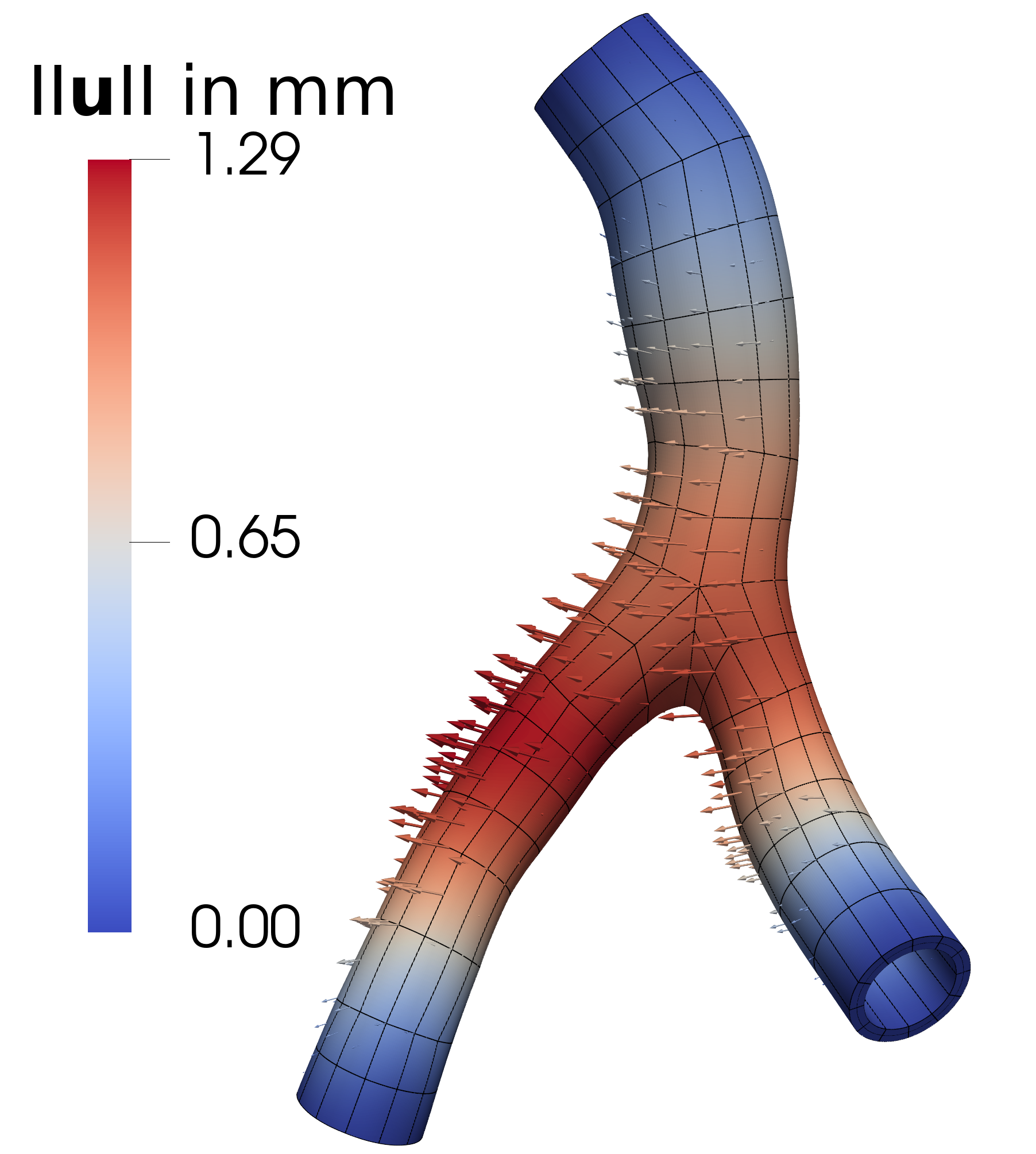}
        \end{overpic}
        \caption{displacement $\ve{u}$}
    \end{subfigure}
   	\hfil
    \begin{subfigure}[b]{0.3275\textwidth}
        \centering
        \begin{overpic}[width=\textwidth,draft=\draftOri]{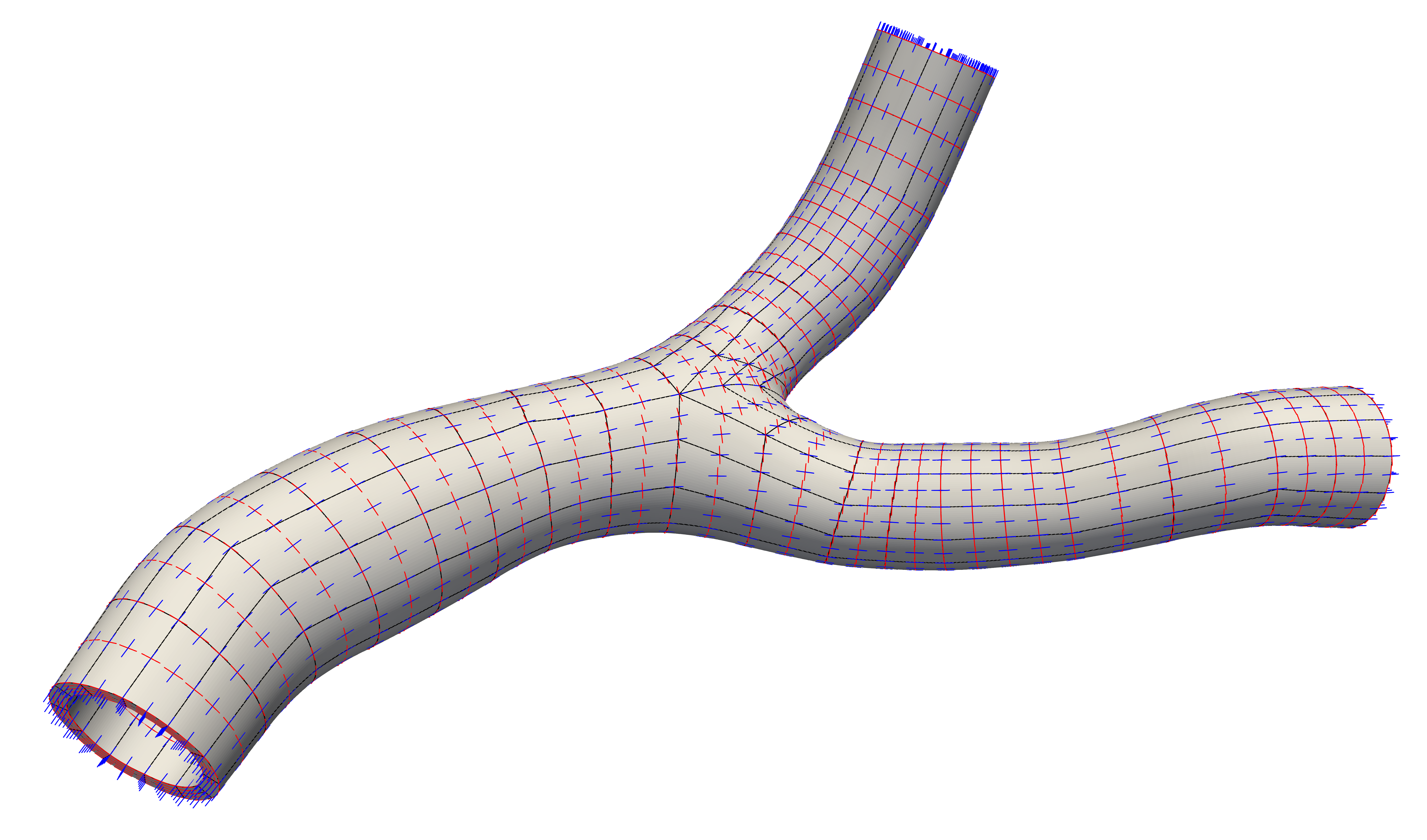}
        \end{overpic}
    	\vspace*{5mm}
        \caption{vectors $\ve{E}_1$ (red) and $\ve{E}_2$ (blue)}
    \end{subfigure}
	\hfil
    \begin{subfigure}[b]{0.3275\textwidth}
        \centering
        \begin{overpic}[width=\textwidth,draft=\draftOri]{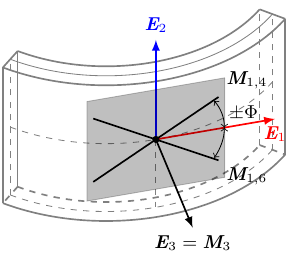}
        \end{overpic}
        \caption{mean fibers $\ve{M}_{1,4}$ and $\ve{M}_{1,6}$}
    \end{subfigure}
    \caption{Iliac bifurcation structural mechanics: tissue displacement due to pressure differential (a), material orientation (b) and sketch for derived mean fiber orientations (c). The circumferential ($\ve{E}_1$, red) and longitudinal direction vectors ($\ve{E}_2$, blue) are used to describe the helical fiber reinforcement on each $h$-level.}
\label{fig:iliac_bifurcation_orientation}
\end{figure}

This test involves an initial boundary value problem, adding an acceleration term, $\rho \ddot{\ve{u}}$, to the linear momentum balance equation employing a standard single-step WBZ-$\alpha$ time integration scheme~\cite{Wood1980} with spectral radius $\rho_\infty=0.8$ and a time step size $\Delta t = 0.1~\text{ms}$. The fiber model is employed with parameters listed in Tab.~\ref{tab:fiber_parameters}, a physiological density of $\rho = 1200~\text{kg/m}^2$, and the local coordinate systems as shown in Fig.~\ref{fig:iliac_bifurcation_orientation}(b) (see~\cite{Schussnig2022_a, Schussnig2021_b} for details). 
Focusing on a simplified structural mechanics problem, the fluid flow's effect on the tissue is roughly approximated by a uniform pressure differential acting on the vessel wall. Naturally, a fluid--structure interaction approach might in fact yield quite different results. In our simplified setup, the in- and outlets of the vessel are fixed; a rough simplification, which should be replaced for a more realistic model. Due to the lack of viscoelastic support on the exterior, the internal pressure ``straightens'' the curved vessel. For the present purposes, it suffices to choose the pressure differential such that a displacement of $\approx1.3~\text{mm}$ is obtained as shown in Fig.~\ref{fig:iliac_bifurcation_orientation}(a), referring the interested reader to
~\cite{Schussnig2022_a, Schussnig2021_a, Schussnig2021_b}.

Regarding the preconditioner settings, we employ 6 smoothing sweeps of the Jacobi-type smoother, see Sec.~\ref{sec:matrix-free_solver}. 
The coarse-level preconditioner uses the \texttt{Amesos-KLU} direct solver on the coarsest algebraic level once the total number of degrees of freedom on the level falls below 2000, which is in this present example reached immediately with 516 DoFs on level $l=0$. However, AMG is also used for comparison with a matrix-based preconditioner on the finest level, simply referred to as AMG approach. It is important to note that the matrix-vector product of both alternatives are realized in matrix-free fashion in the Krylov solver, but the AMG preconditioner utilizes the system matrix for setup and matrix-vector products internally.
If not stated otherwise, the matrix-free preconditioner operates in single precision, while the AMG preconditioner operates in double precision.
When used on the fine level, the AMG preconditioner adopts a single V-cycle with Chebyshev smoother (degree 6). All these settings were tuned to yield shortest runtimes in the present scenario with a polynomial degree of $p=5$, while increasing the number of sweeps or changing the smoother type in the AMG case did not speed-up runtimes significantly for the other polynomial degrees employed. In fact, a time-dependent problem was chosen for this test due to the AMG preconditioner encountering convergence issues for higher-order finite element discretizations.
With this numerical setup, we investigate the wall time per Newton solve averaged over 5 time steps and record the number of average FGMRES iterations (absolute tolerance of $10^{-12}$, relative tolerance of $10^{-3}$, maximal Krylov space dimension of 30 before a restart) to solve the arising Newton update steps in the nonlinear solver, which considers an absolute tolerance of $10^{-8}$ and relative tolerance of $10^{-3}$. Due to the small time step size, convergence is reached in one to two Newton steps per time step due to the rather loose convergence criteria. 

\begin{figure}
	\centering
	\begin{subfigure}[b]{.245\textwidth}
		\centering	
		\begin{overpic}[height=4.5cm, draft=\draftIliac]{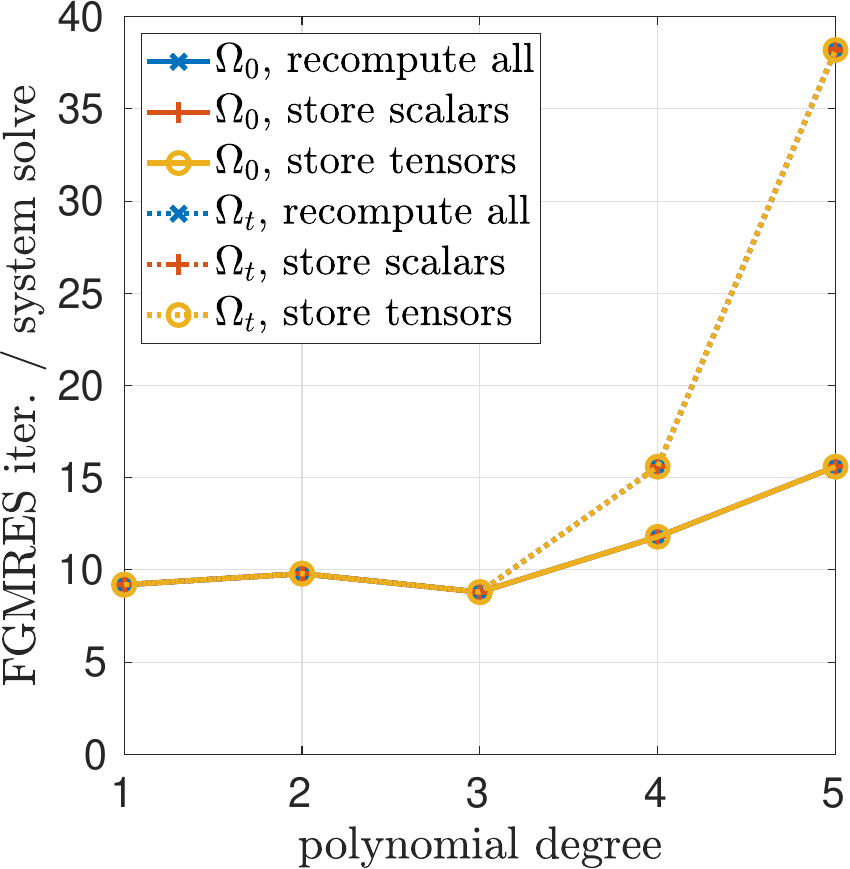}
			\put(60,20){\tiny$\OO$}
			\put(65,25){\vector(1,2){5}}
			\put(54,54){\tiny$\Ot$}
			\put(62,52){\vector(2,-1){10}}
		\end{overpic}	
		\caption{iterations: AMG}
	\end{subfigure}%
	\hfil
	\begin{subfigure}[b]{.245\textwidth}
		\centering
		\begin{overpic}[height=4.5cm, draft=\draftIliac]{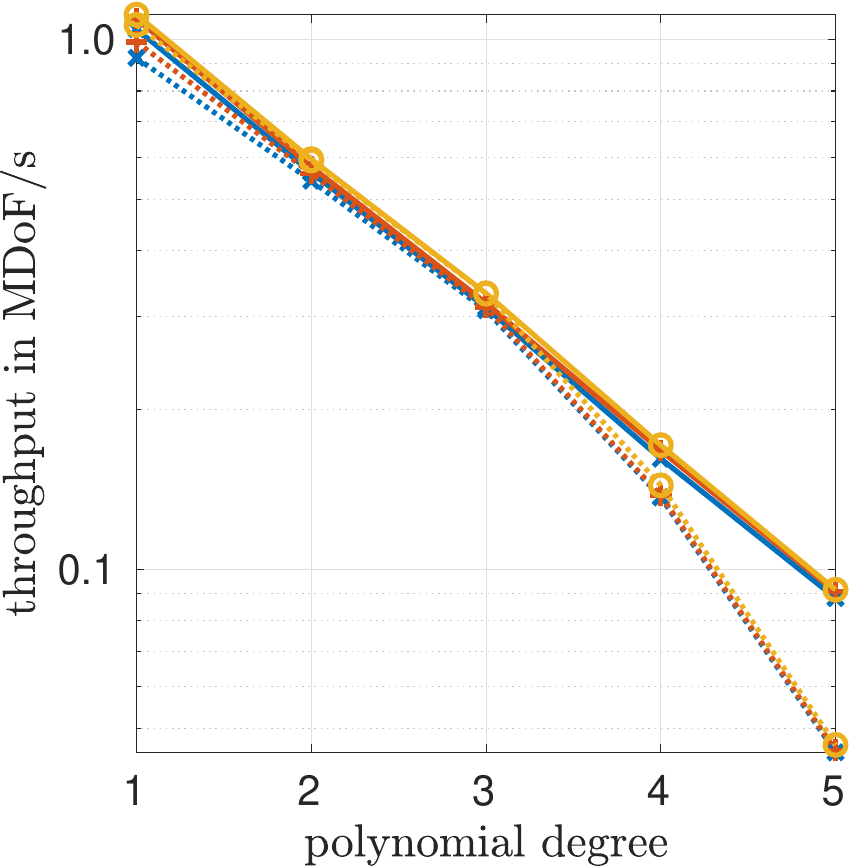}
			\put(65,22){\tiny$\Ot$}
			\put(72,25){\vector(2,1){10}}
			\put(85,54){\tiny$\OO$}
			\put(87,52){\vector(0,-1){10}}
		\end{overpic}	
		\caption{throughput: AMG}
	\end{subfigure}
	\hfil
	\begin{subfigure}[b]{.245\textwidth}
		\centering		
        \includegraphics[height=4.5cm, draft=\draftIliac]{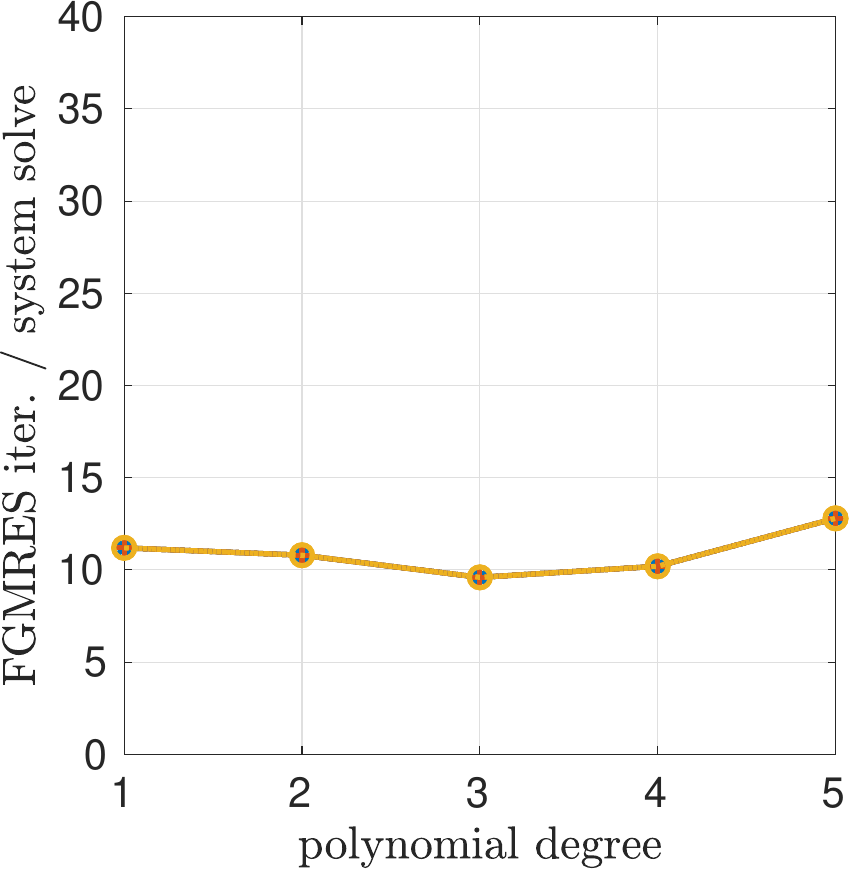}
		\caption{iterations: $hp$-multigrid}
	\end{subfigure}%
	\hfil
	\begin{subfigure}[b]{.245\textwidth}
		\centering
		\includegraphics[height=4.5cm, draft=\draftIliac]{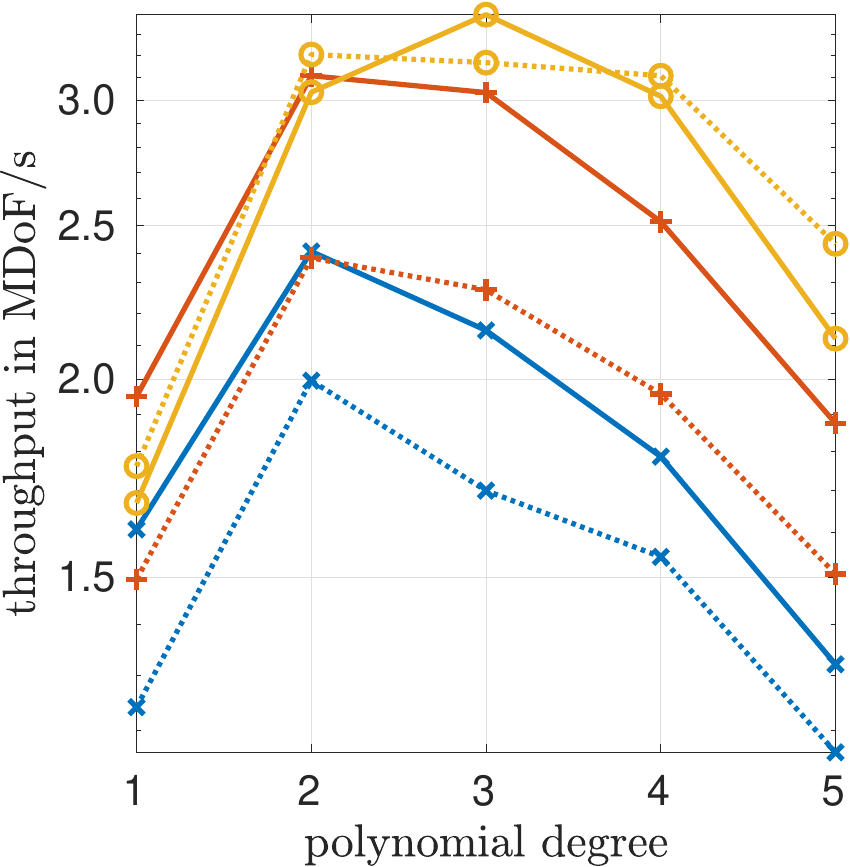}
		\caption{throughput: $hp$-multigrid}
	\end{subfigure}
	\caption{FGMRES iterations until convergence adopting the AMG preconditioner on the fine grid directly in variants AMG-$\OO$/$\Ot$-none/scalar/tensor (a) or matrix-free geometric multigrid in variants $hp$MG-MF-MP-$\OO$/$\Ot$-none/scalar/tensor (c) and respective average throughput for a single system solve (b, d).}
	\label{fig:iliac_bifurcation_performance}
\end{figure}

Results comparing integration over the material or spatial configuration and the different storage strategies are depicted in Fig.~\ref{fig:iliac_bifurcation_performance}. The related speed-up of the matrix-free over the matrix-based preconditioner is reported in Tabs.~\ref{tab:iliac_performance_comparison_mf_vs_AMG}--\ref{tab:iliac_performance_comparison_material_vs_spatial}. 
The considered variants summarized in Tab.~\ref{tab:iliac_algorithm_variants} encompass matrix-free and matrix-based preconditioners, single-precision and double-precision floating point arithmetic, integration over the spatial configuration ($\Ot$) or the material configuration ($\OO$), and precomputing strategies storing scalars or tensors, or recomputing all quantities. 
\begin{table}
	\centering
	\caption{Overview of the employed variants of the finite strain elasticity solver.}
	{
		\scriptsize
		\label{tab:iliac_algorithm_variants} 
		\begin{tabular}{|| l | c c c c c c||}
			\hline
			&&&&&&\\[-1.5ex]
			\multirow{2}{*}{variant}
			& \multirow{2}{*}{preconditioner}
			& preconditioner
			& operator
			& floating point
			& domain of
			& precomputing
			\\[0.75ex]
			& 
			& evaluation
			& evaluation
			& precision
			& integration
			& strategy\\[0.75ex]
			\hline\hline
			&&&&&&\\[-1.25ex]
			$hp$MG-MF-MP-$\OO$-none&$hp$-multigrid&matrix-free&matrix-free&mixed&material config. $\OO$&recompute all\\[0.75ex]
			$hp$MG-MF-MP-$\OO$-scalar&$hp$-multigrid&matrix-free&matrix-free&mixed&material config. $\OO$&store scalars\\[0.75ex]
			$hp$MG-MF-MP-$\OO$-tensor&$hp$-multigrid&matrix-free&matrix-free&mixed&material config. $\OO$&store tensors\\[0.75ex]
			$hp$MG-MF-MP-$\Ot$-none&$hp$-multigrid&matrix-free&matrix-free&mixed&spatial config. $\Ot$&recompute all\\[0.75ex]
			$hp$MG-MF-MP-$\Ot$-scalar&$hp$-multigrid&matrix-free&matrix-free&mixed&spatial config. $\Ot$&store scalars\\[0.75ex]
			$hp$MG-MF-MP-$\Ot$-tensor&$hp$-multigrid&matrix-free&matrix-free&mixed&spatial config. $\Ot$&store tensors\\[0.75ex]
			%
			%
			AMG-$\OO$-none&AMG&matrix-based&matrix-free&double&material config. $\OO$&recompute all\\[0.75ex]
			AMG-$\OO$-scalar&AMG&matrix-based&matrix-free&double&material config. $\OO$&store scalars\\[0.75ex]
			AMG-$\OO$-tensor&AMG&matrix-based&matrix-free&double&material config. $\OO$&store tensors\\[0.75ex]
			AMG-$\Ot$-none&AMG&matrix-based&matrix-free&double&spatial config. $\Ot$&recompute all\\[0.75ex]
			AMG-$\Ot$-scalar&AMG&matrix-based&matrix-free&double&spatial config. $\Ot$&store scalars\\[0.75ex]
			AMG-$\Ot$-tensor&AMG&matrix-based&matrix-free&double&spatial config. $\Ot$&store tensors\\[0.75ex]
			%
			%
			$hp$MG-MF-DP-$\OO$-tensor&$hp$-multigrid&matrix-free&matrix-free&double&material config. $\OO$&store tensors\\[0.75ex]
			$hp$MG-MB-DP-$\OO$-tensor&$hp$-multigrid&matrix-based&matrix-based&double&material config. $\OO$&store tensors\\[0.75ex]
			\hline
		\end{tabular}
	}
\end{table}

Based on this data, we make the following observations:
\begin{enumerate}[i)]
    \item The linearized operators formed by integration of the respective weak forms over the spatial and material configurations yield different approximations of the same operator. Once the nonlinear solver converges, both operators are the same up to round-off. The performance of the AMG preconditioner for higher polynomial degrees and integrating over $\Ot$ suffers remarkably. This is due to the AMG smoother settings aimed at fastest time to solution, which is not the most robust choice. That is, more expensive smoothers and/or multiple V-cycles in the AMG case yield nearly identical iteration counts irrespective of the domain of integration, similar to the matrix-free approach. The remaining combinations result in low and almost $p$-independent iteration counts.
    Here, the matrix-free $hp$-multigrid has been found to be more robust than the AMG variant used in a black-box fashion.
    
    \item The overall throughput of the proposed matrix-free approach is 
    1.34--52 times 
    higher than the throughput of the AMG approach (Tab.~\ref{tab:iliac_performance_comparison_mf_vs_AMG}, top). 
    Factoring out bad preconditioner performance, we note improvements of 
    1.32--19 times
    , respectively (Tab.~\ref{tab:iliac_performance_comparison_mf_vs_AMG}, bottom).
    
    \item Storing tensors is the fastest option in most cases, see Tab.~\ref{tab:iliac_performance_comparison_tensor_vs_scalar_or_recompute_all}. For the matrix-free approach and integrating over the spatial configuration, speed-ups of 1.18--2.09 are observed, while integrating over $\OO$, we note a relative speed of 0.86--1.64 (storing scalars is faster for $p=1,2$). For the AMG approach, however, integration of the operators is no longer the dominating part of the algorithm, such that only mild improvements stemming from faster integration are observed.
    
    \item Integrating over the material configuration is in almost all cases faster than integrating over the spatial configuration with the only exception being the tensor-storing matrix-free variant, see Tab.~\ref{tab:iliac_performance_comparison_material_vs_spatial}.
    For the matrix-free variant, speed-up factors are 0.95--1.33,
    while with the AMG preconditioner leads to factors of 1.00--1.99, with higher values related to increased iteration counts.
    
    \item In summary, storing tensorial quantities barely pays off compared to storing scalars for the present constitutive models, since the most complex operations involve the stored scalars and many tensorial quantities cannot be precomputed as they depend on the solution vector. Updating the mapping data for the approach integrating over $\Ot$ is costly and this variant is thus not preferable over the alternative integrating over $\OO$ despite the more involved integrals. This depends on the constitutive law and ratio of linear iterations required per Newton step. However, both formulations perform similar here when precomputing tensors.
\end{enumerate}

Note that especially point iii) in the above list is opposed to the throughput example from Sec.~\ref{sec:results_performance_block}, where storing scalars was faster than recomputing all quantities, which was faster than storing tensors. In the present example, storing tensors is the fastest option, followed or tied with the variant storing scalars, while the variant recomputing all terms was consistently the slowest.
This is due to the small problem size compared to the available 200~MB of combined L2 and L3 cache, 
which fits potentially large portions of the (integration point) data depending on the approach chosen. 
Recomputing all data is the most compute intense approach, while storing scalar or tensorial data 
trades arithmetic operations for loading data from main memory. If a potentially large fraction 
of the integration point data resides in cache, as is the case for some of the variants in this example, 
loading precomputed data comes with a smaller penalty with regards to achievable throughput. 
In the present example, trends are hence partially different from the previously presented 
throughput results. The results in Sec.~\ref{sec:results_performance_block} show trends for saturated caches, while the results here show trends for engineering-size problems with plenty compute resources used to reduce solver turnaround times.
\begin{table}
	\centering
	\caption{Overall (top) and per iteration (bottom) speed-up of the matrix-free $hp$-multigrid ($hp$-MG) preconditioner ($hp$MG-MF-MP-$\OO$/$\Ot$-none/scalar/tensor) over double-precision matrix-based AMG preconditioner (AMG-$\OO$/$\Ot$-none/scalar/tensor). Both variants use matrix-free operator evaluation in the Krylov solver. Computed from data displayed in Fig.~\ref{fig:iliac_bifurcation_performance}.}
	{
		\scriptsize
		\label{tab:iliac_performance_comparison_mf_vs_AMG} 
		\begin{tabular}{|| c c | l || c c c c c||}
			\hline
			\multicolumn{3}{||r||}{}&&&&&\\[-2.0ex]
			\multicolumn{3}{||r||}{polynomial degree $p$}&1&2&3&4&5\\
			\hline\hline
			&&&&&&&\\[-1.25ex]
			\hspace*{-5mm}
   			\multirow{6}{*}{\begin{turn}{90}\hspace{-4mm}speed-up\end{turn}}
   			\hspace*{-5mm}
   			&
   			\hspace*{-3mm}
   			\multirow{6}{*}{\begin{turn}{90}\hspace{-4mm}$hp$-MG vs. AMG\end{turn}}
   			 &$\OO$, recompute all & 1.54 &   4.29 &   6.88 &  11.05 &  14.97\\[0.75ex]
            &&$\OO$, store scalars & 1.78 &   5.40 &   9.53 &  15.00 &  20.68\\[0.75ex]
            &&$\OO$, store tensors & 1.50 &   5.10 &  10.23 &  17.59 &  23.18\\[0.75ex]
   			&&$\Ot$, recompute all & 1.34 &   3.68 &   5.52 &  11.30 &  25.76\\[0.75ex]
            &&$\Ot$, store scalars & 1.51 &   4.27 &   7.28 &  14.20 &  33.08\\[0.75ex]
            &&$\Ot$, store tensors & 1.65 &   5.42 &   9.56 &  21.58 &  52.22\\[0.75ex]
            \hline\hline
			&&&&&&&\\[-1.25ex]
			\hspace*{-5mm}
	        \multirow{6}{*}{\begin{turn}{90}\hspace{-4mm}speed-up/iter.\end{turn}}
	        \hspace*{-5mm}
	        &
	        \hspace*{-3mm}
	        \multirow{6}{*}{\begin{turn}{90}\hspace{-4mm}$hp$-MG vs. AMG\end{turn}}
     		 &$\OO$, recompute all & 1.87 &   4.73 &   7.51 &   9.55 &  12.29\\[0.75ex]
            &&$\OO$, store scalars & 2.17 &   5.95 &  10.40 &  12.97 &  16.97\\[0.75ex]
            &&$\OO$, store tensors & 1.82 &   5.62 &  11.16 &  15.20 &  19.02\\[0.75ex]
   			&&$\Ot$, recompute all & 1.64 &   4.06 &   6.02 &   7.39 &   8.63\\[0.75ex]
            &&$\Ot$, store scalars & 1.84 &   4.71 &   7.94 &   9.28 &  11.08\\[0.75ex]
            &&$\Ot$, store tensors & 2.01 &   5.97 &  10.42 &  14.11 &  17.50\\[0.75ex]
			\hline
		\end{tabular}
%
	}
\end{table}
\begin{table}
	\centering
	\caption{Speed-up of the tensor-storing variants ($hp$MG-MF-MP-$\OO$/$\Ot$-tensor or AMG-$\OO$/$\Ot$-tensor) over the scalar-storing and recompute all variants ($hp$MG-MF-MP-$\OO$/$\Ot$-none/scalar or AMG-$\OO$/$\Ot$-none/scalar). Computed from data displayed in Fig.~\ref{fig:iliac_bifurcation_performance}.}
	{
		\scriptsize
		\label{tab:iliac_performance_comparison_tensor_vs_scalar_or_recompute_all} 
		\begin{tabular}{|| c c | l || c c c c c||}
			\hline
			\multicolumn{3}{||r||}{}&&&&&\\[-2.0ex]
			\multicolumn{3}{||r||}{polynomial degree $p$}&1&2&3&4&5\\
			\hline\hline
			&&&&&&&\\[-1.25ex]
			\hspace*{-5mm}
			\multirow{6}{*}{\begin{turn}{90}\hspace{-4mm}speed-up\end{turn}}
			\hspace*{-5mm}
			&
			\hspace*{-3mm}
			\multirow{6}{*}{\begin{turn}{90}\hspace{-4mm}AMG\end{turn}}
			&$\OO$, recompute all  & 1.07 & 1.06 & 1.06 & 1.06 & 1.04\\[0.75ex]
			&&$\OO$, store scalars & 1.02 & 1.03 & 1.04 & 1.02 & 1.01\\[0.75ex]
			&&$\OO$, store tensors & 1.00 & 1.00 & 1.00 & 1.00 & 1.00\\[0.75ex]
			&&$\Ot$, recompute all & 1.15 & 1.09 & 1.08 & 1.05 & 1.03\\[0.75ex]
			&&$\Ot$, store scalars & 1.08 & 1.06 & 1.06 & 1.04 & 1.02\\[0.75ex]
			&&$\Ot$, store tensors & 1.00 & 1.00 & 1.00 & 1.00 & 1.00\\[0.75ex]
			\hline\hline
			&&&&&&&\\[-1.25ex]
			\hspace*{-5mm}
			\multirow{6}{*}{\begin{turn}{90}\hspace{-4mm}speed-up\end{turn}}
			\hspace*{-5mm}
			&
			\hspace*{-3mm}
			\multirow{7}{*}{\begin{turn}{90}\hspace{-1mm}$hp$-MG\end{turn}}
			&$\OO$, recompute all  & 1.04 & 1.26 & 1.58 & 1.69 & 1.61\\[0.75ex]
			&&$\OO$, store scalars & 0.86 & 0.98 & 1.12 & 1.20 & 1.13\\[0.75ex]
			&&$\OO$, store tensors & 1.00 & 1.00 & 1.00 & 1.00 & 1.00\\[0.75ex]
			&&$\Ot$, recompute all & 1.42 & 1.61 & 1.86 & 2.01 & 2.09\\[0.75ex]
			&&$\Ot$, store scalars & 1.18 & 1.34 & 1.39 & 1.59 & 1.62\\[0.75ex]
			&&$\Ot$, store tensors & 1.00 & 1.00 & 1.00 & 1.00 & 1.00\\[0.75ex]            
			\hline
		\end{tabular}
		%
	}
\end{table}
\begin{table}
	\centering
	\caption{Speed-up integrating over the material configuration (variants indicated with $\OO$) rather than over the spatial configuration (variants indicated with $\Ot$). Computed from data displayed in Fig.~\ref{fig:iliac_bifurcation_performance}.}
	{
		\scriptsize
		\label{tab:iliac_performance_comparison_material_vs_spatial} 
		\begin{tabular}{|| c c | l || c c c c c||}
			\hline
			\multicolumn{3}{||r||}{}&&&&&\\[-2.0ex]
			\multicolumn{3}{||r||}{polynomial degree $p$}&1&2&3&4&5\\
			\hline\hline
			&&&&&&&\\[-1.25ex]
			\hspace*{-5mm}
			\multirow{6}{*}{\begin{turn}{90}\hspace{-4mm}speed-up $\OO$/$\Ot$ \end{turn}}
			\hspace*{-5mm}
			&
			\hspace*{-3mm}
			\multirow{3}{*}{\begin{turn}{90}\scalebox{1.00}{\hspace{0mm}$hp$-MG~\,}\end{turn}}
			&recompute all  & 1.29 & 1.21 & 1.26 & 1.16 & 1.14\\[0.75ex]
			&&store scalars & 1.30 & 1.30 & 1.33 & 1.28 & 1.24\\[0.75ex]
			&&store tensors & 0.95 & 0.95 & 1.07 & 0.97 & 0.87\\[0.75ex]
			\cline{3-8}
			&
			\hspace*{-3mm}
			\multirow{4}{*}{\begin{turn}{90}\hspace{-2mm}AMG\end{turn}}
			&&&&&&\\[-1.5ex]
			&&recompute all & 1.13 & 1.04 & 1.01 & 1.18 & 1.95\\[0.75ex]
			&&store scalars & 1.10 & 1.03 & 1.02 & 1.22 & 1.99\\[0.75ex]
			&&store tensors & 1.05 & 1.01 & 1.00 & 1.19 & 1.96\\[0.75ex]
			\hline
		\end{tabular}	
	}
\end{table}

For the present scenario, using the standard formulations with impaired numerical stability in the small strain limit does not impact the results significantly in terms of iteration counts (results omitted for brevity), hinting at the effectiveness of the smoothers at reducing the fine-scale errors introduced by numerical round-off. 
This observation, however, does not extend to other scenarios straight-forwardly and cannot be generalized. Since performance is not affected tremendously, stable reformulations are thus to be preferred, while further benefits in the light of low-precision arithmetic are to be investigated in the future.

In a final comparison, we investigate the relative throughput of double-precision matrix-free and matrix-based $hp$-multigrid variants, the latter of which also consider matrices for the matrix-vector products on all levels for this comparison, taking the matrix-free mixed-precision $hp$-multigrid solver as baseline. Integration over the reference configuration and storing tensorial quantities are considered. As indicated in Tab.~\ref{tab:iliac_performance_matrix_based_matrix_free}, iterations are independent of the choice of mixed/double precision (and of course for matrix-free/matrix-based variants). The throughput using a matrix-free double precision variant is roughly 0.60 of the mixed-precision equivalent, while a matrix-based implementation achieves 0.40 of the throughput for $p=1$, which drastically decreases 
for higher polynomial degrees 
in the present example. These factors match the ones presented in Sec.~\ref{sec:results_performance_block}. For a mixed-precision matrix-based variant, similar trends as shown here for double precision are to be expected. Again, this comparison underlines that for lower order finite elements, matrix-free and matrix-based solution strategies are competitive. 
These results indicate that for higher polynomial degrees, assembled sparse matrices are a poor format for achieving high performance in this application.
Combining Tab.~\ref{tab:iliac_performance_matrix_based_matrix_free} and Fig.~\ref{fig:iliac_bifurcation_performance}, we deduce that for a problem with the same number of DoFs 
using $p=1$ and a matrix-based approach (1.1 MDoF/s $\times$ 0.4) vs. $p=2,3,4$ and a matrix-free approach (3.3 MDoF/s $\times$ 0.6), the latter is roughly 4.5 times faster, and using a single-precision preconditioner for the latter (3.3 MDoF/s $\times$ 1.0), this factor increases to 7.5.
\begin{table}
	\centering
	\caption{Relative throughput of $hp$-multigrid solvers using double precision in matrix-free and matrix-based settings, taking the matrix-free mixed-precision version as baseline (see Tab.~\ref{tab:iliac_algorithm_variants}). Integration over the reference configuration and storing tensorial quantities are considered. Iteration counts are identical for all variants.}
	{
		\scriptsize
		\label{tab:iliac_performance_matrix_based_matrix_free} 
		\begin{tabular}{||l || c c c c c||}
			\hline
			&&&&&\\[-1.5ex]
			\multicolumn{1}{||r||}{polynomial degree $p$}
			&1&2&3&4&5\\[0.75ex]
			\hline\hline
			&&&&&\\[-1.25ex]
			rel. throughput, mixed precision, matrix-free
			($hp$MG-MF-MP-$\OO$-tensor)
			&1.00&1.00&1.00&1.00&1.00
			\\[0.75ex]
			rel. throughput, double precision, matrix-free
			($hp$MG-MF-DP-$\OO$-tensor)
			&0.57&0.57&0.62&0.58&0.58
			\\[0.75ex]
			rel. throughput, double precision, matrix-based
			($hp$MG-MB-DP-$\OO$-tensor)
			&0.40&0.14&0.07&0.05&0.04
			\\[0.75ex]
			avg. FGMRES iterations per system solve
			&11.2&10.8&9.6&10.2&12.8
			\\[0.75ex]
			\hline
		\end{tabular}	
	}
\end{table}

Closing the discussion of this practical application, it should be noted that the numerical results presented here strongly depend on the tuning of the preconditioner and problem setup, where difficulties were encountered reaching convergence for the AMG-preconditioned solver and higher polynomial degrees, as this preconditioner is not particularly well suited for this scenario. The AMG-preconditioned variant also adopts the matrix-free operator evaluation in the FGMRES solver, such that only the impact of the matrix-free vs.~matrix-based preconditioners can be evaluated. Results for matrix-free vs.~matrix-based approaches are given in Tab.~\ref{tab:iliac_performance_matrix_based_matrix_free}. Hence, all other results do not show the full speed-ups achievable via matrix-free solution strategies. Furthermore, the AMG preconditioner operates in double precision, 
while the matrix-free $hp$-multigrid alternative operates in single precision. Assuming that an AMG preconditioner delivering constant iteration counts exists, factoring out the increasing iteration counts reveals that especially for higher polynomial degrees, the matrix-free variants are significantly faster. Due to the limitations of this comparison, we refrain from extending these observations towards $p=1$, as more involved or better tuned AMG preconditioners might outperform the matrix-free variant for this case, while factors of larger than 4 and up to 19 (for $p=2,3,4,5$) might be much harder to compensate.
Overall, this example represents practically relevant scenarios, and the results shown are hence very promising beyond academic setups. Furthermore, we want to emphasize at this point that choosing a Poisson's ratio $\nu \rightarrow 0.5$ (fully incompressible case) and hence $\kappa_b\rightarrow\infty$ impacts performance significantly through ill-conditioning of the linear system, 
which is a well-studied limitation of the purely displacement-based formulations in general. For such scenarios, the problem formulation itself, the restriction and prolongation operators, and the smoothers might need to be adapted, while the presented combination of methods is an excellent starting point as demonstrated.

\section{Summary and Concluding Remarks}
\label{sec:summary_conclusion}

This work presents a matrix-free finite element solver for finite-strain elasticity, where several variants for fast and numerically stable integration are discussed. 
We devise stable reformulations of the classical continuum mechanical relations for the anisotropic hyperelastic model for vascular tissue by Holzapfel~et~al.~\cite{Holzapfel2015b}, which are demonstrated in a simplified forward stability test to not suffer from excessive round-off errors in the small strain limit. 
The results also encompass stable reformulations of the directional derivatives for this fiber model as well as for compressible and nearly incompressible neo-Hookean models extending the work by~\cite{Shakeri2024} in this regard. 

We further discuss variants integrating over the material configuration or, alternatively, over the spatial configuration and precompute and store data in integration points extending work by~\cite{Davydov2020} towards more complex constitutive models. 
Changing the domain of integration alters the integrals to be evaluated, where, depending on the material model and precomputing strategy considered, any of the two integration approaches might turn out to be favorable. 

In the presented tests, integrating over the reference configuration and storing scalars turned out to be faster in most cases, as the constitutive models used here feature complex terms for scalar quantities and many tensor-valued operations involve the current iterate and thus cannot be precomputed. 
Comparing matrix-free and matrix-based preconditioners, we observe increased robustness with respect to the polynomial degree $p$ and significantly reduced time to solution for the matrix-free approach using higher polynomial degrees $p>1$ due to increased iteration counts using an AMG preconditioner in a black-box fashion.

Even for linear elements, where the matrix-based AMG preconditioner is found to perform well, the matrix-free approach was found to yield speed-ups of 1.34--1.78 in the present setup. For higher polynomial degrees, speed-up factors of 3.68--19.02 have been recorded. These results are primarily due to faster operator evaluation, i.e., matrix-vector products, in the matrix-free case.
Here, cache effects might play a central role in the comparison as well, as the problem is purposely chosen to be of engineering size. However, the presented results showcase one particular problem size, while optimized caching strategies go beyond the scope of this contribution.
The presented results have not exhaustively analyzed AMG settings tailored to finite-strain hyperelasticity. For example, more sophisticated smoothers or coarsening strategies, not available in the matrix-free $hp$-multigrid solver, have not been considered. In consequence, the present results have to be interpreted with caution, and further research would be necessary to quantify these options.
Nonetheless, factors of 3.68--19.02 for $p=2,3,4,5$ might be difficult to overcome.

Altogether, the matrix-free finite element solvers for finite-strain problems presented in this work show excellent properties also when tackling challenging real-world applications. 
The additional implementation effort of matrix-free methods and the different variants discussed in the present work appears justified in light of the significant speed-up that can be achieved over matrix-based methods, in particular when considering higher-order finite element methods, which additionally counteracts locking effects~\cite{Heisserer2008, Suri1995, Suri1996, Radtke2017}. 

While the present results were recorded on a particular hardware, the computational models quantifying the memory transfer and arithmetic work
allow for predictions also on evolving hardware: In general, moderate polynomial degrees $p=2,\ldots,6$ give the most favorable memory access per DoF. In consequence, the best performance can also be expected in this regime for memory-limited cases. As the proposed models are relatively heavy on operations at quadrature points with many register spills to local memory, architectures with limited cache sizes such as GPUs can be expected to benefit even more from the storage option of tensors than the results presented here.

Ongoing developments are centered around further improving the routines to update quadrature point data to experiment with Jacobian-free Newton--Krylov methods, which avoid the formulation and optimization of the linearized operator, but require fast residual evaluation (see, e.g.~\cite{Munch2024}). The multigrid solver can be improved following many ideas: First, resolving the incompressibility constraint via mixed formulations instead of the penalty-based approach prevents locking for coarse discretizations and may hence better approximate the solution on coarser multigrid levels. Second, structure-preserving strategies regarding the finite element mapping and restriction/prolongation operators might significantly improve robustness of the solver when facing large deformations. Similar concepts might also be used to derive tailored smoothers for the fully incompressible case. Third, the outer FGMRES solver might be replaced by a CG method once the preconditioner is constant, 
which is the case for the specific iliac bifurcation example shown here, 
but the principal solver design allows for rather large, general coarse-level problems, 
without relying too much on problem-dependent tuning or a fixed number of AMG V-cycles, 
a fixed number of AMG-preconditioned Krylov solver iterations or low-tolerance preconditioned iterative solvers.

\section{Acknowledgements}
The authors gratefully acknowledge the scientific support and HPC resources provided by the Erlangen National High Performance Computing Center (NHR@FAU) of the Friedrich-Alexander-Universit{\"a}t Erlangen-N{\"u}rnberg (FAU). NHR funding is provided by federal and Bavarian state authorities. NHR@FAU hardware is partially funded by the German Research Foundation (DFG)~--~440719683. Moreover, this work was partially supported by the German Ministry of Education and Research through project ``PDExa: Optimized software methods for solving partial differential equations on exascale supercomputers'', grant agreement no. 16ME0637K. 

\appendix

\section{Material Models: Directional Derivatives}
\label{sec:appendix}

Providing the strain energy density suffices to define a hyperelastic material model. Therefore, we present further derivations here to not clutter the main document. For the compressible neo-Hookean model~\eqref{eqn:NH_compressible}, the directional derivative of the second Piola--Kirchoff stress reads
\begin{gather*}
	\DDu\te{S}_\mathrm{cNH} 
	= 
	-
	\left(
	\mu-2\lambda\ln {J}
	\right)
	\DDu \inv{\te{C}}
	+
	2 \lambda
	\left(\nicefrac{1}{{J}} \, \DDu J\right)
	\inv{{\te{C}}}
	,
\end{gather*}
with the following terms not being specific to the material model:
\begin{gather*}
	\DDu\inv{\te{C}}
	=
	\left(\DDu \inv{\te{F}} \right) \invt{{\te{F}}}
	+
	\inv{{\te{F}}} \DDu \invt{\te{F}}
	=
	2
	\left[
	\left(
	\DDu \inv{{\te{F}}}
	\right)
	\invt{\te{F}} 
	\right]^\mathrm{S}
	,
	\nonumber
	\\
	\DDu \te{F} 
	= 
	\Grad\Delta\ve{u}
	,
	\quad
	\DDu\inv{\te{F}}
	=
	-\inv{{\te{F}}} \left(\Grad \Delta \ve{u}\right) \inv{{\te{F}}}
	,	
	\quad
	\nicefrac{1}{{J}}
	\,
	\DDu J = \tr\left(\inv{{\te{F}}} \Grad \Delta \ve{u}\right)
	.
\end{gather*}
Likewise, for nearly incompressible neo-Hookean materials~\eqref{eqn:NH_nearly_incompressible}, we obtain
\begin{align}
	\DDu \te{S}_\mathrm{iNH}
	&
	=
	-
	\nicefrac{2\mu}{3} \Jpow \left(\nicefrac{1}{{J}} \, \DDu J\right)
	\, 
	\id
	+
	\left[
	\nicefrac{\kappa_b}{2}
	\left({J}^2-1\right)
	-
	\nicefrac{\mu}{3} \Jpow \, {I}_1
	\right]
	\DDu
	\inv{\te{C}}
	\nonumber
	\\
	&
	\phantom{=}\,\,
	+
	\left[
	\left(
	\nicefrac{1}{{J}}
	\,
	\DDu J
	\right)
	\left(
	\nicefrac{2\mu}{9} \Jpow {I}_1
	+
	\kappa_b {J}^2
	\right)
	-
	\nicefrac{\mu}{3} \Jpow \DDu I_1		
	\right]
	\inv{{\te{C}}}
	,
	\label{eqn:NH_DDuS}
\end{align}
with $
	\DDu I_1 
	= 
	\tr
	\left( 
	\DDu \te{F}^\top {\te{F}} 
	+ 
	{\te{F}}^\top \DDu \te{F} 
	\right)  
	= 
	2 \,\tr 
	\left( 
	{\te{F}}^\top \Grad\Delta\ve{u} 
	\right)
$. 
For the fiber model~\cite{Holzapfel2015b}, we sum contributions from the nearly incompressible neo-Hookean ground material~\eqref{eqn:NH_DDuS} and collagen fibers to obtain
\begin{gather}
	\DDu \te{S}_\mathrm{fiber}
	=
	\DDu \te{S}_\mathrm{iNH}
	+
	\DDu \te{S}_\mathrm{c}
	=
	\DDu \te{S}_\mathrm{iNH}
	+
	\sum_{i=4,6}
	2 k_1 \exp\left(k_2 E_i^2\right) \left( 2 k_2 E_i^2 + 1 \right) 
	\left(
	\te{H}_i : \DDu \te{C} 
	\right) \te{H}_i 
	,
	\label{eqn:DduS_fiber}
\end{gather}
where we use
\begin{gather*}
	\DDu E_i = \DDu (\te{H}_i : \te{C}) = \te{H}_i : \DDu\te{C}
	,
	\quad
	\text{and}
	\quad
	\DDu\te{C} = 2 \left( \te{F}^\top \Grad \Delta \ve{u}\right)^\mathrm{S}.
\end{gather*}
Turning our attention to the approach integrating over the spatial configuration, we have
\begin{equation*}
	J\tefour{c}
	\eqq 
	\chi
	\left( 
	4
	\frac{\partial^2 \Psi(\te{C})}{\partial\te{C}\motimes\partial\te{C}} 
	\right)
	,
\end{equation*}
where the contravariant push-forward of a fourth-order tensor is defined as
$
\chi(\cdot)_{ijkl}
\eqq
\te{F}_{iA}
\te{F}_{jB}
(\cdot)_{ABCD}
\te{F}_{kC}
\te{F}_{lD}
$ using Einstein's summation convention. 
For the compressible neo-Hookean model~\eqref{eqn:NH_compressible}, we have
\begin{align*}
	\te{\tau}_\mathrm{cNH}
	=
	\mu \, \te{b} - \left( \mu - 2\lambda\ln J \right) \id
	,
	\quad
	\text{and}
	\quad
	\frac{\partial^2 \Psi_\mathrm{cNH}(\te{C})}{\partial\te{C}\motimes\partial\te{C}}
	=
	\nicefrac{1}{2}
	\left(
	\mu - 2 \lambda \ln J
	\right)
	\inv{\te{C}} \modot \inv{\te{C}}
	+
	\nicefrac{\lambda}{2}
	\,
	\inv{\te{C}}\motimes\inv{\te{C}}
	,
\end{align*}
with the left Cauchy--Green tensor $\te{b}\eqq\te{F}\,\te{F}^\top$ and using $\frac{\partial\inv{\te{C}}}{\partial \te{C}} = - \inv{\te{C}} \modot \inv{\te{C}}$ (see \cite{Holzapfel2000}), 
\begin{equation*}
	\left(
	\inv{\te{C}} \modot \inv{\te{C}}
	\right)_{ABCD}
	\eqq
	- \frac{\partial\te{C}_{AB}^{-1}}{\partial \te{C}_{CD}} 
	= 
	\nicefrac{1}{2}
	\left(
	\te{C}_{AC}^{-1}
	\te{C}_{BD}^{-1}
	+
	\te{C}_{AD}^{-1}
	\te{C}_{BC}^{-1}
	\right)
	,
\end{equation*}
such that we further have
\begin{align}
	J\tefour{c}_\mathrm{cNH} : (\cdot)^\mathrm{S}
	&=
	2
	\left(
	\mu - 2 \lambda \ln J
	\right)
	\chi 
	\left( 
	\inv{\te{C}} \modot \inv{\te{C}}
	\right)
	:
	(\cdot)^\mathrm{S}
	+
	2 \lambda \,
	\chi 
	\left( 
	\inv{\te{C}} \motimes \inv{\te{C}} 
	\right)
	:
	(\cdot)^\mathrm{S}
	\nonumber
	\\
	&=
	2
	\left(
	\mu - 2 \lambda \ln J
	\right)
	\tefour{S}
	:
	(\cdot)^\mathrm{S}
	+
	2 \lambda \,
	\left( 
	\id \motimes \id
	\right)
	:
	(\cdot)^\mathrm{S}
	\nonumber
	\\
	&=
	2
	\left(
	\mu - 2 \lambda \ln J
	\right)
	(\cdot)^\mathrm{S}
	+
	2 \lambda \,
	\tr(\cdot)
	\,
	\id
	,	
	\label{eqn:JC_contract_compressible_NH}
\end{align}
using
$
	\chi\left( \inv{\te{C}} \modot \inv{\te{C}} \right)_{ijkl} 
	= 
	\tefour{S}_{ijkl} \eqq \nicefrac{1}{2}\left(\delta_{ik}\delta_{jl} + \delta_{il}\delta_{jk}\right)
	,
	\quad
	\text{and}
	\quad
	\chi \left(\inv{\te{C}}\motimes \inv{\te{C}}\right) 
	= 
	\id \motimes \id
$
, and where the symmetry of the argument is exploited in the inner product with the fourth-order tensor $\tefour{S}$. For the nearly incompressible neo-Hookean model~\eqref{eqn:NH_nearly_incompressible}, similar steps lead to 
\begin{align*}
	\te{\tau}_\mathrm{iNH}
	&=
	\mu \Jpow \te{b}	
	+ 
	\left[ \nicefrac{\kappa_b}{2} (J^2-1) - \nicefrac{\mu}{3} \Jpow I_1 \right] \id
	,
	\\
	\text{and}
	\quad
	\frac{\partial^2 \Psi_\mathrm{iNH}(\te{C})}{\partial\te{C}\motimes\partial\te{C}}
	&=
	-
	\nicefrac{\mu}{3} \Jpow  
	\id \motimes \inv{\te{C}} 
	+
	\left[
	\nicefrac{\mu}{6} \Jpow I_1 - \nicefrac{\kappa_b}{4} (J^2-1)
	\right]
	\inv{\te{C}} \modot \inv{\te{C}}
	+
	\left(
	\nicefrac{\mu}{18} \Jpow I_1 + \nicefrac{\kappa_b}{4} J^2
	\right)
	\inv{\te{C}}\motimes\inv{\te{C}}
	,
\end{align*}
which together with
$
	\chi
	\left( 
	\id \motimes \inv{\te{C}} 
	\right) 
	: 
	(\cdot) 
	= 
	\left(
	\te{C} \motimes \id
	\right) 
	: 
	(\cdot) 
	= 
	\te{C}
	\,
	\left[
	\id:(\cdot)
	\right] 
	=
	\tr(\cdot)
	\, \te{C}
$ 
then finally gives a compact expression similar to Eqn.~\eqref{eqn:JC_contract_compressible_NH} only involving second-order tensors,
\begin{align*}
	J \tefour{c}_\mathrm{iNH} : (\cdot)^\mathrm{S}
	&=
	-
	\nicefrac{4\mu}{3} \Jpow  
	\te{C} \, \tr (\cdot)
	+
	\left[
	\nicefrac{2\mu}{3} \Jpow I_1 - \kappa_b (J^2-1)
	\right]
	(\cdot)^\mathrm{S}
	+
	\left(
	\nicefrac{2\mu}{9} \Jpow I_1 + \kappa_b J^2
	\right)
	\tr(\cdot) \, \id
	,
\end{align*}
again exploiting symmetry of the argument in the inner product with the fourth-order symmetric identity tensor $\tefour{S}$. For the model including non-symmetrically dispersed fibers~\cite{Holzapfel2015b}, we have
\begin{align*}
	\te{\tau}_\mathrm{fiber}
	&=
	\te{\tau}_\mathrm{iNH}
	+
	\te{\tau}_\mathrm{c}
	=
	\te{\tau}_\mathrm{iNH}
	+
	\sum_{i=4,6}
	2 k_1 \exp\left(k_2 E_i^2\right) E_i \te{F} \, \te{H}_i \, \te{F}^\top
	,
	\\
	\text{and}
	\quad
	\frac{\partial^2 \Psi_\mathrm{fiber}(\te{C})}{\partial\te{C}\motimes\partial\te{C}}
	&=
	\frac{\partial^2 \Psi_\mathrm{iNH}(\te{C})}{\partial\te{C}\motimes\partial\te{C}}
	+
	\sum_{i=4,6}
	k_1 \exp\left(k_2 E_i^2\right) \left(2 k_2 E_i^2 + 1\right)
	\te{H}_i \otimes \te{H}_i
	.
\end{align*}
The directional derivative of the fiber contribution is easiest found by pushing forward the fiber contribution in~\eqref{eqn:DduS_fiber} and using $\Grad\Delta \ve{u} = (\grad \Delta \ve{u}) \, \te{F}$, yielding
{
	\small
	\begin{align*}
		&
		\sum_{i=4,6}
		2 k_1 \exp\left(k_2 E_i^2\right) \left( 2 k_2 E_i^2 + 1 \right) 
		\left[
		\te{H}_i :
		\left(
		\te{F}^\top (\grad \Delta \ve{u}) \, \te{F}
		+
		\te{F}^\top (\grad \Delta \ve{u})^\top \, \te{F}
		\right) 
		\right] \te{F} \, \te{H}_i \, \te{F}^\top
		\\
		=&
		\sum_{i=4,6}
		2 k_1 \exp\left(k_2 E_i^2\right) \left( 2 k_2 E_i^2 + 1 \right) 
		2
		\left[
		\left(
		\te{F} \te{H}_i \, \te{F}^\top
		\right)
		:
		\grad \Delta \ve{u}
		\right] \te{F} \, \te{H}_i \, \te{F}^\top
	\end{align*}
}%
For symmetric arguments, this then finally leads to
\begin{gather*}
	J \tefour{c}_\mathrm{fiber} : (\cdot)^\mathrm{S}
	=
	J \tefour{c}_\mathrm{iNH} : (\cdot)^\mathrm{S}
	%
	+
	\sum_{i=4,6}
	4 k_1 \exp\left(k_2 E_i^2\right) \left(2 k_2 E_i^2 + 1\right)
	\left[
	\left(
	\te{F} \te{H}_i \, \te{F}^\top
	\right)
	:
	(\cdot)
	\right] \te{F} \, \te{H}_i \, \te{F}^\top
	.
\end{gather*}

 \bibliographystyle{unsrtnat} 
 \bibliography{bibliography}

\end{document}